%% file: arxiv-final.tex
\documentclass[acmtog,screen,nonacm]{acmart}

\newcommand{\Z}{\mathbb{Z}}

\newcommand{\rev}[1]{#1}

\newcommand{\norm}[1]{\left\|#1\right\|}
\newcommand{\abs}[1]{\left|#1\right|}
\newcommand{\inner}[1]{\left\langle#1\right\rangle}

\newcommand{\ppd}[1]{\frac{\partial}{\partial #1}}

\definecolor{lightgreen}{rgb}{.01,.64,.73}

\usepackage[normalem]{ulem}
\usepackage{tikz}
\usepackage{caption}
\usepackage{wrapfig}
\usepackage{bm}
\usepackage{subcaption}
\usepackage{multirow}
\usepackage{booktabs}
\usepackage{array}
\usepackage{pdfpages}
\usepackage{dblfloatfix}
\usepackage{xcolor}
\usepackage[ruled,linesnumbered,vlined]{algorithm2e}

\AtBeginDocument{%
  }

\setcopyright{acmlicensed}
\copyrightyear{2026}
\acmYear{2026}
\acmDOI{XXXXXXX.XXXXXXX}

\acmISBN{978-1-4503-XXXX-X/18/06}

\acmSubmissionID{128}

\begin{document}

\title{Efficient B-Spline Finite Elements for Cloth Simulation}

\author{Yuqi Meng}
\affiliation{%
  \institution{Carnegie Mellon University}
  \country{USA}}
\affiliation{%
  \institution{University of Utah}
  \country{USA}}

\author{Yihao Shi}
\affiliation{%
  \institution{Carnegie Mellon University}
  \country{USA}}
\affiliation{%
  \institution{Zhejiang University}
  \country{China}}

\author{Kemeng Huang}
\affiliation{%
  \institution{Carnegie Mellon University}
  \country{USA}}
\affiliation{%
  \institution{University of Hong Kong}
  \country{China}}

\author{Zixuan Lu}
\affiliation{
  \institution{University of Utah}
  \country{USA}}
  
\author{Ning Guo}
\affiliation{%
  \institution{Zhejiang University}
  \country{China}}
  
\author{Taku Komura}
\affiliation{%
  \institution{University of Hong Kong}
  \country{China}}

\author{Yin Yang}
\affiliation{%
  \institution{University of Utah}
  \country{USA}}

\author{Minchen Li}
\affiliation{%
  \institution{Carnegie Mellon University}
  \country{USA}}
\affiliation{%
  \institution{Genesis AI}
  \country{USA}}
\email{minchernl@gmail.com}


\begin{abstract}
  We present an efficient B-spline finite element method (FEM) for cloth simulation. While higher-order FEM has long promised higher accuracy, its adoption in cloth simulators has been limited by its larger computational costs while generating results with similar visual quality. Our contribution is a full algorithmic pipeline that makes cloth simulation using quadratic B-spline surfaces faster than standard linear FEM in practice while consistently improving accuracy and visual fidelity. Using quadratic B-spline basis functions, we obtain a globally $C^1$-continuous displacement field that supports consistent discretization of both membrane and bending energies, effectively reducing locking artifacts and mesh dependence common to linear elements. To close the performance gap, we introduce a reduced integration scheme that separately optimizes quadrature rules for membrane and bending energies, an accelerated Hessian assembly procedure tailored to the spline structure, and an optimized linear solver based on partial factorization. Together, these optimizations make high-order, smooth cloth simulation competitive at scale, yielding an average $2\times$ speedup over linear FEM in our tests. Extensive experiments demonstrate improved accuracy, wrinkle detail, and robustness, including contact-rich scenarios, relative to linear FEM and recent higher-order approaches. Our method enables realistic wrinkling dynamics across a wide range of material parameters and supports practical garment animation, providing a new promising spatial discretization for high-quality cloth simulation.
\end{abstract}



\begin{teaserfigure}
  \includegraphics[width=0.99\linewidth]{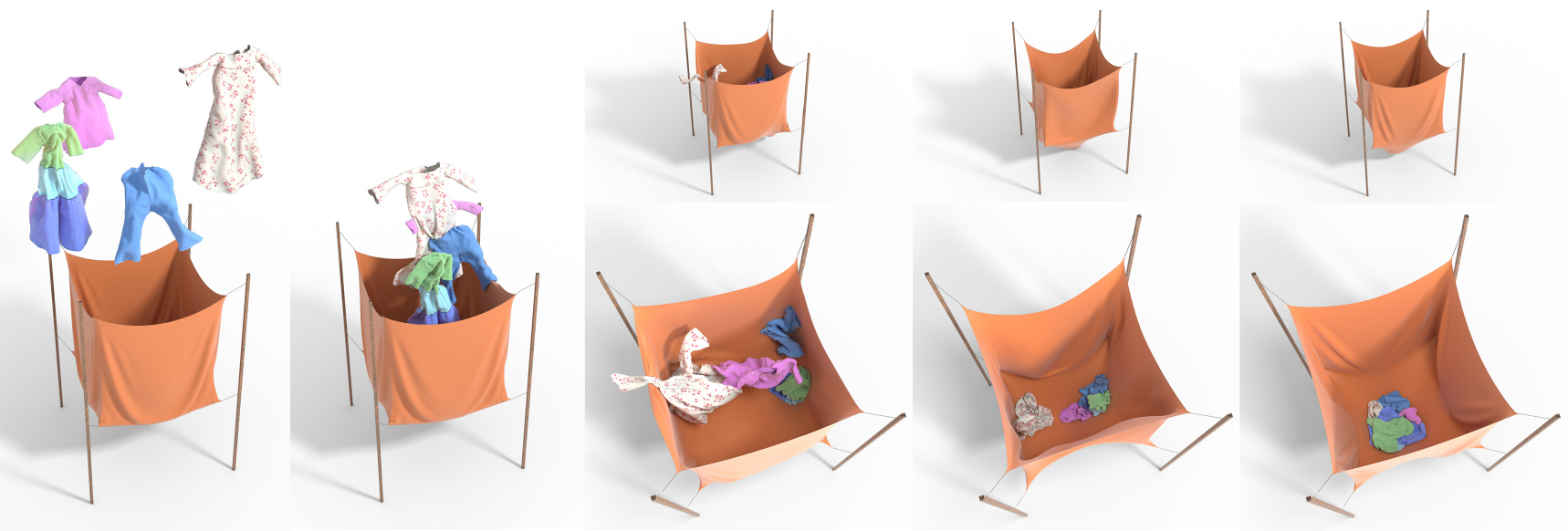}
  \vspace{-0.2cm}
  \caption{\textbf{Cloth basket.} Four garments are launched toward a basket with fixed corners, collide with each other, and accumulate inside. This scene contains 51K control points and is simulated with timestep $\Delta t = 0.01$~s. Both the garments and the basket faces are modeled as B-spline surfaces. Our method captures the rich wrinkling behavior that emerges as garments clash and fold against each other, on both rectangular and non-rectangular meshes, while IPC integration ensures that contact is properly handled throughout the simulation.}
  \label{fig: teaser}
\end{teaserfigure}


\maketitle

\section{Introduction}
\label{sec: introduction}

In computer graphics, cloth simulation has witnessed remarkable advancements over the past decades since the pioneering work of~\citet{terzopoulos1987elastically}. These developments span a wide range of fundamental techniques, including stable time integration, efficient and realistic simulation of bending deformations, robust contact handling, fast computation on GPUs, and parameter estimation from real-world captures. These efforts have further enabled various practical applications such as character animation and computer-aided fashion design. Among the many approaches explored, triangle meshes combined with linear finite element methods (FEM) have been established as the predominant spatial discretization scheme, owing to their simplicity and accessibility.

Although recent improvements to this classic model have demonstrated its ability to capture rich wrinkles under various mechanical conditions, including stretching, shearing, and complex frictional contact, linear FEM inherently struggles to accurately model complex curved surfaces. This limitation stems from its piecewise linear surface representation, which only has $C^0$-continuity across triangle boundaries. As a result, fine mesh resolution is often needed to properly model smooth and/or high-frequency wrinkles. Additionally, issues can arise when discretizing membrane energy: even under isometric surface deformations, mesh elements can still distort locally, introducing spurious resistance forces and leading to artificially stiffened bending behavior -- a phenomenon known as \textit{membrane locking} from which linear FEM suffers significantly. Addressing this often requires modifying the energy model (e.g., softening membrane stiffness and adding strain-limiting energies) and, in some cases, even remeshing.

In addition, on a piecewise linear triangle mesh, second- and higher-order spatial derivatives of the displacement field are either undefined or $0$. This necessitates special treatments for discretizing bending energies, such as relying on dihedral angles, which complicates the computation of forces and their derivatives, and can introduce configurations with singularities. These issues can further cause large deviations from analytic solutions and poor convergence rates under spatial refinement. Moreover, discretizing bending energy onto mesh edges can lead to mesh dependency problems. While increasing the mesh resolution can reduce discretization error and better handle complex configurations, it also significantly increases computational costs.

Recently, to address the limitations arising from piecewise linear surface representations, researchers in graphics have explored using higher-order shape functions for displacement field interpolation in cloth or thin shell simulations. \citet{SecondOrderFEM2023} applied quadratic triangle elements, achieving reduced mesh dependency and improved wrinkle simulation at comparable cost to linear FEM. However, their quadratic elements remain only $C^0$-continuous across element boundaries, still requiring special treatments for bending energy discretization. In contrast, \citet{ni2024simulating} employed bicubic Hermite elements and explicitly enforced $C^1$-continuity at element boundaries to achieve a globally smooth displacement field. Their method demonstrated improved efficiency for simulating comparable wrinkling behaviors but requires displacement derivatives to serve as additional degrees of freedom (DOF), leading to a much denser Hessian (due to larger local stencils) and more quadrature points ($4\times4$ per element) for force evaluation.

We propose using quadratic B-spline basis functions for displacement interpolation, which naturally provide global smoothness while introducing only a modest increase in stencil size and quadrature points. This leads to a unique balance between efficiency and accuracy. The globally smooth displacement field enables consistent and accurate discretization of both membrane and bending energies throughout the simulation domain, effectively mitigating locking artifacts and mesh dependency.
To further improve performance, we introduce a novel reduced integration scheme that separately optimizes quadrature placement for membrane and bending energies, applying alternating patterns for membrane and a one-point quadrature for bending. Combined with a fast, parallel Hessian assembly strategy, our method achieves high computational efficiency while maintaining accuracy, under scenarios with or without contact.

We validate the effectiveness of our method through a comprehensive set of experiments, including comparisons against linear FEM and \citet{SecondOrderFEM2023} in terms of accuracy, results quality, and efficiency. On average, our method is up to $2\times$ faster than linear FEM for comparable visual quality, and orders of magnitude faster when targeting the same level of accuracy. In comparison with other high-order FEM methods, it achieves substantially better performance and convergence; for example, it is on average $5\times$ faster than \citet{SecondOrderFEM2023} for the same number of DOFs, and $2\times\!\sim\!20\times$ faster than~\citet{ni2024simulating}. We also conduct detailed ablation studies on our efficient reduced integration scheme and parallel Hessian assembly strategy. Additionally, we demonstrate the ability of our method to simulate complex wrinkling dynamics across a range of material parameters, highlighting the potential of our approach to advance cloth simulation with a new paradigm in spatial discretization. Source code and data for our paper is available at \url{https://simulation-intelligence.github.io/BS-Cloth/}.

Our main contributions are summarized as follows:
\begin{itemize}
    \item A {performant B-spline FEM framework} for cloth simulation with a globally $C^1$-continuous displacement field, enabling consistent discretization of both membrane and bending energies. This effectively mitigates locking and mesh dependency while achieving {higher computational efficiency compared to previous works}.
    \item A novel reduced integration scheme that separately optimizes quadrature rules for membrane and bending energies, minimizing quadrature points and stencil sizes to further improve efficiency without sacrificing accuracy.
    \item {An efficient Hessian assembly scheme with a faster linear solver with partial factorization, ensuring simulation efficiency under varying contact conditions.}
\end{itemize}

\section{Related Work}
\label{sec:related works}

\paragraph{Traditional Cloth Simulation}
Since the pioneering work of~\citet{terzopoulos1987elastically}, deformable shell simulation for materials such as cloth, paper, and flexible metals has received significant attention in the graphics community~\cite{carignan1992dressing,lafleur1991cloth,grinspun2003discrete,chen2023multi,chen2018physical,li2020codimensional,narain2013folding,jiang2017anisotropic,guo2018material,weidner2018eulerian}. Among these efforts, efficiently and realistically modeling the material behavior of cloth remains an enduring research topic. Mass-spring models~\cite{breen1994predicting,provot1995deformation,liu2013fast,jin2017inequality} are fast and simple but often suffer from severe mesh dependency and poor performance in capturing Poisson effects and accurate bending behaviors. In contrast,~\citet{terzopoulos1988deformable} and~\citet{volino2009simple} developed general finite element frameworks for cloth simulation based on elasticity theory, enabling realistic modeling of bending, folding, wrinkling, and interactions with solid objects. To effectively simulate bending behaviors of thin shells on piecewise linear triangle meshes, a series of hinge-based models~\cite{bridson2005simulation,grinspun2003discrete,liang2025corotational} were introduced, utilizing dihedral angles at mesh edges to measure bending deformation. Under the assumptions of a flat rest shape and (near) isometric in-plane deformation, these approaches led to the popular quadratic bending model with linear bending forces~\cite{QuadraticBending2006,wardetzky2007discrete}, which offers greater computational efficiency compared to hinge-based models. In our work, we discretize the quadratic bending model using the B-spline basis and further develop a one-point quadrature rule to enable efficient yet accurate force evaluation.

For time integration, the simplest approach is to use explicit methods \cite{harmon2009asynchronous}; however, due to the highly nonlinear mechanical behavior of cloth and its complex contact interactions, explicit schemes require relatively small time steps to maintain stability. To ensure robustness, implicit integration schemes have been applied \cite{baraff1998large,english2008animating,bridson2005simulation}. These methods allow for larger time steps while maintaining stability, at the cost of solving a nonlinear system at each time step. Many existing approaches adopt penalty-based methods for contact handling due to their simplicity~\cite{guan2012drape,provot1997collision,baraff2003untangling}. However, penalty methods inevitably permit a certain amount of interpenetration, with penetration depth controlled by the penalty stiffness -- a parameter that must be carefully tuned to avoid ill-conditioning and is often problem-specific. More recently, incremental potential contact (IPC)~\cite{Li2020IPC} has been introduced, which uses a barrier energy formulation to guarantee penetration-free elastodynamic simulation, and has been extended to support arbitrary codimensional geometries~\cite{li2020codimensional}. In this work, we base our time integration on the optimization time integration framework used in IPC and apply its contact formulation on an embedded proxy triangle mesh to achieve robust simulation. \rev{Similar to \citet{Witemeyer2021QLB}, we avoid re-factorizing the contact Hessian at every Newton iteration by treating its diagonal and off-diagonal components separately.}

\paragraph{Discontinuous Galerkin}
\rev{Discontinuous Galerkin (DG) methods provide an alternative approach to locking: rather than strictly enforcing inter-element continuity, DG allows discontinuities in the displacement field across element boundaries and recovers compatibility weakly via penalty terms~\cite{arnold2002unified}. By relaxing inter-element continuity, DG methods compute strains locally within each element~\cite{hansbo2002discontinuous}, which avoids the spurious coupling between bending and parasitic membrane or shear modes that causes locking in linear FEM. However, decoupling the displacement field across element boundaries introduces additional degrees of freedom and requires careful tuning of penalty parameters, resulting in larger systems and extra boundary treatment compared to conforming discretizations.}

\paragraph{High-Order FEM}
FEM with linear shape functions relies heavily on high mesh resolution to capture curved surfaces and fine wrinkles. Insufficient resolution can also lead to locking issues, where strain-limiting methods~\cite{provot1995deformation,bridson2002robust} and reduced integration techniques~\cite{kim2005resultant,schwarze2011reduced,cardoso2008enhanced,chen2023multi} have been applied as remedies. {As an extension to linear FEM,~\cite{Longva2020Embedded} embeds the geometry into a coarse background mesh composed of higher-order elements, decoupling geometric detail from mesh resolution, and leveraging classical high-order FEM for improved simulation accuracy. While embedding-based methods enable coarse high-order meshes, they do not fully preserve the original boundary geometry, potentially limiting exact shape representation. Alternatively,} high-order discretization schemes have been explored for cloth animation, achieving higher fidelity and effectively mitigating locking while maintaining computational costs comparable to linear discretizations.
Several early works focused primarily on curved surface modeling~\cite{celniker1991deformable,dey1999curvilinear,witkin1992variational}. {~\citet{Bargteil2014BezierFE} uses quadratic B\'{e}zier elements, which, while providing $C^1$-continuity within each element, do not automatically guarantee global smoothness}. \rev{More recently, large-scale FEM benchmarks~\cite{Schneider2022LargeScale} confirm that quadratic spline elements on structured lattices reduce locking and improve solver efficiency over linear elements.} \citet{SecondOrderFEM2023} adapted hinge-based bending models for quadratic triangle meshes, demonstrating improved accuracy and expressiveness. \citet{loschner2024curved} introduced three-director Cosserat shells for graphics animation, while~\citet{montes2024q3t} proposed simulating elastoplastic surfaces using quadratic through-the-thickness (Q3T) solid shell elements. Despite the more accurate geometry description within elements, these methods still suffer from nonsmooth displacement fields at element boundaries.
Instead,~\citet{ni2024simulating} proposed a framework based on bicubic Hermite elements, achieving global $C^1$-continuity across patches. However, their method requires storing additional derivative information at each node for interpolation, and adopts a 16-point Gauss-Legendre quadrature rule for force integration, resulting in increased computational costs for both Hessian assembly and linear solving. 
B-spline FEM offers an alternative, requiring only field variables for interpolation and fewer quadrature points for force evaluation. Prior work on B-spline FEM has primarily focused on static problems~\cite{kagan1998new} and has not been applied to dynamic cloth simulation. {Oftentimes these methods are also not performant enough, which is the major concern preventing applications from adopting high-order methods.} Inspired by these advances, we propose a B-spline FEM framework for cloth simulation {and design several optimization methods, including a reduced integration scheme, parallel barrier Hessian construction, and a Hessian splitting iterative solver, which together achieve superior performance compared to linear FEM, and strike} a unique balance between accuracy and efficiency.

\paragraph{Subdivision FEM}
Subdivision-based FEM was introduced by \citet{Cirak2000Subdivision} as a way to obtain globally smooth basis functions on general triangle meshes. Using the Loop subdivision scheme, the resulting limit surface yields basis functions that are $C^1$ everywhere and $C^2$ away from extraordinary vertices, without requiring any special parametrization beyond the input triangle mesh. In regular regions these basis functions reproduce three-direction box splines, while neighborhoods of extraordinary vertices are handled through subdivision eigenanalysis.

Subsequent work extended the approach with adaptive refinement and multiresolution techniques~\citep{Grinspun2002AdaptiveSubdivision}, and applied subdivision surfaces to cloth simulation~\citep{Thomaszewski2006Cloth}. Despite their attractive smoothness and geometric flexibility, subdivision elements incur nontrivial computational overhead. Each basis function has wide support, and even a regular Loop triangle involves twelve control points. Near extraordinary vertices or boundaries, basis evaluation requires local dense transformations (e.g., Stam's eigenbasis evaluation), increasing the per-quadrature cost and expanding the effective stencil. As a result, subdivision FEM produces significantly denser stiffness/Hessian matrices than linear FEM, and its computational cost depends on the distribution of extraordinary vertices rather than being strictly mesh-independent.

\paragraph{Isogeometric Analysis}
Isogeometric analysis (IGA) has been extensively studied for simulating geometrically nonlinear shell structures, including the dynamic simulation of cloth~\cite{BSCloth2014,FastProj2023,NonlinearIGA2023,nakashino2020geometrically}, cars~\cite{zhang2017nurbs,kuraishi2022space}, and deformation analysis of membranes~\cite{chen2014explicit,leonetti2018isogeometric,tepole2015isogeometric}. Originally, IGA employed non-uniform rational B-splines (NURBS) as basis functions for geometry representation~\cite{hughes2005isogeometric}, with later developments introducing NURBS variants that enable local mesh refinement~\cite{bazilevs2010isogeometric}. Like B-spline functions, NURBS bases provide flexible geometric modeling and high-order continuity, allowing for smooth solutions with fewer degrees of freedom compared to traditional FEM.
Leveraging these advantages, Kirchhoff-Love (KL) and Reissner-Mindlin shell elements have been rapidly developed within the IGA framework~\cite{kiendl2009isogeometric,benson2010isogeometric}. Both linear and nonlinear elastic deformations of single patches can be accurately and efficiently captured, even under large deformations~\cite{kiendl2009isogeometric,leonetti2018efficient,hosseini2014isogeometric}. Compared to B-splines, NURBS introduce an additional weight for each control point, allowing for more flexible modeling of rest shapes. In this work, we focus on developing an efficient and robust cloth simulator with globally smooth surface representations, and therefore simply adopt B-spline basis functions. Extending our framework to support NURBS would be straightforward.


\section{Spatial and Temporal Discretization}
\label{sec:primitive formalization}

We first introduce the B-spline surface representation in~\autoref{sec:b-spline_elements}, and then derive the incremental potential \cite{kane2000variational} for our B-spline FEM based on implicit Euler time integration within a Lagrangian mechanics framework in~\autoref{sec:govern_eq}.

\subsection{B-spline Surface Representation} \label{sec:b-spline_elements}

\subsubsection{B-spline Basis}

\label{subsubsec: B-spline basis}

Given a parametric space $P \subset \mathbb{R}$, a \emph{knot vector} $\Xi = (\xi_1, \dots, \xi_{n+p+1})$ with $\xi_1 \leq \xi_2 \leq \cdots \leq \xi_{n+p+1}$ is a non-decreasing sequence of coordinates in $P$, where $n$ and $p$ denote the number of B-spline control points and the polynomial order of the B-spline basis functions, respectively. A knot vector is said to be \emph{open} if the first $p+1$ and last $p+1$ knots are repeated, and \emph{uniform} if all non-repeated knots are evenly spaced (e.g., \autoref{fig: open/non-open B-spline curves}). A \emph{knot span} of $\Xi$ is the closed interval $[\xi_i, \xi_{i+1}] \subset P$ for some non-repeated knots $\xi_i, \xi_{i+1} \in \Xi, \ \xi_i \neq \xi_{i+1}$.

A $p$-th order \emph{B-spline} basis function $N_{i,p}: [\xi_1, \xi_{n+p+1}] \to [0,1]$ is defined recursively with respect to the polynomial order, starting with zeroth order (piecewise constant)~\cite{piegl2012nurbs}:
\begin{align}\label{eq: B-spline basis}
\begin{split}
N_{i,0}(\xi) & =
\begin{cases}
1 & \text{if } \xi_i \leq \xi < \xi_{i+1}, \\
0 & \text{otherwise,}
\end{cases} \\
N_{i,p}(\xi) & = \frac{\xi - \xi_i}{\xi_{i+p} - \xi_i} N_{i,p-1}(\xi) + \frac{\xi_{i+p+1} - \xi}{\xi_{i+p+1} - \xi_{i+1}} N_{i+1,p-1}(\xi).
\end{split}
\end{align}

\begingroup
\setlength{\columnsep}{10pt}
\begin{wrapfigure}[24]{l}{0.5\linewidth}
  \vspace*{-12pt}
  \includegraphics[width=\linewidth]{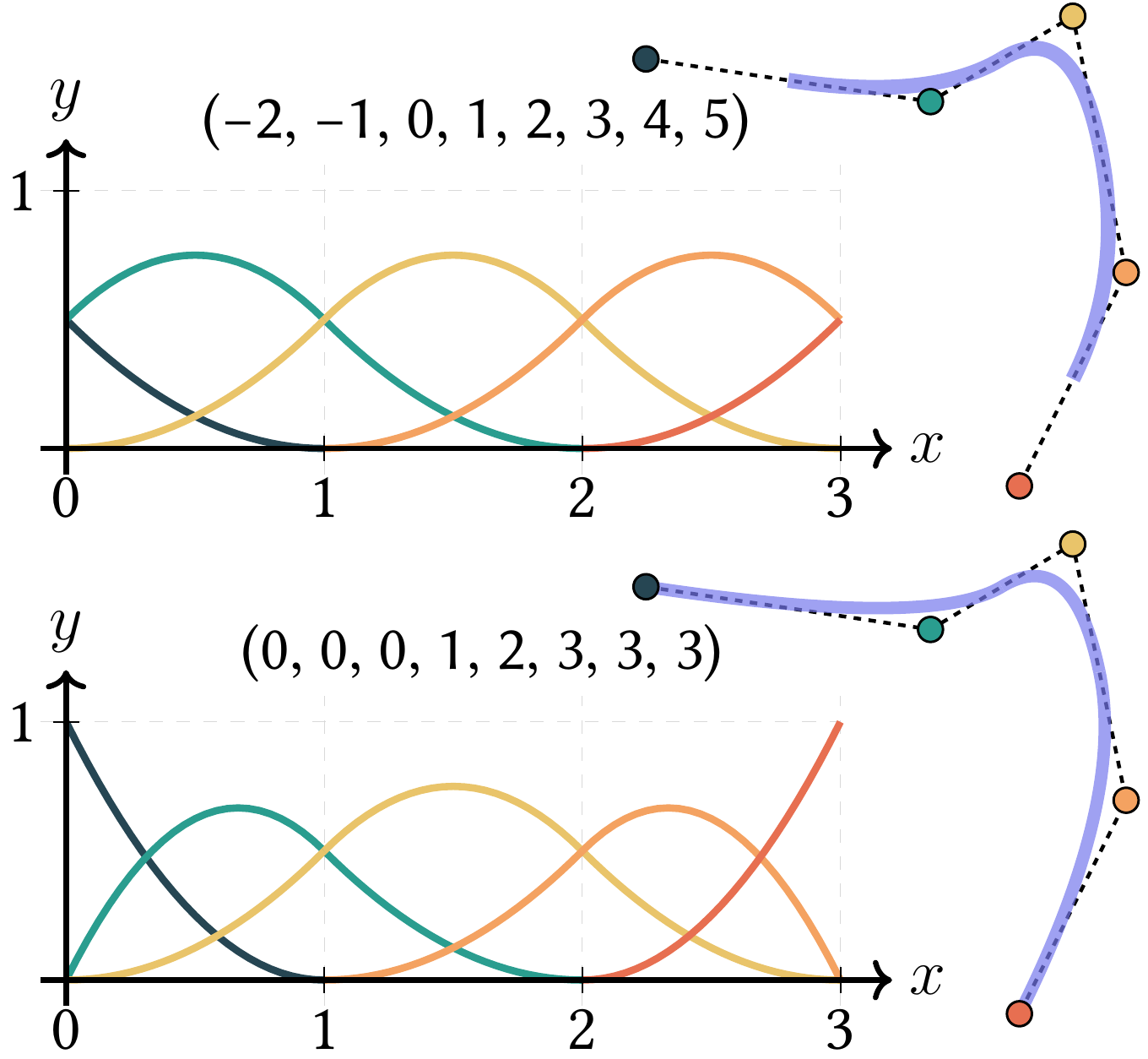}
  \caption{ \textbf{Open vs. non-open B-spline curves.} Uniform quadratic B-spline curves and basis functions with non-open (top) and open (bottom) knot vectors. The associated knot vector is specified besides each curve. Control points and their corresponding basis functions (plotted in the parametric space) are color-matched. An open knot vector ensures the curve passes through the control points at the endpoints, while a non-open one does not.}\label{fig: open/non-open B-spline curves}
\end{wrapfigure}

\vspace*{5pt}
When applying the second formula, any terms involving division by zero due to repeated knots is set to zero. Throughout our method, we employ second-order (quadratic) B-spline bases with uniform open knot vectors, providing global $C^1$-continuity over the surface and ensuring the surface's boundary is determined only by boundary control points. For simplicity, we omit the subscript indicating polynomial order and write $N_i(\cdot)$ instead of $N_{i,2}(\cdot)$.

Two-dimensional B-spline basis functions are constructed via the tensor product of one-dimensional bases. Given two $p$-th order open knot vectors $\Xi = (\xi_1, \dots, \xi_{n+p+1})$ and $\Theta = (\eta_1, \dots, \eta_{m+p+1})$, the 2D knot positions are defined as $(\xi_i, \eta_j)$ for $i \in \{1, \dots, n+p+1\}$ and $j \in \{1, \dots, m+p+1\}$. Correspondingly, the 2D knot spans are defined as $[\xi_i, \xi_{i+1}]\times[\eta_j, \eta_{j+1}]$, based on the 1D spans $[\xi_i, \xi_{i+1}]$ and $[\eta_j, \eta_{j+1}]$.
A two-dimensional basis function $N_{i,j}(\cdot, \cdot)$ is then constructed by
\begin{align}\label{eq: tensor product of B-spline basis}
N_{i,j}(u,v) := N_i(u) N_j(v).
\end{align}
\endgroup

\subsubsection{Quadratic B-spline Surface}
\label{subsubsec: quadratic B-spline surface}

\begin{figure}[htbp]
    \centering
    \includegraphics[width=\linewidth]{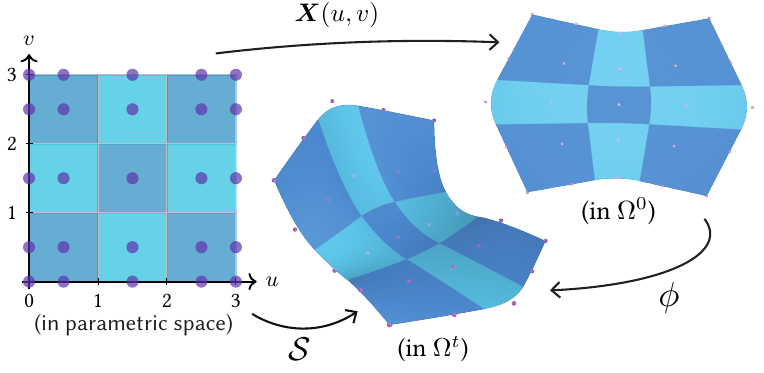}
    \vspace*{-0.5cm}
    \caption{\textbf{B-spline surfaces.} Left: Equidistant knots placed at integer lattice points in parametric space. Top right: 2D B-spline surface in material space $\Omega^0$, allowing for curvy rest shapes. Bottom right: 3D B-spline surface undergoing deformation in world space $\Omega^t$. Here, knot spans are drawn as grid cells with alternating color, and control points are shown as purple dots.}
    \label{fig: B-spline surface}
\end{figure}

Consider two second order open knot vectors $\Xi = (\xi_1, \dots, \xi_{n+3})$ and $\Theta = (\eta_1, \dots, \eta_{m+3})$, and a set of control points $\bm{C}^{i,j} \in \mathbb{R}^3$ indexed by their corresponding 2D B-spline basis functions for $i \in \{1, \dots, n\}$ and $j \in \{1, \dots, m\}$.  A \emph{quadratic B-spline surface} $\mathcal{S}: [\xi_1, \xi_{n+3}]\times[\eta_1, \eta_{m+3}] \to \mathbb{R}^3$ (\autoref{fig: B-spline surface}) is then defined as

\begin{align}\label{eq: general B-spline patch}
\mathcal{S}(u,v) := \sum_{(i,j) \in \mathcal{I}} N_{i,j}(u,v) \bm{C}^{i,j}.
\end{align}
For simplicity, we require that knot vectors start at 0, i.e. $\xi_1 = \eta_1 = 0$; and that all knot coordinates lie on integer lattice points, i.e. $\xi_i, \eta_j \in \Z$ for all $i \in \{1, \dots, n+p+1\}$ and $j \in \{1, \dots, m+p+1\}$.
In the following sections, we refer to $[\xi_1, \xi_{n+3}]\times[\eta_1, \eta_{m+3}]$ as the \emph{parametric space}, and $(u,v) \in [\xi_1, \xi_{n+3}]\times[\eta_1, \eta_{m+3}]$ the \emph{parametric coordinates}.

\subsection{Governing Equation} \label{sec:govern_eq}
Let $\Omega^{t} \subset \mathbb{R}^3$ denote the world space domain at time $t$, and $\Omega^0 \subset \mathbb{R}^2$ the \emph{material space}, where the rest configuration is defined. For any point $\bm{X} \in \Omega^0$, we denote its material coordinates by $(X_1, X_2)$, and define the deformation map at time $t$ as $\bm{\phi}(\cdot, t): \Omega^0 \to \Omega^t$.

{We discretize both the material space and the embedded surfaces in world space using quadratic B-spline surfaces} (\autoref{fig: B-spline surface}), enabling each simulated cloth piece to have \rev{general} disk-topology rest shape potentially with curved boundaries, provided a \rev{sufficiently regular} mapping from its parametric space. The control points serve as the degrees of freedom (DOFs). Let $\bm{X}^\alpha$ and $\bm{C}^\alpha$ denote the material and world space coordinates of the control points, indexed by knot vectors (for brevity, we denote $(i,j)$ by $\alpha$). For an arbitrary parametric coordinate $(u,v)$, its material and world space coordinates are given by B-spline interpolation:
\begin{equation}
\bm{X}(u,v) = \sum_{\alpha} N_\alpha(u,v) \bm{X}^\alpha, \quad
\boldsymbol{\phi}(\bm{X}(u,v)) = \sum_{\alpha} N_\alpha(u,v) \bm{C}^\alpha,
\end{equation}
where we omit the explicit time dependence in $\boldsymbol{\phi}$ and $\bm{C}$ for clarity.

Given a potential energy $V = \int_{\Omega^0} \Psi(\bm{X}, \boldsymbol{\phi})\, \mathrm{d}\bm{X}$ (e.g., from gravity, elasticity, contact, where $\Psi$ is the energy density function) and kinetic energy $T = \frac{1}{2} \int_{\Omega^0} R^0 \dot{\boldsymbol{\phi}}^T \dot{\boldsymbol{\phi}}\, \mathrm{d}\bm{X}$ for the semi-discrete system, where $R^0$ is the rest mass density and $\dot{\boldsymbol{\phi}}$ is the velocity (omitting the dependency on $\bm{X}$ and $t$ for brevity), we can form the Lagrangian $L = T - V$ and derive the Euler-Lagrange equations:
$
\frac{\mathrm{d}}{\mathrm{d}t} \left( \frac{\partial L}{\partial \dot{\bm{C}}^\gamma} \right) = \frac{\partial L}{\partial \bm{C}^\gamma}
$.
Expanding this, we obtain:
\begin{multline}
\frac{1}{2} \frac{\mathrm{d}}{\mathrm{d}t} \left( \ppd{\dot{\bm{C}}^\gamma} \int_{\Omega^0} R^0 \inner{\sum_\alpha N_\alpha \dot{\bm{C}}^\alpha, \sum_\beta N_\beta \dot{\bm{C}}^\beta} \mathrm{d}\bm{X} \right) \\
= -\frac{\partial}{\partial \bm{C}^\gamma} \int_{\Omega^0} \Psi(\bm{X}, \boldsymbol{\phi})\, \mathrm{d}\bm{X}.
\end{multline}
This equation must hold at any time and for every control point $\gamma$.

Defining the consistent mass matrix with entries:
\begin{equation}
M_{\alpha\beta} := \int_{\Omega^0} R^0 N_\alpha N_\beta \, \mathrm{d}\bm{X},
\end{equation}
the governing equation simplifies to:
\begin{equation}
\frac{1}{2} \frac{\mathrm{d}}{\mathrm{d}t} \left( \ppd{\dot{\bm{C}}^\gamma} \sum_{\alpha, \beta} M_{\alpha\beta} \inner{\dot{\bm{C}}^\alpha, \dot{\bm{C}}^\beta}\right)
= -\frac{\partial}{\partial \bm{C}^\gamma} \int_{\Omega^0} \Psi(\bm{X}, \boldsymbol{\phi})\, \mathrm{d}\bm{X}.
\end{equation}

Next, discretizing time into uniform intervals of size $\Delta t$ and approximating time derivatives using backward differences, we obtain:
\begin{multline}
\frac{\partial}{\partial \bm{C}^\gamma} \sum_{\alpha, \beta} \frac{M_{\alpha\beta}}{2\Delta t^2} \inner{ (\bm{C}^\alpha)^{n+1} - (\hat{\bm{C}}^\alpha)^n, (\bm{C}^\beta)^{n+1} - (\hat{\bm{C}}^\beta)^n} \\
\quad + \frac{\partial}{\partial \bm{C}^\gamma} \int_{\Omega^0} \Psi(\bm{X}, \boldsymbol{\phi}^{n+1})\, \mathrm{d}\bm{X} = 0,
\end{multline}
which is essentially implicit Euler time integration, where $(\hat{\bm{C}}^\alpha)^n = (\bm{C}^\alpha)^n + \Delta t\, (\dot{\bm{C}}^\alpha)^n$ is the forward-extrapolated position from the previous time step. Here, superscripts indicate time step indices.

Thus, solving for the system's evolution is equivalent to minimizing the following incremental potential:
\begin{equation}
E(C^{n+1}) = \frac{1}{2\Delta t^2} \| C^{n+1} - \hat{C}^n \|_M^2 + \int_{\Omega^0} \Psi(\bm{X}, \boldsymbol{\phi}^{n+1})\, \mathrm{d}\bm{X}, \label{eq:incremental_potential}
\end{equation}
where $C^{n+1}$ denotes the stacked vector of all control point positions at the next time step, similarly for $\hat{C}^n$, and $\| \cdot \|_M$ denotes the norm induced by the mass matrix $M$.
Here, thanks to the smoothness of $\boldsymbol{\phi}$, both the mass matrix entries and the potential energy integrals can be numerically approximated using standard quadrature rules, such as Gaussian quadrature applied over each local region. In the following section, we discuss how the energy density function $\Psi$ of the potential energy is instantiated and numerically integrated for cloth simulation under our B-spline FEM setting.

\section{Energy Formulation}
\label{sec: energy discretization}

We model membrane elasticity~(\autoref{sec:membrane_energy}) and bending elasticity~(\autoref{sec:bending_energy}) as potential energies in our system, and introduce their reduced numerical integration based on B-spline basis functions. For frictional contact, we adopt the discretization from IPC~\cite{li2023convergent}, applied to a piecewise linear triangle mesh embedded within our B-spline surface for rendering~(\autoref{sec:contact_energy}). Throughout the following subsections, we implicitly assume the mapping between material space coordinates $(X_1, X_2) \in \Omega^0$ and parametric space coordinates $(u,v) \in [\xi_1, \xi_{n+3}]\times[\eta_1, \eta_{n+3}]$ (via the B-spline basis and control points), and for simplicity, we write $u$ instead of $u(X_1, X_2)$, and $N_{i, j}$ instead of $N_{i, j}(u, v)$.

\subsection{Membrane}
\label{subsec: stretching and shearing energy}
\label{sec:membrane_energy}

We adopt the FEM Baraff-Witkin (FBW) model~\cite{Kim2020} for membrane elasticity due to its ability to capture anisotropic behaviors and the availability of analytic eigenanalysis for efficient Hessian computation. FBW defines the energy with a stretching and a shearing component, using the in-plane \emph{deformation gradient} $\bm{F}: (\Omega^0, t) \to \mathbb{R}^{3 \times 2}$. Let $\bm{e}_1$ and $\bm{e}_2$ denote the unit vectors in the material space along the $X_1$ and $X_2$ coordinate directions, respectively. The \emph{anisotropic invariants} are given by
\begin{align}\label{eq: anisotropic invariants}
    I_5(\bm{e}_1) = \bm{e}_1^T \bm{F}^T\!\bm{F} \bm{e}_1, \  I_5(\bm{e}_2) = \bm{e}_2^T \bm{F}^T\!\bm{F} \bm{e}_2, \  I_6(\bm{e}_1, \bm{e}_2) = \bm{e}_1^T \bm{F}^T\!\bm{F} \bm{e}_2.
\end{align}
The FBW stretching and shearing energy densities are defined as
\begin{align}\label{eq: FBW stretching and shearing energy}
\begin{split}
    \Psi_{\text{shear}}   & = \mu_\text{sh} \, I_6^2, \\
    \Psi_{\text{stretch}} & = \mu_\text{st1} \left(\sqrt{I_5(\bm{e}_1)} - 1\right)^2 + \mu_\text{st2} \left(\sqrt{I_5(\bm{e}_2)} - 1\right)^2,
\end{split}
\end{align}
where $\mu$ denotes the material stiffness parameters.

To evaluate the energy density at any point, it suffices to compute the deformation gradient $\bm{F}$ on the B-spline surface, treating the parametric coordinates $(u,v)$ as intermediates:
\begin{align}\label{eq: deformation gradient}
\begin{split}
    F_{\alpha \beta}
    := \frac{\partial \phi_{\alpha}}{\partial X_{\beta}}
    = \!\!\sum_{(i,j) \in \mathcal{I}} \frac{\partial N_{i,j}}{\partial u} \, \bm{C}_{\alpha}^{i,j} \frac{\partial u}{\partial X_{\beta}}
    + \!\!\sum_{(i,j) \in \mathcal{I}} \frac{\partial N_{i,j}}{\partial v} \, \bm{C}_{\alpha}^{i,j} \frac{\partial v}{\partial X_{\beta}}.
\end{split}
\end{align}
Here, $\alpha \in \{1,2,3\}$ and $\beta \in \{1,2\}$, and the subscript on $\bm{C}^{i,j}$ denotes the component index.
Since we generally do not have an analytic expression for the inverse mapping from $\bm{X}$ to $(u,v)$, the derivatives $\partial u / \partial X_{\beta}$ and $\partial v / \partial X_{\beta}$ must be computed by applying the inverse function theorem. Detailed derivations, including the computation of energy derivatives, 
are provided in our supplemental document.

\paragraph{Reduced Integration}
To evaluate the total membrane energy, the most straightforward approach is to apply standard quadrature rules on each knot span and accumulate the contributions across the surface. However, this results in each local stencil involving $3\times3=9$ control points, significantly increasing the density of the global Hessian compared to linear FEM, which only has a stencil size of 3. 
To mitigate this, we partition the surface covered by interior knot spans according to the grid formed by control points, named \textit{dual grid}, and alternate between $1 \times 2$ and $2 \times 1$ quadrature patterns within each grid cell. This design ensures that each quadrature point is associated with only 2 basis functions in one of the directions, reducing the local stencil size to $2\times3=6$. In addition to reducing Hessian density, this strategy also halves the number of quadrature points relative to the standard $2 \times 2$ scheme typically used for quadratic shape functions. Note that one-point quadrature is not used here to avoid null spaces such as the hour-glass modes \cite{zienkiewicz2005finite}.
For knot spans along the surface boundary, where the basis functions become asymmetric, we increase the number of quadrature points to prevent numerical instability (see our supplemental document). Specifically, we apply a $3 \times 3$ quadrature rule at the corners, and $3 \times 2$ or $2 \times 3$ on the sides (\autoref{fig: quadrature}).

\begin{figure}[tbp]
    \centering
    \hspace{-0.04\textwidth}
    \begin{minipage}[b]{0.26\textwidth}
        \begin{tikzpicture}[scale=0.6, transform shape]
            \input{tikz-scripts/Quadrature/MembraneQuadrature}
        \end{tikzpicture}
    \end{minipage}
    \hspace{0.02\textwidth}
    \begin{minipage}[b]{0.16\textwidth}
        \begin{tikzpicture}[scale=0.6, transform shape]
            \input{tikz-scripts/Quadrature/BendingQuadrature}
        \end{tikzpicture}
    \end{minipage}
    \caption{\textbf{Reduced integration.} Illustration of reduced quadrature schemes on a B-spline surface over the parametric space $[0, 5] \times [0, 5]$. Quadrature points are marked with ``$\bm{\times}$''. \textbf{Left:} Quadrature points for membrane energy. In interior knot spans, we apply alternating $2 \times 1$ and $1 \times 2$ Gaussian quadrature on the dual grid. For boundary knot spans (shaded in gray), the quadrature pattern follows that of the corresponding reference cell, shown on the right with matching colored borders. \textbf{Right:} Quadrature points for bending energy. Points are placed at the center of each knot span along the boundary, and at the center of each dual grid cell in the interior.}
    \label{fig: quadrature}
    \vspace*{-0.5em}
\end{figure}

\subsection{Bending}
\label{subsec: bending energy}
\label{sec:bending_energy}

Since the rest shape of cloth is flat and its in-plane deformation is nearly isometric, we adopt a simple quadratic bending energy model~\cite{QuadraticBending2006}. The energy density $\Psi_{\text{bd}}$ is defined as
\begin{align}\label{eq: general bending energy}
\Psi_{\text{bd}} = \frac{1}{2} \mu_\text{bd} H^2,
\end{align}
where $H$ denotes the mean curvature, which can be computed using the induced Laplace-Beltrami operator $\Delta$ on the surface:
\begin{align}\label{eq: mean curvature}
H^2 = \langle \Delta \boldsymbol{\phi}, \Delta \boldsymbol{\phi} \rangle = \left\langle \left(\frac{\partial^2 \boldsymbol{\phi}}{\partial X_1^2} + \frac{\partial^2 \boldsymbol{\phi}}{\partial X_2^2}\right), \left(\frac{\partial^2 \boldsymbol{\phi}}{\partial X_1^2} + \frac{\partial^2 \boldsymbol{\phi}}{\partial X_2^2}\right) \right\rangle.
\end{align}

As with the membrane energy, we leverage the parametric space as an intermediate representation to evaluate spatial derivatives. With shorthand notations for derivatives (e.g., $u_1 = \partial u/\partial X_1$, $u_{11} = \partial^2 u/\partial X_1^2$, and similarly for $v$ and $X_2$), we express the Laplacian as:
\begin{multline}\label{eq: Laplacian in parametric space}
\!\!\!\!\Delta \boldsymbol{\phi}
= \!\!\sum_{(i,j) \in \mathcal{I}}\!\left( (u_1^2 + u_2^2) \frac{\partial^2 N_{i,j}}{\partial u^2} + (v_1^2 + v_2^2) \frac{\partial^2 N_{i,j}}{\partial v^2} + (u_{11} + u_{22}) \frac{\partial N_{i,j}}{\partial u} \right. \\ 
+ (v_{11} + v_{22}) \frac{\partial N_{i,j}}{\partial v} + 2(u_1 v_1 + u_2 v_2) \left. \frac{\partial^2 N_{i,j}}{\partial u \partial v} \right) \bm{C}^{i, j}.
\end{multline}
With quadratic B-spline bases $N_{i,j}$, both the first- and second-order derivatives are well-defined across the entire domain. This enables the use of standard quadrature rules for numerical integration, eliminating the need for edge-based computations.

Note that all partial derivatives of the parametric coordinates with respect to the material space coordinates (as well as those appearing in the in-plane deformation gradient $\bm{F}$) are defined in the material space and depend only on the rest shape. These terms can be precomputed to reduce the computational overhead of evaluating derivatives via the inverse function theorem. Detailed derivations are provided in our supplemental document.

\paragraph{Reduced Integration}
We evaluate the total bending energy using quadrature over the dual grid of knot spans, following a similar strategy to that used for membrane energy. Given that bending forces are typically an order of magnitude smaller than membrane forces, we adopt a one-point quadrature scheme per dual grid cell in the interior and knot span on the boundary (\autoref{fig: quadrature}).
In the interior, each quadrature point is placed at the center of a dual grid cell, which coincides with the vanishing points of adjacent basis functions. As a result, each point is influenced by only the 4 control points of the cell, effectively reducing Hessian density. Despite the minimal quadrature, our experiments show that this scheme still yields convergence to the analytic solution under spatial refinement on a standard benchmark (\autoref{fig: bending calibration}), and it can also capture diverse wrinkling behaviors with varying bending stiffness (\autoref{fig: vary bending stiffness}).




\subsection{Contact}
\label{subsec: contact barrier}
\label{sec:contact_energy}

Since we rely on sampling points from the B-spline surface to form a triangle mesh for rendering, it is critical to ensure that the rendered triangle mesh remains penetration-free, so that no visual interpenetration artifacts occur. To this end, we apply IPC~\cite{Li2020IPC} to handle contact, integrating the contact energy density function over the rendered triangle mesh \cite{li2023convergent} embedded within the B-spline surface. The resolution of this triangle mesh can be chosen independently of the underlying B-spline surface's resolution. \rev{In practice, we set the contact mesh to have twice the number of DOFs of the B-spline surface, except under extreme contact scenarios where equal resolutions are adopted for improved simulation quality.}

Specifically, consider any vertex $\bm{x}^\alpha$ of the linear triangle mesh with a fixed parametric coordinate $(u_{x^\alpha}, v_{x^\alpha})$ determined during initialization. The world-space position of each vertex is then interpolated from the control points using the B-spline basis functions:
\begin{align}\label{eq: trig mesh node from control points}
    \bm{x}^{\alpha} = \sum_{(i,j) \in \mathcal{I}} c_{ij}^{\alpha} \bm{C}^{i,j}, \quad c_{ij}^{\alpha} = N_{i,j}(u_{x^{\alpha}}, v_{x^{\alpha}}).
\end{align}
As a result, the barrier energy ultimately becomes a function of the control points, which are the degrees of freedom driving the motion of the rendered mesh via the B-spline surface.

Using the chain rule, we compute the gradient and Hessian of the barrier energy $B$ with respect to (w.r.t.) the control points based on its gradient and Hessian w.r.t. the linear triangle mesh vertices:
\begin{align}\label{eq: gradient and hessian of contact barrier}
\begin{split}
    \frac{\partial B}{\partial \bm{C}^{i,j}} 
    &= \ \sum_{\alpha \in \Gamma} \frac{\partial B}{\partial \bm{x}^{\alpha}} \frac{\partial \bm{x}^{\alpha}}{\partial \bm{C}^{i,j}}, \\
    \frac{\partial^2 B}{\partial \bm{C}^{i,j} \partial \bm{C}^{k,\ell}}
    &= \sum_{\alpha, \beta \in \Gamma} \left( \frac{\partial^2 B}{\partial \bm{x}^{\alpha} \partial \bm{x}^{\beta}} \frac{\partial \bm{x}^{\alpha}}{\partial \bm{C}^{i,j}} \frac{\partial \bm{x}^{\beta}}{\partial \bm{C}^{k,\ell}} + \frac{\partial B}{\partial \bm{x}^{\alpha}} \frac{\partial^2 \bm{x}^{\alpha}}{\partial \bm{C}^{i,j} \partial \bm{C}^{k,\ell}} \right),
\end{split}
\end{align}
where $\Gamma$ denotes the set of vertices in the linear triangle mesh, and $\alpha$, $\beta$ index mesh vertices.
The second term in the Hessian expression vanishes because $\bm{x}^\alpha$ is a linear function of $\bm{C}^{i,j}$. The treatment of frictional contact follows similarly.

\section{Performance Optimization}
\label{sec: performance optimization}

We describe the key strategies used to optimize the runtime performance of our method. In~\autoref{sec: solver details}, we detail our Newton solver setup and analyze the runtime breakdown, highlighting the trade-offs introduced by switching from linear FEM to B-spline FEM. Then, in~\autoref{sec: fast hessian assembly}, we present our optimized Hessian assembly strategy, which ensures the efficiency of each Newton iteration despite the increased stencil size from B-spline discretization{, and minimizes the overhead of introducing an extra linear triangle mesh for contact. To avoid the potentially expensive Hessian merging operation, and improve the conditioning of the linear system, in~\autoref{sec: partial factorization for accelerated linear solves} we present a linear solve optimization scheme with partial factorization, maintaining efficiency under various contact scenarios.} 

\subsection{Solver Details}\label{sec: solver details}

We minimize our incremental potential (IP) in~\autoref{eq:incremental_potential} using a projected Newton method with backtracking line search to ensure global convergence~\cite{Li2020IPC}. After solving for the search direction on the control point DOFs using Cholesky factorization in each Newton iteration, we propagate the search direction to the linear triangle mesh vertices using the B-spline basis functions. Continuous collision detection (CCD) is then performed to determine \rev{the largest} feasible step size to initialize the backtracking line search, ensuring a monotonic decrease in the IP. The solver terminates once the norm of the Newton search direction becomes sufficiently small. We refer to~\cite{Li2020IPC} for more algorithmic and implementation details.

To ensure positive definiteness and reduce computational cost, we lump the mass matrix by summing each row:
\begin{align}\label{eq: lumped mass matrix}
    M_{(i,j)(i,j)}^{\text{lump}} = \sum_{(k,\ell) \in \mathcal{I}} M_{(i,j)(k,\ell)},
\end{align}
while setting all off-diagonal entries to zero. Gravity forces and boundary conditions are handled similarly to standard FEM treatments~\cite{li2024physics} and are omitted here for brevity.

In traditional linear FEM frameworks, such as~\citet{li2020codimensional,huang2024gipc}, solving the linear system typically dominates the runtime. Our B-spline FEM approach exhibits a different runtime profile, due to a new trade-off between system size and matrix sparsity: quadratic B-spline surfaces can capture comparable wrinkling behaviors with significantly fewer DOFs compared to linear triangle meshes (see comparison in~\autoref{sec: comparison on square cloth sheet}), thereby reducing the linear system size. 
However, each quadrature point on the B-spline surface is influenced by up to 9 control points, resulting in larger local stencils and a denser global Hessian matrix. Moreover, multiple quadrature points are required per element, further increasing the cost of matrix assembly. In \autoref{sec: ablation} we show the timing breakdown per Newton iteration, illustrating that, without careful optimization, Hessian assembly can become even more expensive than the linear solve.

\subsection{Fast Hessian Assembly}\label{sec: fast hessian assembly}

Constructing IP's global Hessian matrix in compressed sparse column (CSC) format for the linear solver involves two main steps: 1) compute local Hessians at quadrature points and collect their entries; 2) assemble these triplets into a CSC matrix.
Our reduced integration schemes already effectively accelerate step 1) by reducing the number of quadrature points and stencil sizes. {For step 2), we handle the Hessian from elasticity and contact separately, due to their inherent difference in sparsity pattern, and the extra conversion phase from triangle mesh to B-spline surface is needed for the contact part. We will detail how to combine two parts of Hessian in~\autoref{sec: partial factorization for accelerated linear solves}.}

The main purpose of the optimization is to avoid using \rev{$\texttt{Eigen}$'s \cite{eigenweb}} $\texttt{setFromTriplets}$ routine, which is slow because of its single-threaded nature and the extra layer of sorting involved~\cite{KimEberle2020Course}. Instead we construct the CSC matrix manually, best utilizing the Hessian structure implied by the geometry or contact pairs.

\paragraph{Elasticity Hessian Assembly}

The sparsity pattern of elasticity Hessian (from membrane and bending energy) depends only on stencil structure, which remains the same throughout the simulation and

\setlength{\columnsep}{10pt}
\begin{wrapfigure}{r}{0.15\textwidth}
  \vspace*{-1.5em}
  \begin{minipage}{0.15\textwidth} 
    \centering
    \begin{tikzpicture}[scale=0.8]
      \input{tikz-scripts/QuadratureSupport/QuadratureSupport}
    \end{tikzpicture}
    \captionsetup{width=\textwidth}
    \caption{\textbf{Mapping control points to quadrature points.} The example control point (purple dot) and its associated quadrature points ($\times$) in an interior $3\times3$ knot span are highlighted.}
    \label{fig: quadrature support}
  \end{minipage}
\end{wrapfigure}

thus can be precomputed. During simulation initialization, we construct a map from each control point to the set of quadrature points inside its support, using which we can fixate the inner and outer indices in the CSC matrix. To determine the $\texttt{data}$ values at each timestep, for each 3-by-3 Hessian block, we find its corresponding control point pair, in parallel query the associated quadrature points for each control point, and sum the weighted contribution of each quadrature point to the block. The weight can also be pre-computed by evaluating the tensor product shape function coefficient at the quadrature point. An example of the correspondence between control points and the quadrature points is shown in~\autoref{fig: quadrature support}.

\begin{wrapfigure}[19]{r}{0.15\textwidth}
  \vspace*{-4pt}
  \begin{minipage}{0.15\textwidth} 
    \centering
    \begin{tikzpicture}[scale=0.8]
      \input{tikz-scripts/QuadratureSupport/VertexSupport}
    \end{tikzpicture}
    \caption{\textbf{Control points influenced by contact mesh DoFs.} A $5 \times 5$ B-spline surface with its contact mesh. A representative contact vertex (black square) and the 9 control points (purple circles) within its support are highlighted.}
    \label{fig: correspondence between vertices and control points}
  \end{minipage}
\end{wrapfigure}

\begin{algorithm}[htbp]
  \caption{Contact Hessian conversion and assembly}
  \label{alg: contact Hessian assembly}
  \KwIn{Set of control points $\mathcal{C}$ and linear mesh vertices $\mathcal{X}$; Set of contact pair with Hessian on linear mesh $\left(\mathcal{X}_{c}, \bm{H}_{c}\right)$; Interpolation weights $\partial \bm{x}^{\alpha}/\partial \bm{C}^i$}
  \KwOut{Array $\texttt{innerPtr}, \texttt{outerPtr}$ and $\texttt{data}$ of CSC matrix}
  $\texttt{B\_lin, B\_bs: Vec[]} \gets \mathsf{CreateSpatialBlocks}(\bm{X})$\;
  $\texttt{L\_lin, L\_bs: HashMap<Int[2], Vec>}$\;
  $\texttt{H\_lin, H\_bs: HashMap<Int[2], Mat<3, 3>\!>}$\;
  \ForEach{$(\mathcal{X}_{c_i}, \bm{H}_{c_i})$ in $\left(\mathcal{X}_{c}, \bm{H}_{c}\right)$ \textbf{\textup{in parallel}}}{
    Take any vertex $\bm{x}^\alpha$ from $\mathcal{X}_{c_i}$\;
    $\texttt{si} \gets \mathsf{QuerySpatialIndex}(\bm{x}^\alpha)$\;
    $\texttt{B\_lin[si].push\_back}(\mathcal{X}_{c_i}, \bm{H}_{c_i})$\;
  }
  \ForEach{\texttt{v\_lin} in \texttt{B\_lin} \textbf{\textup{in parallel}}} {
    \ForEach{$(\mathcal{X}_{c_i}, \bm{H}_{c_i})$ in \texttt{v\_lin}} {
      \ForEach{$(\bm{x}^\alpha, \bm{x}^\beta)$ in $\mathcal{X}_{c_i}$} {
        $\texttt{L\_lin}(\alpha, \beta).\texttt{push\_back}(c_i)$\;
      }
    }
  }
  \ForEach{$((\bm{x}^\alpha, \bm{x}^\beta), \texttt{v})$ in \texttt{L\_lin} \textbf{\textup{in parallel}}}{
    \ForEach{$c_i$ in \texttt{v}}{
      $\texttt{H\_lin}(\alpha, \beta) \mathrel{+}= \bm{H}_{c_i}$\;
    }
  }
  \ForEach{$((\alpha, \beta), \bm{H}_{\alpha, \beta})$ in $\texttt{H\_lin}$ \textbf{\textup{in parallel}}}{
    $\texttt{sj} \gets \mathsf{QuerySpatialIndex}(\bm{x_i})$\;
    $\texttt{B\_bs[sj].push\_back}((\alpha, \beta), \bm{H}_{\alpha, \beta})$\;
  }
  \ForEach{\texttt{v\_bs} in \texttt{B\_bs} \textbf{\textup{in parallel}}} {
    \ForEach{$((\alpha, \beta), \bm{H}_{\alpha, \beta})$ in \texttt{v\_bs}} {
      \ForEach{$(C^i, C^j)$ in $\mathcal{C}$} {
        \If{$\partial \bm{x}^{\alpha}/\partial \bm{C}^i \neq 0$ and $\partial \bm{x}^{\beta}/\partial \bm{C}^j \neq 0$}{
          $w^{\alpha, \beta}_{i, j} \gets \partial \bm{x}^{\alpha}/\partial \bm{C}^i \cdot \partial \bm{x}^{\beta}/\partial \bm{C}^j$\;
          $\texttt{L\_bs}(i, j).\texttt{push\_back}((\alpha, \beta), w^{\alpha, \beta}_{i, j})$\;
        }
      }
    }
  }
  \ForEach{$((\bm{x}^\alpha, \bm{x}^\beta), \texttt{v})$ in \texttt{L\_bs} \textbf{\textup{in parallel}}}{
    \ForEach{$((\alpha, \beta), w^{\alpha, \beta}_{i, j})$ in \texttt{v}}{
      $\texttt{H\_bs}(\alpha, \beta) \mathrel{+}= w^{\alpha, \beta}_{i, j} \bm{H}_{\alpha, \beta}$\;
    }
  }
  $(\texttt{innerPtr}, \texttt{outerPtr}, \texttt{data}) \gets \mathsf{HashMapToCSC}(\texttt{B\_bs})$\;
  \Return $(\texttt{innerPtr}, \texttt{outerPtr}, \texttt{data})$\;
\end{algorithm}

\paragraph{Contact Barrier Hessian Assembly}
  Assembling the contact Hessian poses a significant challenge, because its sparsity pattern depends on the set of active contact pairs, which varies across timesteps. Under a linear FEM framework, the prevailing practice falls back to the triplet (row, column, value) approach, with optional parallelization during the conversion from triplets to a CSC matrix. However, this approach does not extend well to B-spline surfaces with a linear triangle contact mesh, due to the additional layer of conversion introduced by the chain rule (see~\autoref{subsec: contact barrier}). Since the resolution of the underlying linear mesh is user-defined, most vertices on the contact mesh lie within the support of 9 control points of the B-spline surface~(\autoref{fig: correspondence between vertices and control points}), causing one triplet entry of the linear mesh to propagate to $9 \times 9 = 81$ triplet entries of the B-spline surface. This drastically increases the assembly cost, and scales poorly under dense contact.

  To reduce the overhead of converting the contact Hessian from linear mesh to the B-spline surface, we propose the following Hessian assembly algorithm (\autoref{alg: contact Hessian assembly}). Following the same strategy used for assembling the elasticity Hessian, we first construct the sparsity pattern of the matrix, establish the mapping between each 3-by-3 Hessian block and the contact pairs contributing to it, and then for each block accumulate the weighted Hessian contribution in parallel. During simulation initialization, we record a bidirectional map between B-spline surface control points and linear mesh vertices, where a vertex $\bm{x}^{\alpha}$ and a control point $\bm{C}^{i, j}$ are linked if $\bm{x}^{\alpha}$ is in the support of $\bm{C}^{i, j}$. The interpolation weight $\partial \bm{x}^{\alpha}/\partial \bm{C}^{i, j}$ is also precomputed as it remains constant throughout the simulation. For parallel Hessian assembly, we use a hash map that maps pairs of control points to their corresponding 3-by-3 block Hessian as an intermediate representation, rather than directly constructing a CSC matrix. At each timestep, we first construct the hash-map representation of the contact Hessian on the linear mesh from contact pairs, and then propagate it to the B-spline surface. The same sparsity pattern construction strategy applies when accumulating per-block Hessians on B-spline surfaces using the precomputed bidirectional mapping between control points and vertices. The contact Hessian in CSC format can then be constructed from the hash map in a straightforward manner.

  This parallelization strategy has overhead, however, as concurrent writes to the hash map may conflict as multiple vertices lie in the support of the same control point. As shown in \autoref{sec: ablation}, randomly assigning accumulation sources to threads leads to severe write conflicts and significant overhead. To minimize such write conflicts, we partition the axis-aligned bounding box of the simulated mesh into spatial blocks, and assign all vertices within the same spatial block to the same thread (\autoref{fig: contact hessian conversion}). As a result, write conflicts are confined to block boundaries. Since vertices within the same contact pair are spatially close, we \rev{use one arbitrarily chosen vertex from the pair to determine its spatial block assignment}. The same strategy is applied when converting the Hessian from linear mesh to B-spline surface.

\setlength{\columnsep}{24pt}

\begin{figure}[htbp]
    \centering
    \includegraphics[width=\linewidth]{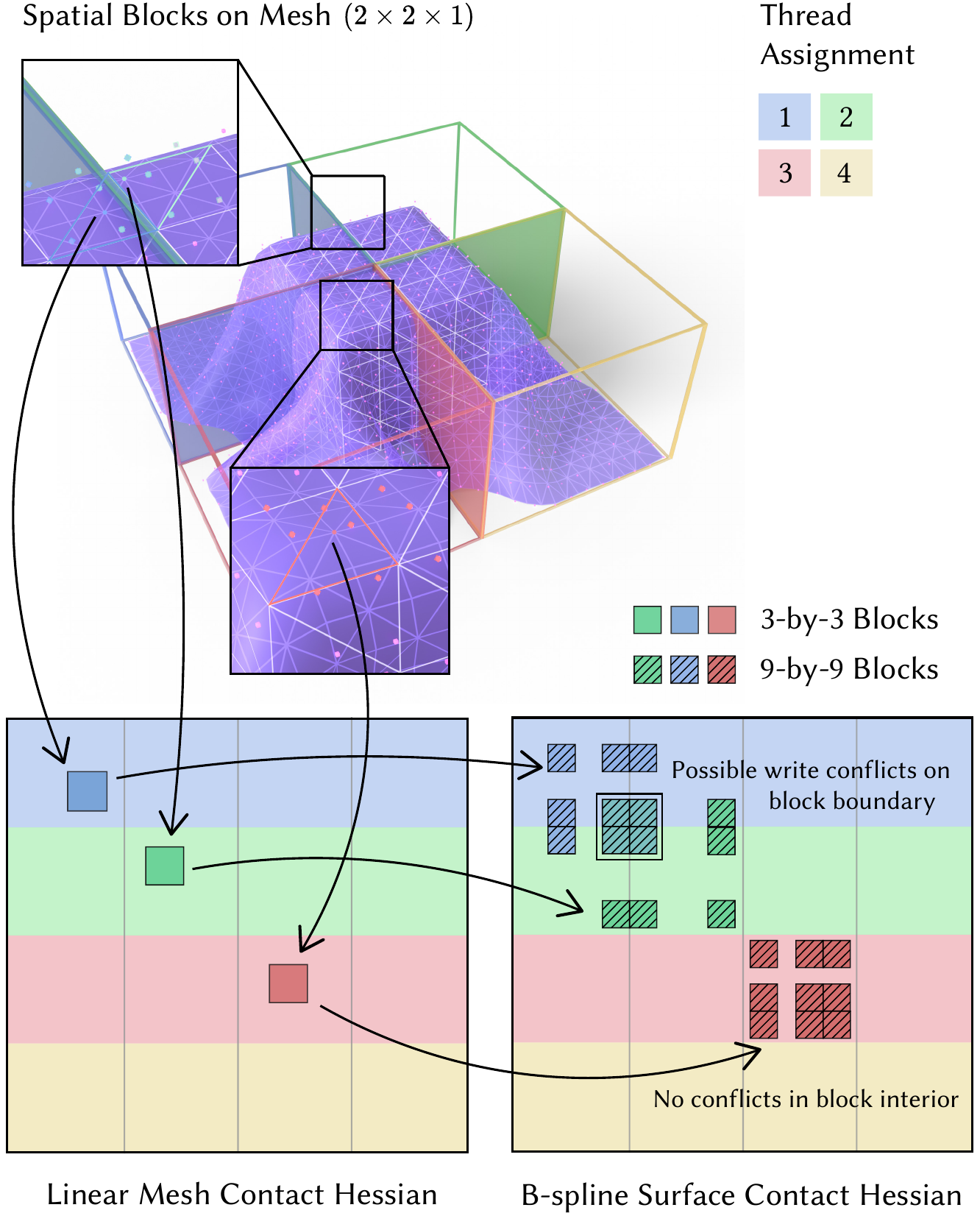}
    \caption{\textbf{Conversion and assembly of contact Hessian.} Example of assembling contact Hessian for a cloth sheet falling onto a cube using a $2 \times 2 \times 1$ spatial block partition with 4 threads. Thread assignments of spatial blocks and Hessian entries are matched by colors. Only writes associated with control points near block boundary may conflict.}
    \label{fig: Hessian assembly figure}
\end{figure}

\subsection{Partial Factorization for Accelerated Linear Solves}\label{sec: partial factorization for accelerated linear solves}
At each projected Newton iteration, once the elastic Hessian and the contact barrier Hessian are assembled, the remaining steps are to combine them and solve the resulting linear system. The sparsity pattern of the contact Hessian changes across linear solves as the contact conditions vary, which incurs extra computational cost in both Hessian merging and reanalysis of the sparsity pattern. We therefore seek further optimization opportunities to reduce this cost.

The update at each iteration is
\begin{equation}\label{eq: per iteration solve}
  \Delta C = -\left( \bm{H}_{\text{inertia}} + \bm{H}_{\text{elasticity}} + \bm{H}_{\text{barrier}} \right)^{-1}\bm{g}_{\text{total}},
\end{equation}
which is computed by solving the corresponding linear system. Every nonzero entry in $\bm{H}_{\text{inertia}}$ lies on the diagonal as we use a lumped mass matrix, and these entries are also nonzero in $\bm{H}_{\text{elasticity}}$. Merging these two Hessians is therefore very fast. To avoid the second costly merge, we apply the Woodbury formula~\cite{Hager1989}
\begin{equation}\label{eq: Woodbury}
  (\bm{D} + \bm{UBV})^{-1} = \bm{D}^{-1} - \bm{D}^{-1}\bm{U}(\bm{B}^{-1} + \bm{VD}^{-1}\bm{U})^{-1}\bm{VD}^{-1}.
\end{equation}
Setting $\bm{U}$ and $\bm{V}$ to be identity matrices and expanding the inverse on the right-hand side recursively gives
\begin{align}\label{eq: Neumann expansion}
  \begin{split}
    (\bm{D} + \bm{B})^{-1} 
    &= \bm{D}^{-1} - \bm{D}^{-1}\bm{B}\bm{D}^{-1} + \bm{D}^{-1}\bm{B}\bm{D}^{-1}\bm{B}\bm{D}^{-1} - \cdots \\
    &= \sum_{k=0}^{\infty} (-1)^k(\bm{D}^{-1}\bm{B})^k\bm{D}^{-1},
  \end{split}
\end{align}
which allows computing only the inverse of $\bm{D}$ (under the context of linear systems, factorizing only $\bm{D}$), and provides fast convergence if the operator norm $\norm{\bm{D}^{-1}\bm{B}}_{\text{op}}$ is small.

To apply this approach to~\autoref{eq: per iteration solve}, we split the barrier Hessian into diagonal and off-diagonal terms $\bm{H}_{\text{barrier}} = \bm{H}_{\text{barrier, d}} + \bm{H}_{\text{barrier, od}}$, and substitute~\autoref{eq: Neumann expansion} with 
\begin{equation}
  \bm{D} = \bm{H}_{\text{inertia}} + \bm{H}_{\text{elasticity}} + \bm{H}_{\text{barrier, d}}, \quad \bm{B} = \bm{H}_{\text{barrier, od}}
\end{equation}
for cases where the convergence rate is estimated to be fast. To efficiently approximate convergence rate, we bound the operator norm from above
\begin{equation}
  \norm{\bm{D}^{-1}\bm{B}}_{\text{op}} \leq \norm{\bm{D}^{-1}}_{\text{op}} \norm{\bm{B}}_{\text{op}} = \frac{\lambda_{\max}(\bm{B})}{\lambda_{\min}(\bm{D})},
\end{equation}
and then estimate the upper bound using a spectrum bound derived from the Gershgorin Circle Theorem~\cite{Horn1985}
\begin{equation}
  \sigma(\bm{B}) \subset \bigcup_{i} D(b_{ii}, R_i), \quad \text{where }b_{ij} \in \bm{B}, \ R_i := \sum_{j \neq i} \abs{b_{ij}}.
\end{equation}
The associated overhead is minimal: the estimate requires only a single sweep through the Hessian along with a second pass over its diagonal entries, and it does not rely on any iterative routines. When the bound is satisfied it could avoid the expensive symbolic analysis by reusing previous analysis of the elasticity Hessian.

\section{Evaluation}

\label{sec: evaluations}

Our framework is implemented in C++, using CHOLMOD~\cite{cholmod2008} compiled with Intel MKL LAPACK and BLAS as linear solver, Eigen~\cite{eigenweb} for linear algebra operations, and Intel TBB for CPU multi-threading. 
All our experiments are performed on a 12th Gen Intel(R) Core i7-12700F (64GB memory). Throughout the evaluations, unless otherwise specified, we apply the parameter set for cotton cloth with density 472.6~kg/$\text{m}^3$, thickness 0.318~mm, and Poisson ratio 0.243, measured by~\citet{penava2014measurement}. We use 2 MPa for Young's modulus, and set the ratio of shearing stiffness to stretching stiffness as $5 \times 10^{-3}$ as is measured in various cloth materials \cite{penava2014measurement}. For handling frictional contact, we use IPC with $\hat{d} = 10^{-3}$m, friction of $\mu=0.1$ with $\epsilon_v = 10^{-3}$, and contact stiffness of $\kappa=10^5 Pa$. Simulation timestep is set to be $\Delta t = 0.01$~s. {We use a fully parallelized CPU IPC framework\footnote{https://github.com/KemengHuang/CPU-IPC/} as the reference linear FEM, and original code from~\citet{SecondOrderFEM2023} (SFEM) and~\citet{ni2024simulating} (BHEM) in the benchmark test problems.} All statistics are listed in \autoref{tab: evaluation_statistics}.

\begin{table*}[t]
\caption{\textbf{Simulation Statistics.}
Simulation statistics of all experiments discussed in the paper. All experiments are setup with $\rho = 472.6$ kg/m$^3$ and timestep $\Delta t = 0.01$~s, with the rest parameters specified in the table. BHEM implementation adopts StVK energy formulation, which does not provide a tunable $E_{\text{sh}}$. For B-spline FEM on cases involving contact, the contact mesh DOF is indicated in the parenthesis. An ``Ada.'' for barrier stiffness $\kappa$ implies that the adaptive barrier stiffness introduced in ~\citet{Li2020IPC} is adopted. All runtimes are reported as averages per timestep.
For cases involving contact, the resolution of the linear mesh used for collision handling is noted in parentheses.
In contact-free simulations, B-spline FEM achieves up to a $2\times$ speedup over linear FEM and a $4\times$ speedup compared to SFEM~\cite{SecondOrderFEM2023}.
In contact-rich scenarios, our method maintains slightly better performance than linear FEM.
}
\label{tab: evaluation_statistics}
\centering
\footnotesize
\setlength{\tabcolsep}{4pt}
\renewcommand{\arraystretch}{1.25}
\begin{tabular}{@{}>{\raggedright\arraybackslash}p{4cm}|cccccccccccc@{}}
\toprule
\noalign{\vskip-0.4ex}
Case & $h$ (m) & $E_{\text{st}}$ (Pa) & $E_{\text{sh}}$ (Pa) & $E_{\text{bd}}$ (Pa) & $\nu$ & $\kappa$ (Pa) & $\hat{d}$ (m) & $\mu$ & $\varepsilon_{v}$ (m/s) & Method & \#DOF & Runtime (s) \\
\hline

\multirow{9}{=}{Upright Hanging Cloth [\autoref{fig: upright hanging cloth}]}
    & $3.18 \times 10^{-4}$ & $2 \times 10^6$ & $10^4$ & $8 \times 10^3$ & 0.243 & N/A & N/A & N/A & N/A
    & B-spline & 17K & 1.431 \\
    & $3.18 \times 10^{-4}$ & $2 \times 10^6$ & $10^4$ & $8 \times 10^3$ & 0.243 & N/A & N/A & N/A & N/A
    & Linear   & 40K & 3.255 \\
\cline{2-13}
    & $3.18 \times 10^{-4}$ & $2 \times 10^6$ & $10^4$ & $8 \times 10^3$ & 0.243 & N/A & N/A & N/A & N/A
    & B-spline & 30K & 3.128 \\
    & $3.18 \times 10^{-4}$ & $2 \times 10^6$ & $10^4$ & $8 \times 10^3$ & 0.243 & N/A & N/A & N/A & N/A
    & Linear   & 92K & 8.294 \\
\cline{2-13}
    & $3.18 \times 10^{-4}$ & $2 \times 10^6$ & $10^4$ & $8 \times 10^3$ & 0.243 & N/A & N/A & N/A & N/A
    & B-spline & 47K & 5.125 \\
    & $3.18 \times 10^{-4}$ & $2 \times 10^6$ & $10^4$ & $8 \times 10^3$ & 0.243 & N/A & N/A & N/A & N/A
    & Linear   & 120K & 12.017 \\
    & $3.18 \times 10^{-4}$ & $2 \times 10^6$ & $10^4$ & $8 \times 10^3$ & 0.243 & N/A & N/A & N/A & N/A
    & SFEM & 47K & 27.261 \\
    & $3.18 \times 10^{-4}$ & $2 \times 10^5$ & N/A & $8 \times 10^2$ & 0.243 & N/A & N/A & N/A & N/A
    & BHEM$^{*}$ & 8K & 116.67$^{\blacktriangle}$ \\
    & $3.18 \times 10^{-4}$ & $2 \times 10^5$ & N/A & $8 \times 10^2$ & 0.243 & N/A & N/A & N/A & N/A
    & BHEM$^{\dagger}$ & 8K & 503.90$^{\blacktriangle}$ \\
\hline

\multirow{7}{*}{Drape Test [\autoref{fig: drape test}]}
    & $3.18 \times 10^{-4}$ & $2 \times 10^6$ & $10^4$ & $8 \times 10^3$ & 0.243 & N/A & N/A & N/A & N/A
    & B-spline & 30K & 2.106 \\
    & $3.18 \times 10^{-4}$ & $2 \times 10^6$ & $10^4$ & $8 \times 10^3$ & 0.243 & N/A & N/A & N/A & N/A
    & Linear   & 68K & 3.912 \\
    & $3.18 \times 10^{-4}$ & $2 \times 10^6$ & N/A & $8 \times 10^2$ & 0.243 & N/A & N/A & N/A & N/A
    & BHEM   & 8K & 13.922 \\
\cline{2-13}
    & $3.18 \times 10^{-4}$ & $2 \times 10^6$ & $10^4$ & $8 \times 10^3$ & 0.243 & N/A & N/A & N/A & N/A
    & B-spline & 47K & 2.919 \\
    & $3.18 \times 10^{-4}$ & $2 \times 10^6$ & $10^4$ & $8 \times 10^3$ & 0.243 & N/A & N/A & N/A & N/A
    & Linear   & 92K & 6.724 \\
    & $3.18 \times 10^{-4}$ & $2 \times 10^6$ & $10^4$ & $8 \times 10^3$ & 0.243 & N/A & N/A & N/A & N/A
    & SFEM     & 47K & 23.162 \\
    & $3.18 \times 10^{-4}$ & $2 \times 10^6$ & N/A & $8 \times 10^2$ & 0.243 & N/A & N/A & N/A & N/A
    & BHEM     & 16K & 43.801 \\
\hline

\multirow{8}{*}{Shear Test [\autoref{fig: shear test}]}
    & $3.18 \times 10^{-4}$ & $2 \times 10^6$ & $2 \times 10^6$ & $4 \times 10^3$ & 0.243 & N/A & N/A & N/A & N/A
    & B-spline & 30K & 3.019 \\
    & $3.18 \times 10^{-4}$ & $2 \times 10^6$ & $2 \times 10^6$ & $4 \times 10^3$ & 0.243 & N/A & N/A & N/A & N/A
    & Linear   & 47K & 5.823 \\
\cline{2-13}
    & $3.18 \times 10^{-4}$ & $2 \times 10^6$ & $2 \times 10^6$ & $4 \times 10^3$ & 0.243 & N/A & N/A & N/A & N/A
    & B-spline & 47K & 9.107 \\
    & $3.18 \times 10^{-4}$ & $2 \times 10^6$ & $2 \times 10^6$ & $4 \times 10^3$ & 0.243 & N/A & N/A & N/A & N/A
    & Linear   & 68K & 16.784 \\
    & $3.18 \times 10^{-4}$ & $2 \times 10^5$ & N/A & $4 \times 10^2$ & 0.243 & N/A & N/A & N/A & N/A
    & BHEM     & 8K & 15.733 \\
\cline{2-13}
    & $3.18 \times 10^{-4}$ & $2 \times 10^6$ & $2 \times 10^6$ & $4 \times 10^3$ & 0.243 & N/A & N/A & N/A & N/A
    & B-spline & 59K & 18.823 \\
    & $3.18 \times 10^{-4}$ & $2 \times 10^6$ & $2 \times 10^6$ & $4 \times 10^3$ & 0.243 & N/A & N/A & N/A & N/A
    & SFEM     & 47K & 75.946 \\
    & $3.18 \times 10^{-4}$ & $2 \times 10^5$ & N/A & $4 \times 10^2$ & 0.243 & N/A & N/A & N/A & N/A
    & BHEM     & 16K & 32.910 \\
\hline

\multirow{2}{*}{Cloth on Sphere [\autoref{fig: cloth on sphere}]}
    & $3.18 \times 10^{-4}$ & $2 \times 10^6$ & $2 \times 10^6$ & 0 & 0.243 & $10^5$ & $10^{-3}$ & 0.1 & $10^{-3}$
    & B-spline & 17K (22K) & 3.128 \\
    & $3.18 \times 10^{-4}$ & $2 \times 10^6$ & $2 \times 10^6$ & 0 & 0.243 & $10^5$ & $10^{-3}$ & 0.1 & $10^{-3}$
    & Linear   & 22K & 4.362 \\
\hline

Cloth on Rotating Sphere [\autoref{fig: cloth on rotating sphere}]
    & $3.18 \times 10^{-4}$ & $2 \times 10^6$ & $4 \times 10^3$ & $8 \times 10^2$ & 0.243 & Ada. & $3 \times 10^{-4}$ & 0.4 & $10^{-2}$
    & B-spline & 120K (270K) & 422.17 \\
\hline

Helicopter [\autoref{fig: helicopter}]
    & $6 \times 10^{-4}$ & $5 \times 10^5$ & $2.5 \times 10^3$ & 10 & 0.243 & Ada. & $10^{-3}$ & 0.5 & $10^{-3}$
    & B-spline & 120K (270K) & 374.32 \\
\hline

Stripes over Armadillo [\autoref{fig: brush over armadillo}]
    & $6 \times 10^{-4}$ & $5 \times 10^5$ & $2.5 \times 10^3$ & 10 & 0.243 & Ada. & $10^{-3}$ & 0.2 & $10^{-3}$
    & B-spline & 90K (90K) & 73.761 \\
\hline

Shuffling [\autoref{fig: character animation} a)]
    & $3.18 \times 10^{-4}$ & $5 \times 10^5$ & $2.5 \times 10^3$ & 10 & 0.243 & Ada. & $3 \times 10^{-4}$ & 0.1 & $10^{-3}$
    & B-spline & 95K (95K) & 85.226 \\
\hline

Hiphop Dance [\autoref{fig: character animation} b)]
    & $3.18 \times 10^{-4}$ & $5 \times 10^5$ & $2.5 \times 10^3$ & $8 \times 10^2$ & 0.243 & Ada. & $3 \times 10^{-4}$ & 0.1 & $10^{-3}$
    & B-spline & 65K (65K) & 30.063 \\
\hline

Skirt Swirl [\autoref{fig: character animation} c)]
    & $3.18 \times 10^{-4}$ & $5 \times 10^5$ & $2.5 \times 10^3$ & 10 & 0.243 & Ada. & $3 \times 10^{-4}$ & 0.1 & $10^{-3}$
    & B-spline & 124K (124K) & 67.292 \\
\hline

Cloth Basket [\autoref{fig: teaser}]
    & $3.18 \times 10^{-4}$ & $5 \times 10^5$ & $2.5 \times 10^3$ & 10 & 0.243 & Ada. & $10^{-4}$ & 0.1 & $10^{-3}$
    & B-spline & 152K (191K) & 211.98 \\

\noalign{\vskip-0.2ex}
\bottomrule
\end{tabular}
\vspace*{2pt}
\parbox{\linewidth}{\raggedright\footnotesize
$^{*}$Newton's method is force terminated after 100 iterations. $^{\dagger}$Newton's method is force terminated after 400 iterations. $^{\blacktriangle}$Fails to converge after reaching maximum iterations.
}
\end{table*}


\subsection{Unit Tests}

\begin{figure}[h]
    \centering
      \includegraphics[width=0.38\linewidth]{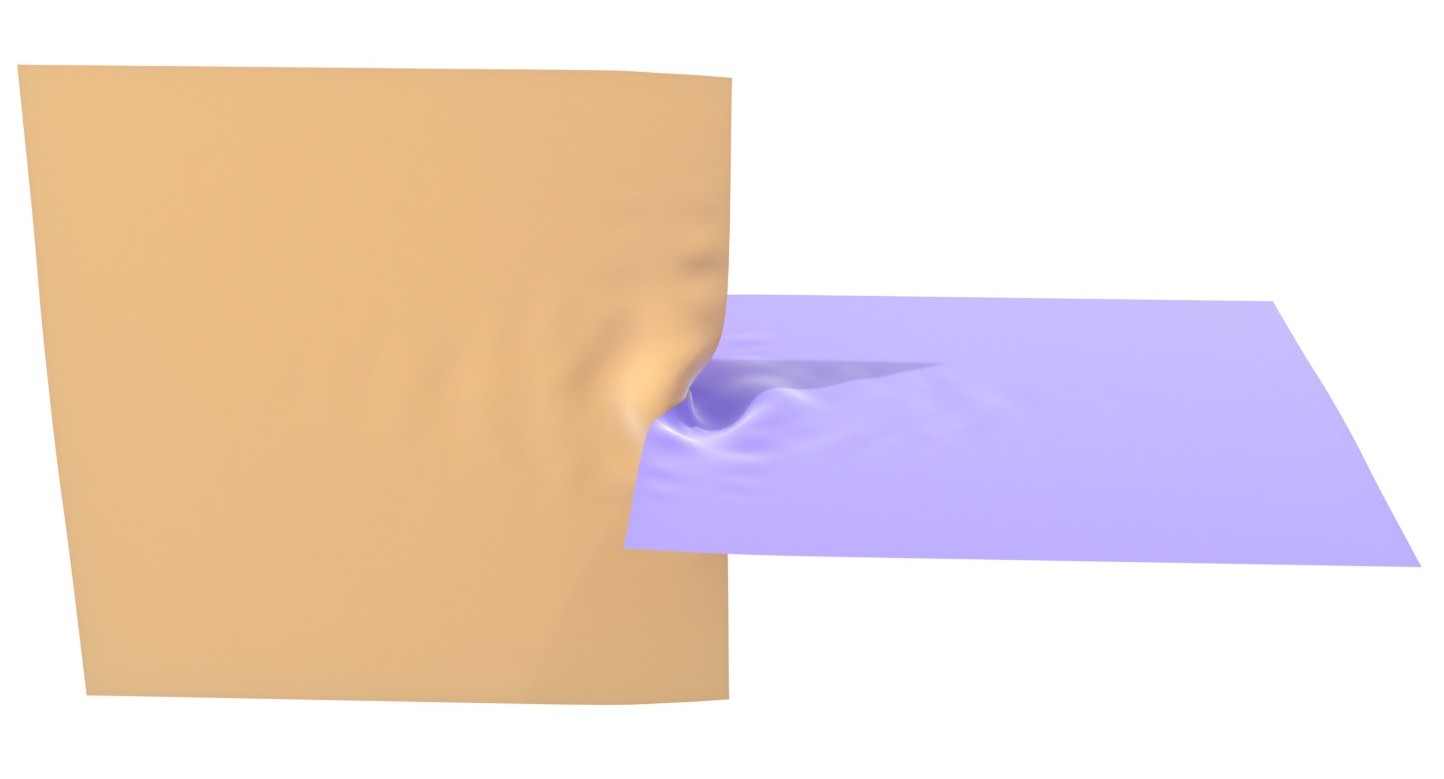}
      \includegraphics[width=0.2\linewidth]{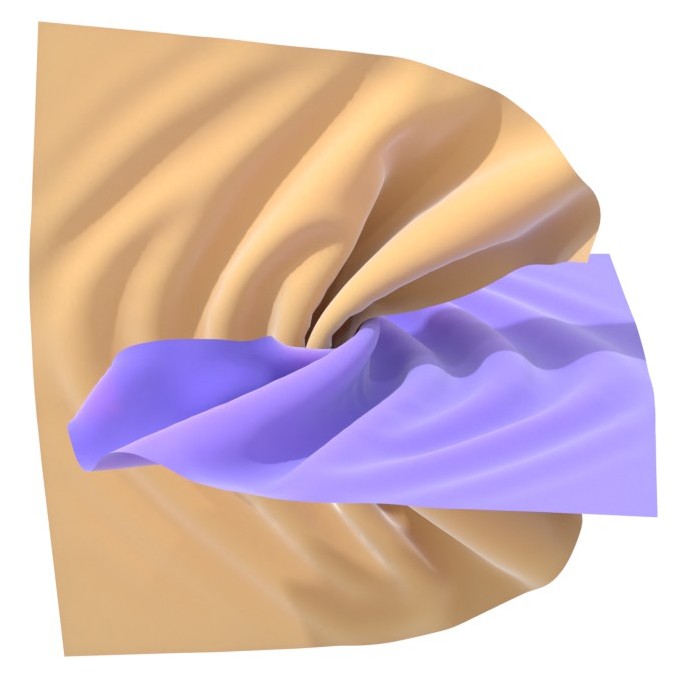}
      \includegraphics[width=0.38\linewidth]{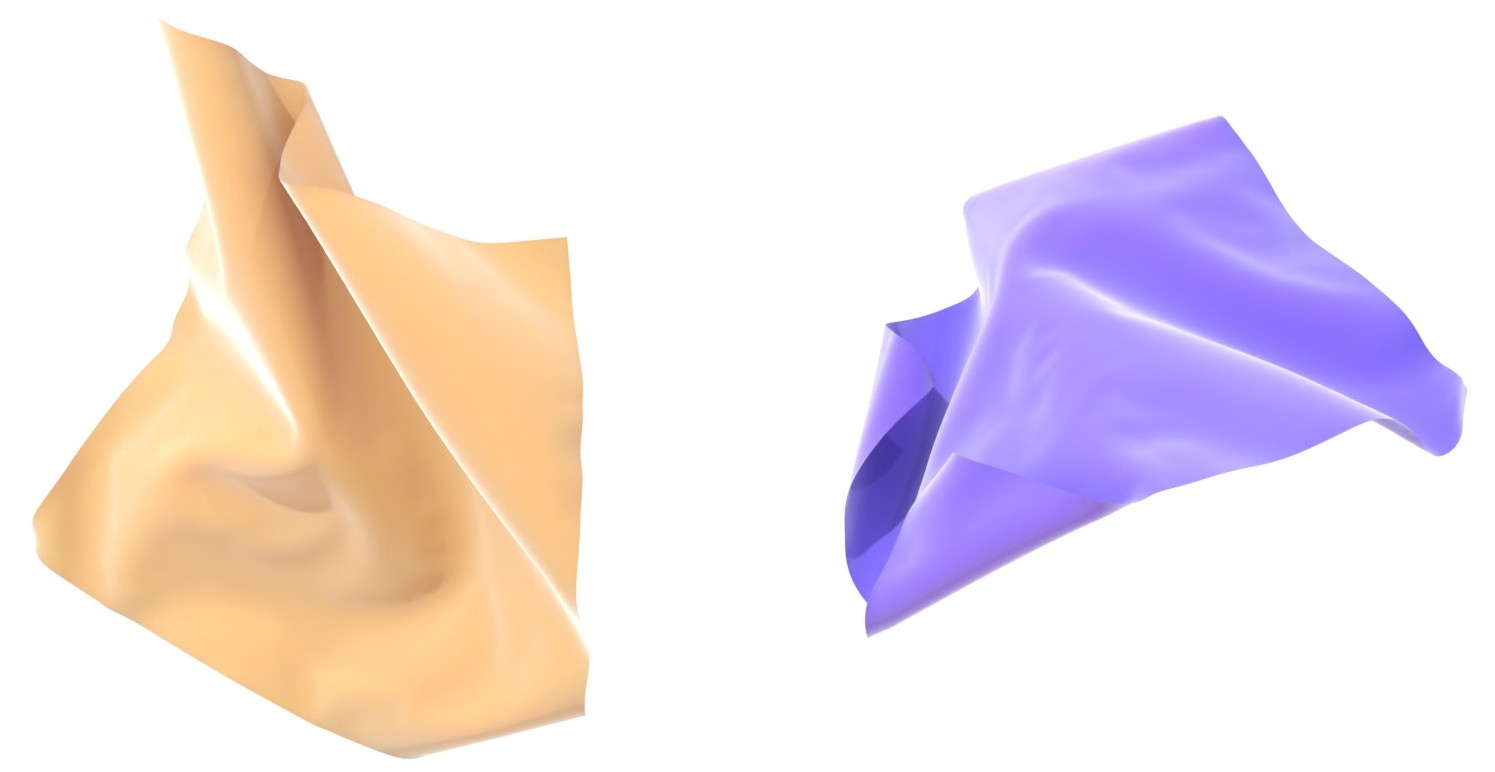}
      \includegraphics[width=\linewidth]{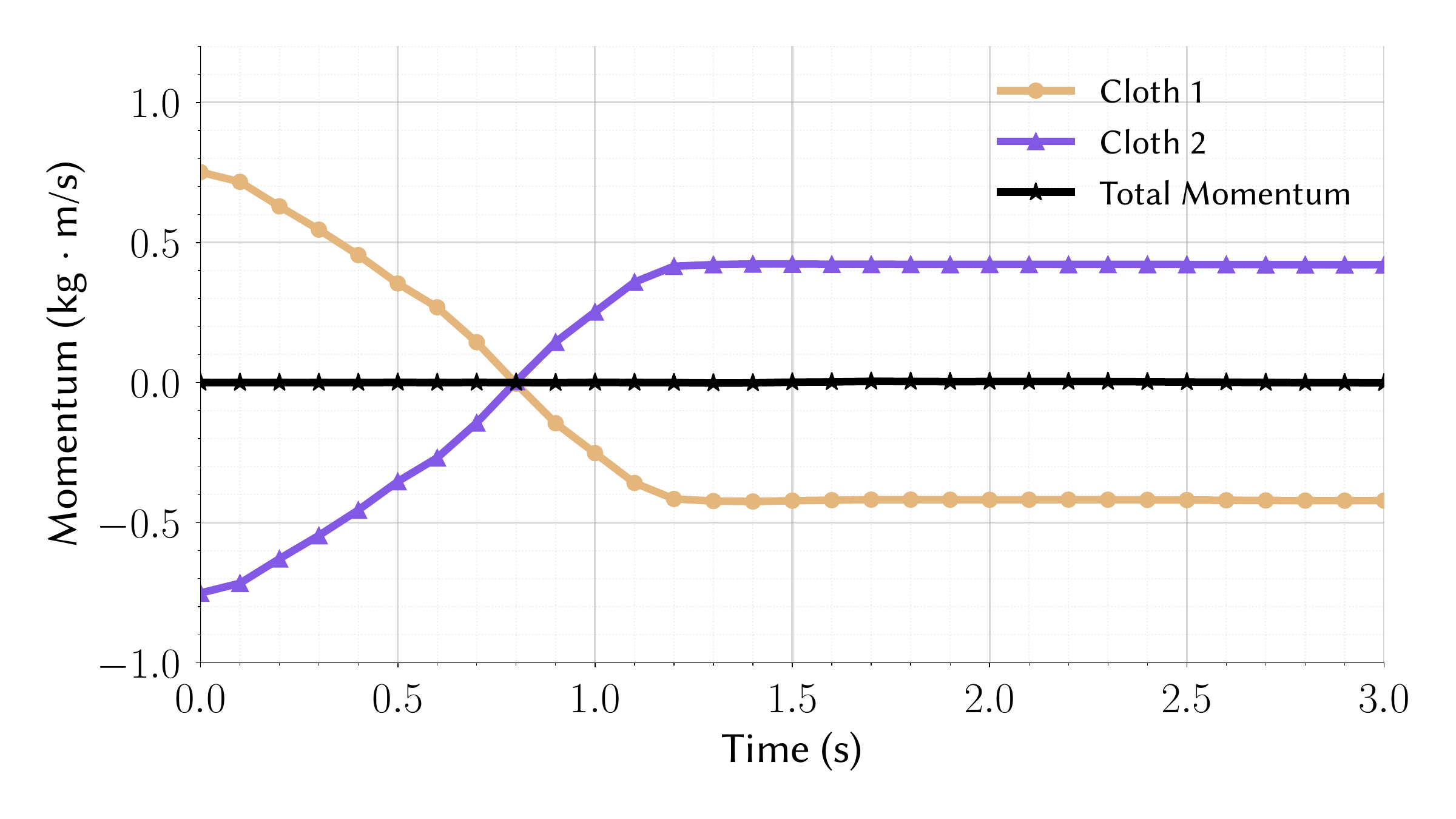}
    \vspace*{-3em}
    \caption{\textbf{Conservation of momentum.} Top: Two pieces of cloth sheet collide and rebound after impact. Bottom: Plot of linear momentum along the $x$-axis for each plate and the total momentum of the system. The total momentum remains constant throughout the simulation.}
    \label{fig: conservation of momentum}
\end{figure}

\paragraph{Conservation of Momentum.}  
We validate the conservation of momentum in our model through a simulation of two colliding elastic plates. As shown in~\autoref{fig: conservation of momentum}, the two plates are initially placed on orthogonal planes, each moving at 0.5~m/s. Upon collision and rebound, the total momentum of the system is accurately preserved.

\paragraph{Varying Bending Stiffness.}

\begin{figure}[h]
    \centering
    \hspace{-0.02\textwidth}
    \begin{minipage}[b]{0.12\textwidth}
        \includegraphics[width=\textwidth]{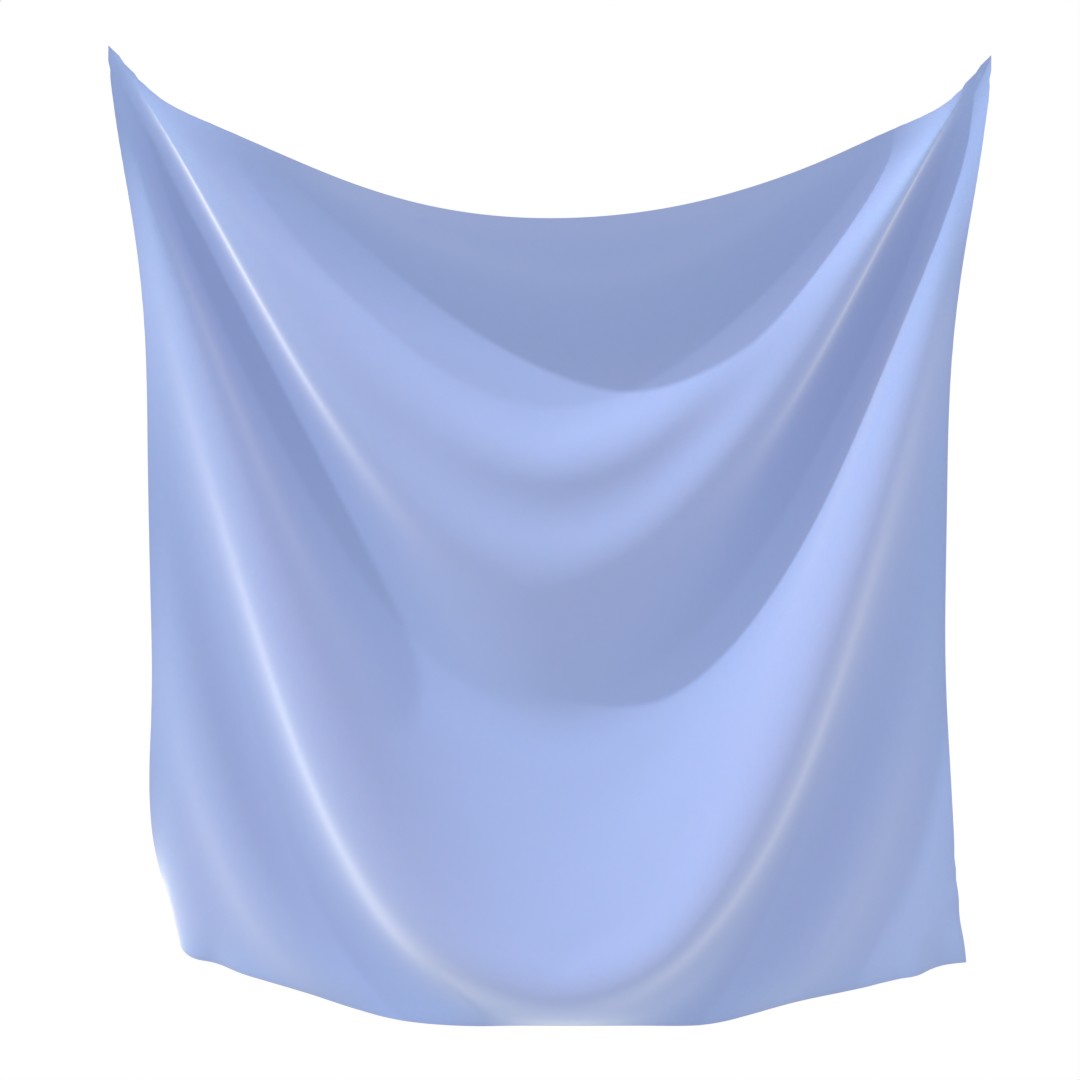}
        \caption*{\centering $E = 8 \times 10^1$}
    \end{minipage}
    \hfill
    \begin{minipage}[b]{0.12\textwidth}
        \includegraphics[width=\textwidth]{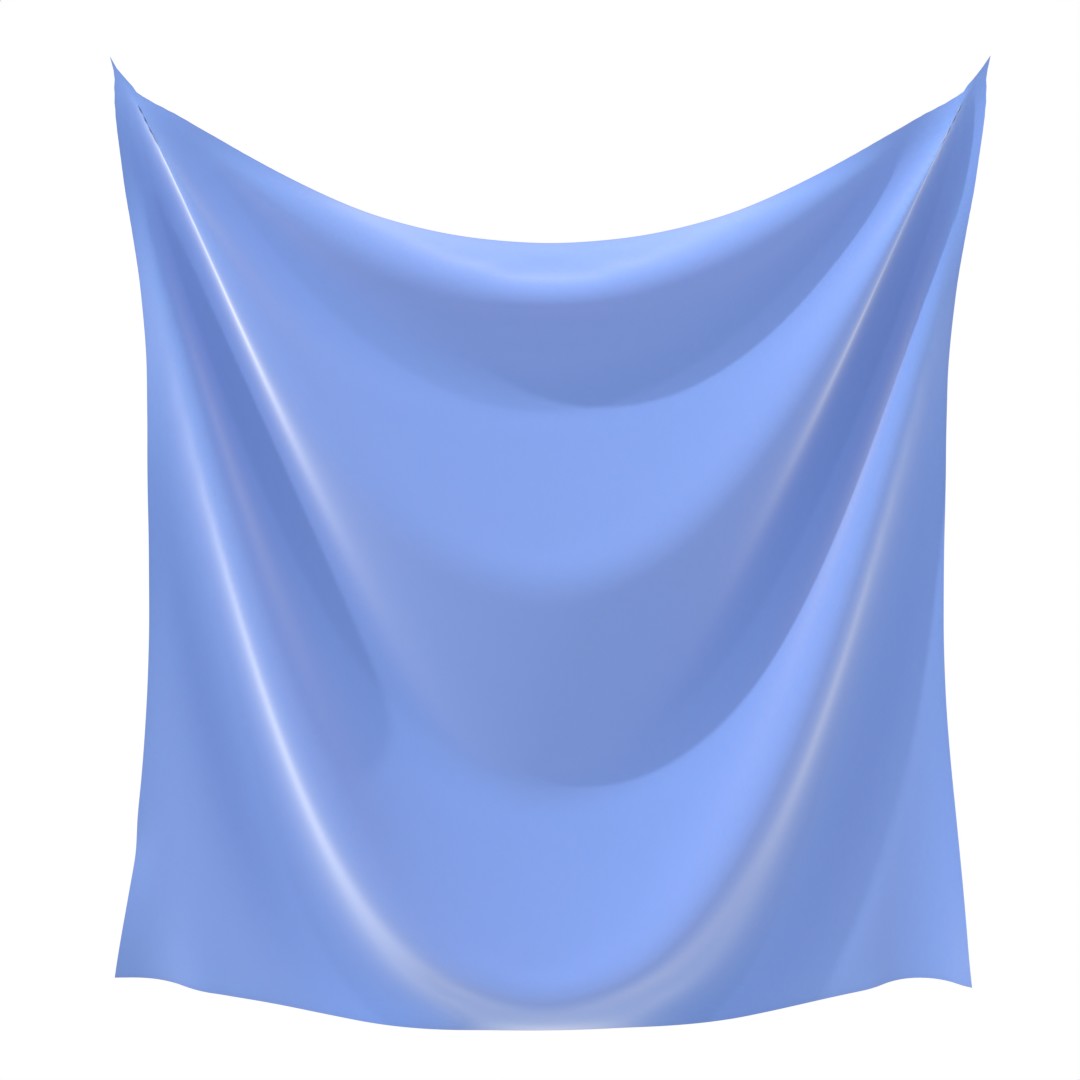}
        \caption*{\centering $E = 8 \times 10^3$}
    \end{minipage}
    \hfill
    \begin{minipage}[b]{0.12\textwidth}
        \includegraphics[width=\textwidth]{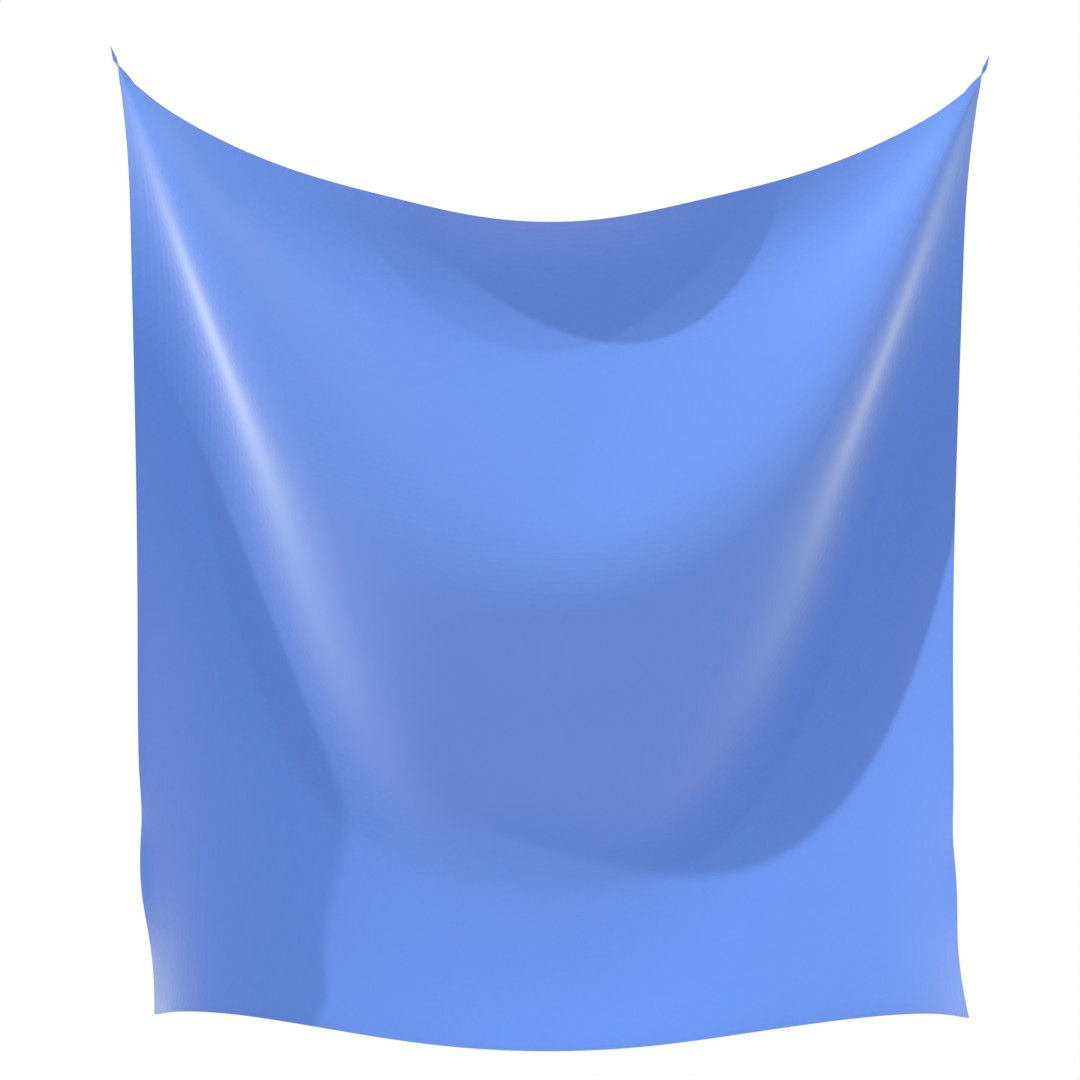}
        \caption*{\centering $E = 8 \times 10^5$}
    \end{minipage}
    \hfill
    \begin{minipage}[b]{0.12\textwidth}
        \includegraphics[width=\textwidth]{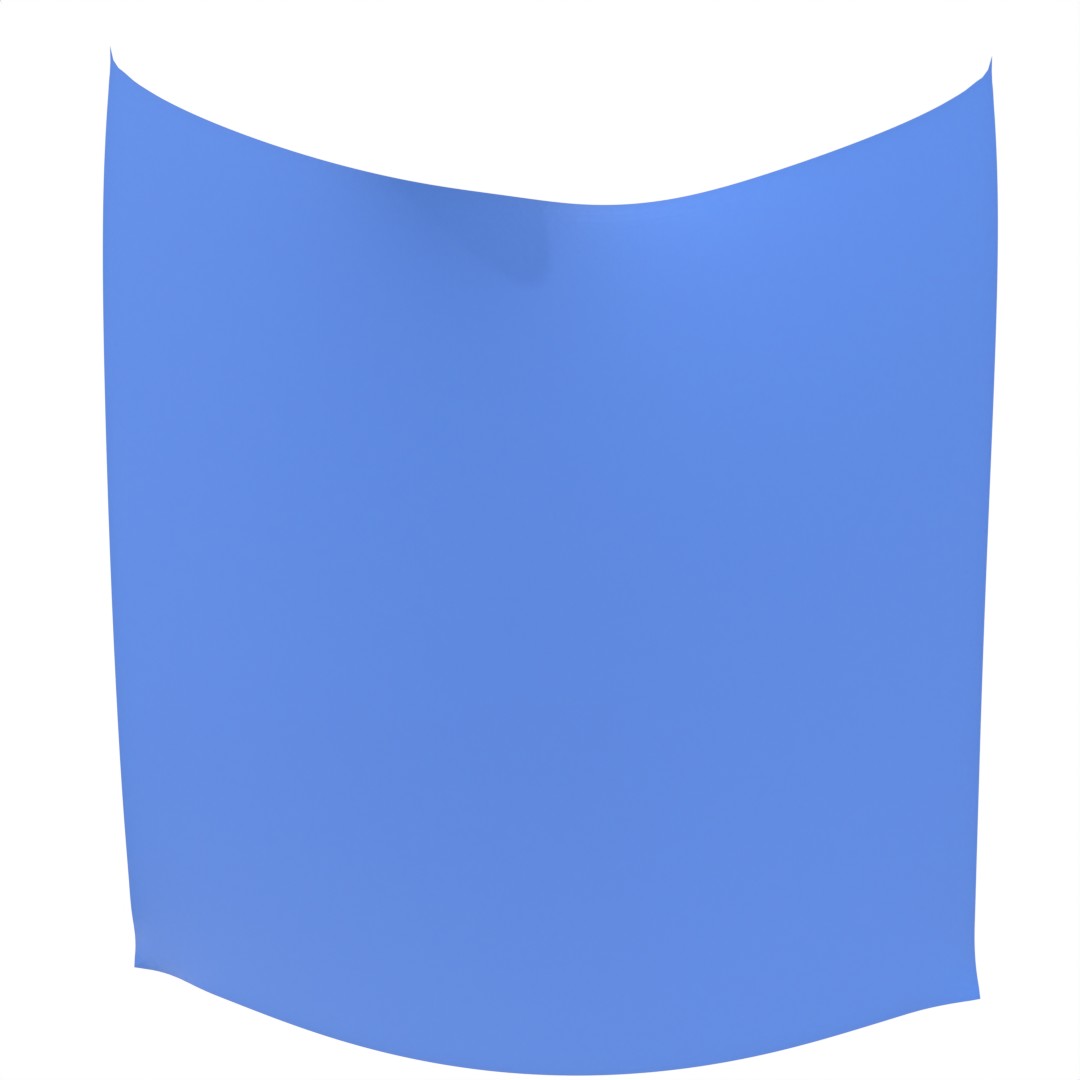}
        \caption*{\centering $E = 8 \times 10^7$}
    \end{minipage}
    \caption{\textbf{Varying bending stiffness.} Simulation of a piece of cloth hanging upright. With larger bending stiffness, the wrinkles gradually smooth out.}
    \label{fig: vary bending stiffness}
\end{figure}

We demonstrate that our bending model can capture different wrinkle behaviors in a controllable way with varying bending stiffness, through a simulation of an upright cloth sheet. The upper corners of the cloth sheet are fixed, and moving towards each other at 0.5~m/s in the first 0.1~seconds. The cloth swings until it comes to rest under gravity. As shown in \autoref{fig: vary bending stiffness}, the wrinkles gradually smooth out as the bending stiffness increases.

\paragraph{Mitigated Membrane Locking}

\begin{figure}[h]
    \centering
    \begin{subfigure}[b]{0.11\textwidth}
      \centering
      \includegraphics[width=\linewidth]{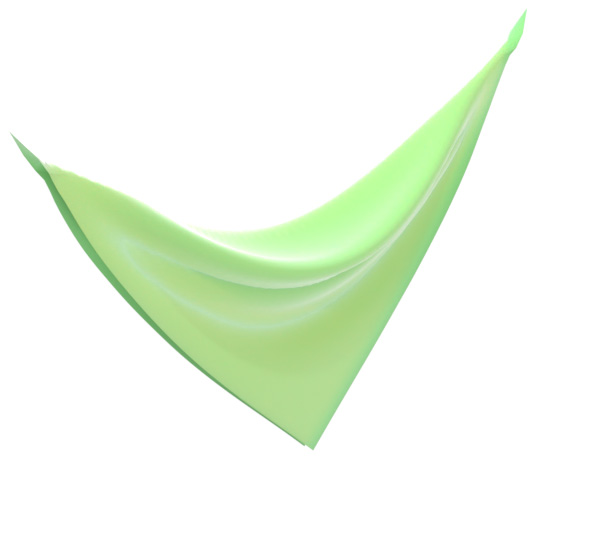}
    \end{subfigure}
    \begin{subfigure}[b]{0.11\textwidth}
      \centering
      \includegraphics[width=\linewidth]{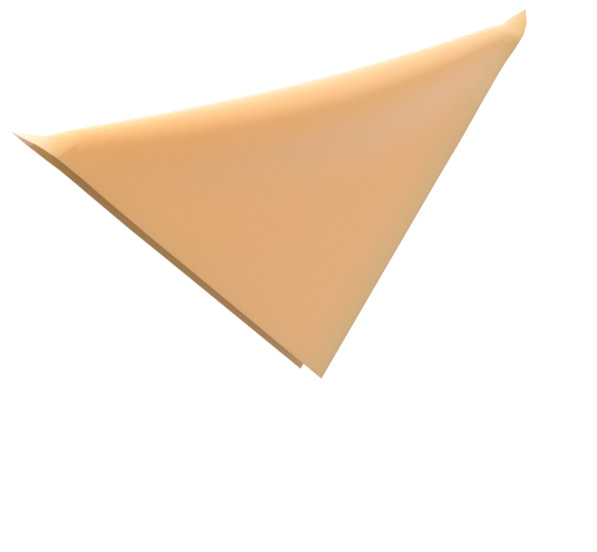}
    \end{subfigure}
    \begin{subfigure}[b]{0.11\textwidth}
      \centering
      \includegraphics[width=\linewidth]{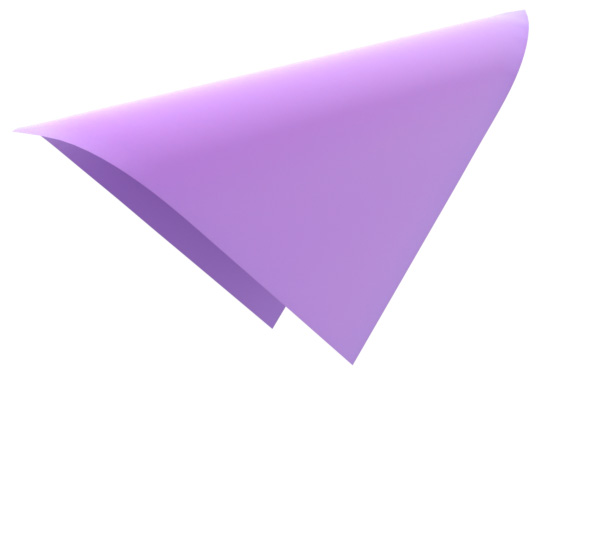}
    \end{subfigure}
    \begin{subfigure}[b]{0.11\textwidth}
      \centering
      \includegraphics[width=\linewidth]{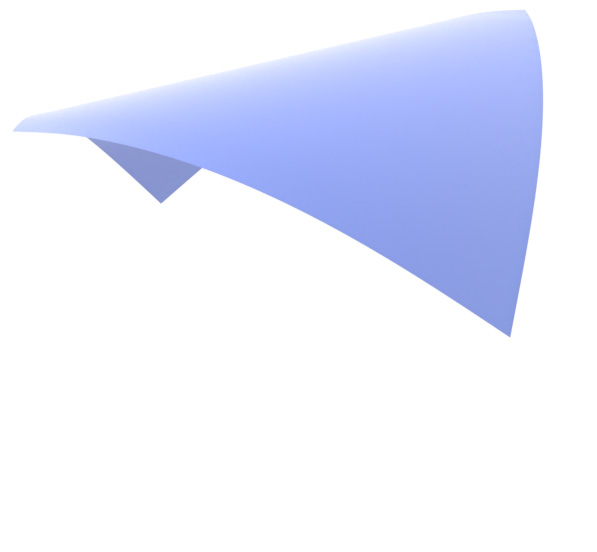}
    \end{subfigure}
    \begin{subfigure}[b]{0.11\textwidth}
      \centering
      \includegraphics[width=\linewidth]{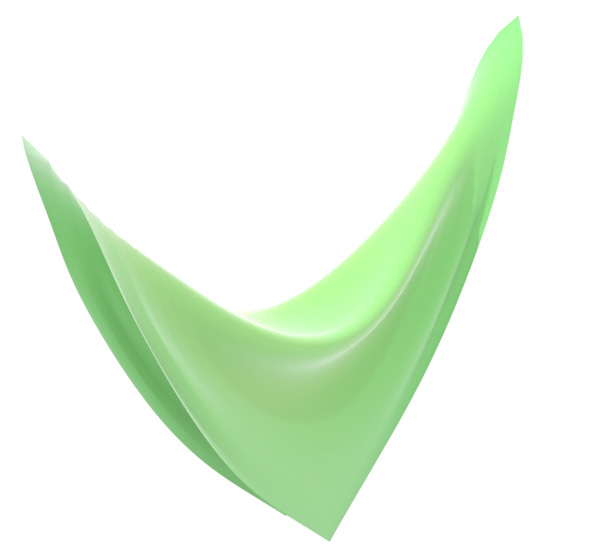}
      \caption*{\centering $E = 2 \times 10^4$}
    \end{subfigure}
    \begin{subfigure}[b]{0.11\textwidth}
      \centering
      \includegraphics[width=\linewidth]{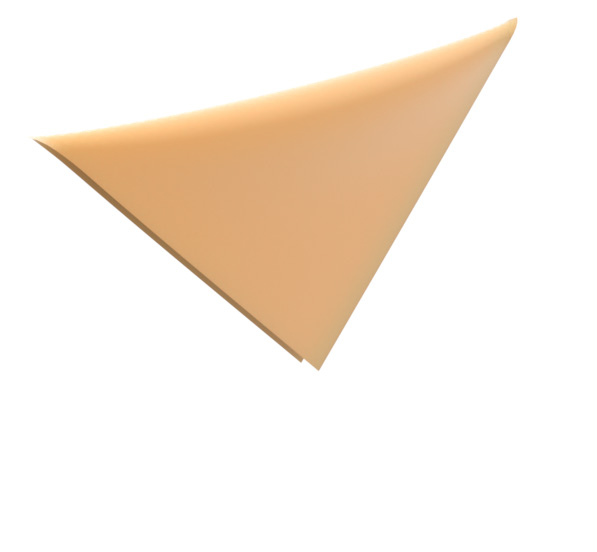}
      \caption*{\centering $E = 2 \times 10^6$}
    \end{subfigure}
    \begin{subfigure}[b]{0.11\textwidth}
      \centering
      \includegraphics[width=\linewidth]{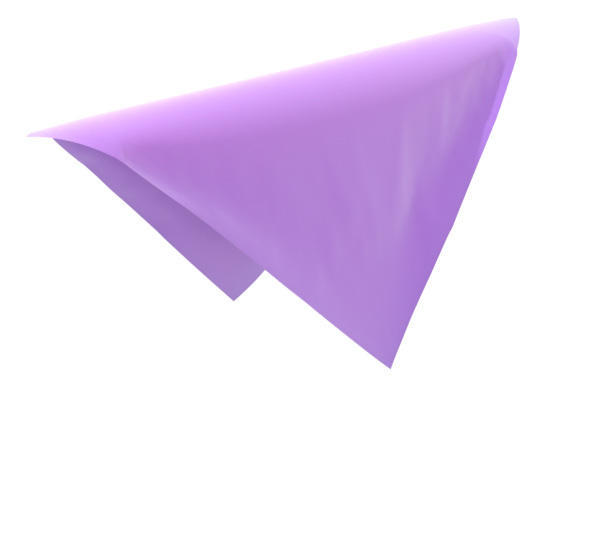}
      \caption*{\centering $E = 2 \times 10^8$}
    \end{subfigure}
    \begin{subfigure}[b]{0.11\textwidth}
      \centering
      \includegraphics[width=\linewidth]{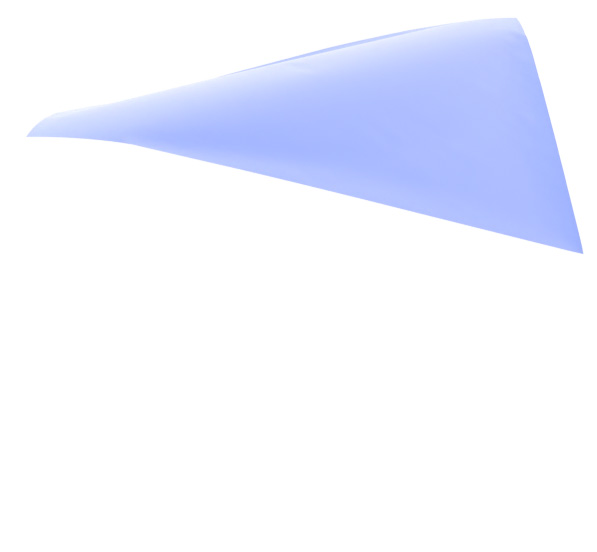}
      \caption*{\centering $E = 2 \times 10^{10}$}
    \end{subfigure}
    \caption{\textbf{Diagonal hanging cloth.}  
Simulation of a square cloth suspended by two diagonal corners under gravity, with zero bending stiffness and varying membrane Young’s modulus $E$.  
B-spline FEM (top row) consistently produces more pronounced bending compared to linear FEM (bottom row), demonstrating reduced membrane locking across stiffness settings.
}\label{fig: locking test}
\end{figure}

To evaluate the membrane locking behavior of our B-spline FEM, we first simulate a standard hanging cloth test. {To eliminate locking artifacts arising from geometry, we generate mesh with square grids with alternating diagonals for linear FEM evaluation. This standard and easily generated mesh layout reduces the severe locking effects associated with uniformly oriented diagonals and serves as a fair benchmark.} The cloth is suspended by fixing two diagonal corner vertices and allowed to deform under gravity, with varying membrane stiffness and zero bending stiffness. In the ideal solution, the cloth would bend freely downward; however, small in-plane deformations in the discrete setting induce artificial resistance due to membrane locking, which becomes more pronounced with increasing membrane stiffness. 
As shown in~\autoref{fig: locking test}, our B-spline FEM consistently produces more pronounced bending deformations compared to linear FEM across all stiffness settings, indicating an effective reduction in membrane locking artifacts.

\begin{figure}[h]
    \centering
    \begin{subfigure}[b]{0.23\textwidth}
      \centering
      \includegraphics[width=\linewidth]{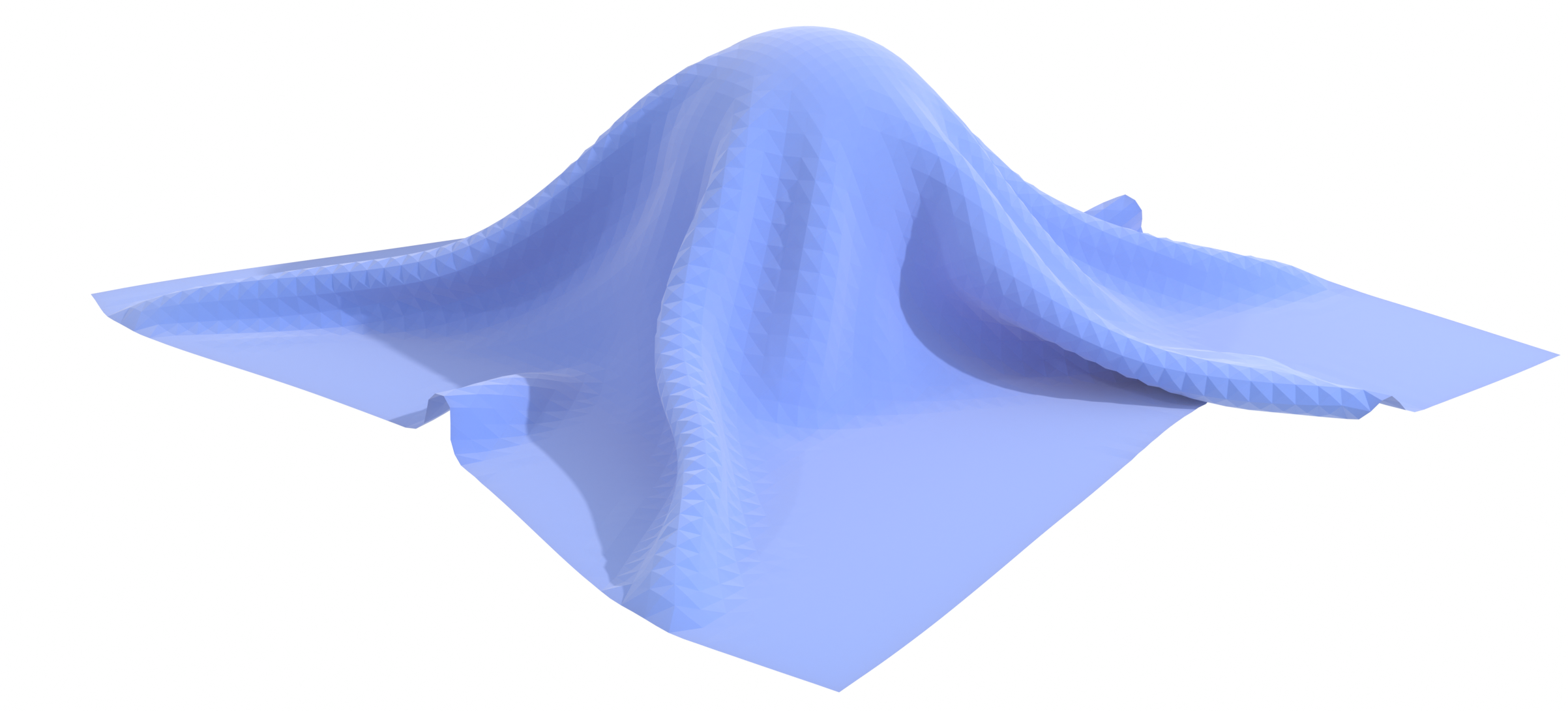}
      \caption*{\centering B-spline, $75 \times 75$}
    \end{subfigure}
    \begin{subfigure}[b]{0.23\textwidth}
      \centering
      \includegraphics[width=\linewidth]{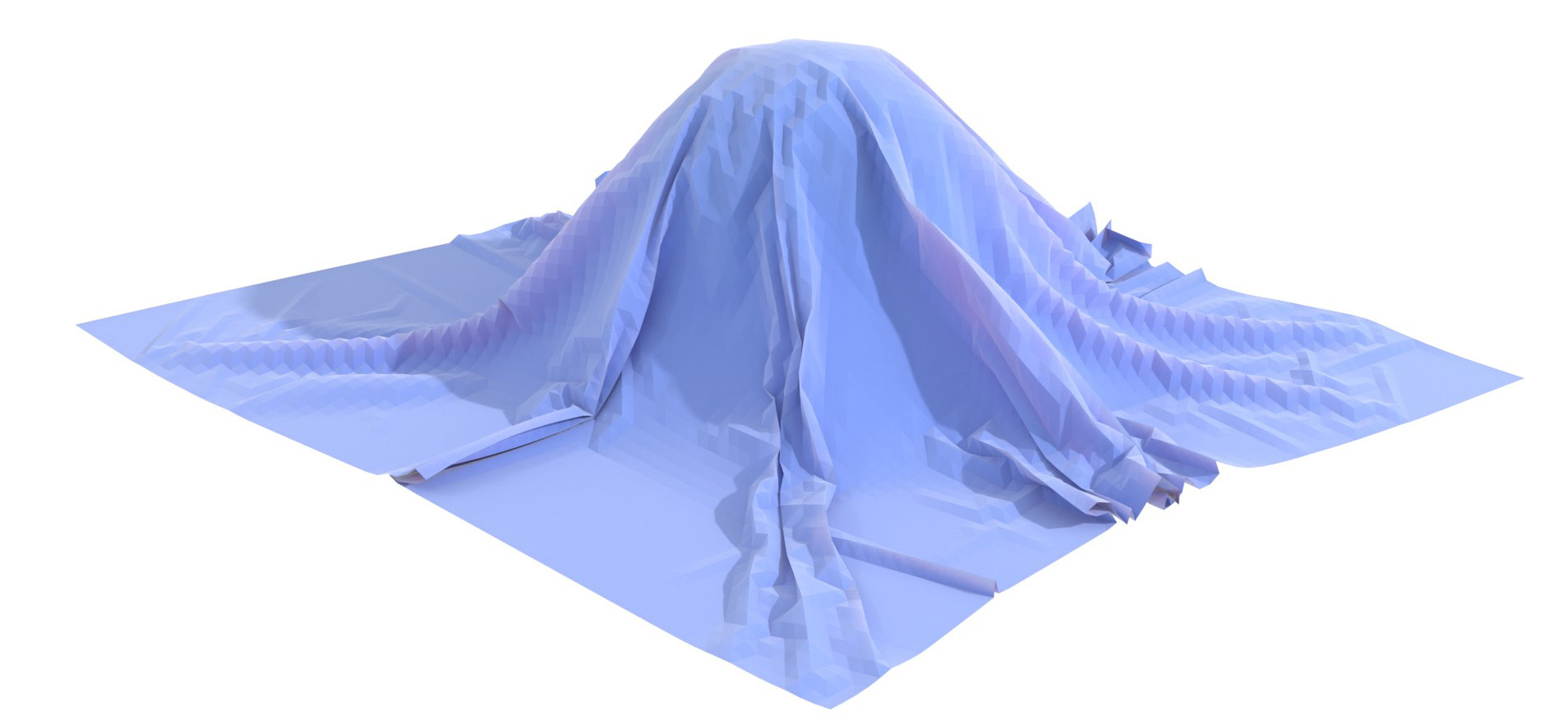}
      \caption*{\centering Linear, $85 \times 85$}
    \end{subfigure}
    \caption{\textbf{Cloth on sphere.} Simulation of a square cloth falling onto a sphere on the ground. Mesh resolution used for handling contact is indicated in parenthesis. Flat shading is applied for clear presentation of geometry. Our method (left) yields smooth wrinkles, while linear FEM (right) suffers from severe membrane locking issue.}
    \label{fig: cloth on sphere}
\end{figure}

Next, we simulate a $1\,\mathrm{m} \times 1\,\mathrm{m}$ square cloth falling onto a static sphere of radius $0.15\,\mathrm{m}$ placed on a ground plane, following the membrane locking benchmark from~\citet{li2020codimensional}. As shown in~\autoref{fig: cloth on sphere}, linear FEM exhibits pronounced locking artifacts, manifesting as unnatural sharp creases. In contrast, our B-spline FEM produces smooth and visually realistic wrinkle patterns using the same number of degrees of freedom.
While some membrane locking behavior remains (as previously illustrated in~\autoref{fig: locking test}), our method effectively avoids the visual artifacts observed in linear FEM and does so without the need for strain-limiting techniques.

\subsection{Validation of Energy Discretization}

\begin{figure}[h]
    \centering
    \hspace{-0.02\textwidth}
    \begin{minipage}[b]{0.24\textwidth}
        \includegraphics[width=\textwidth]{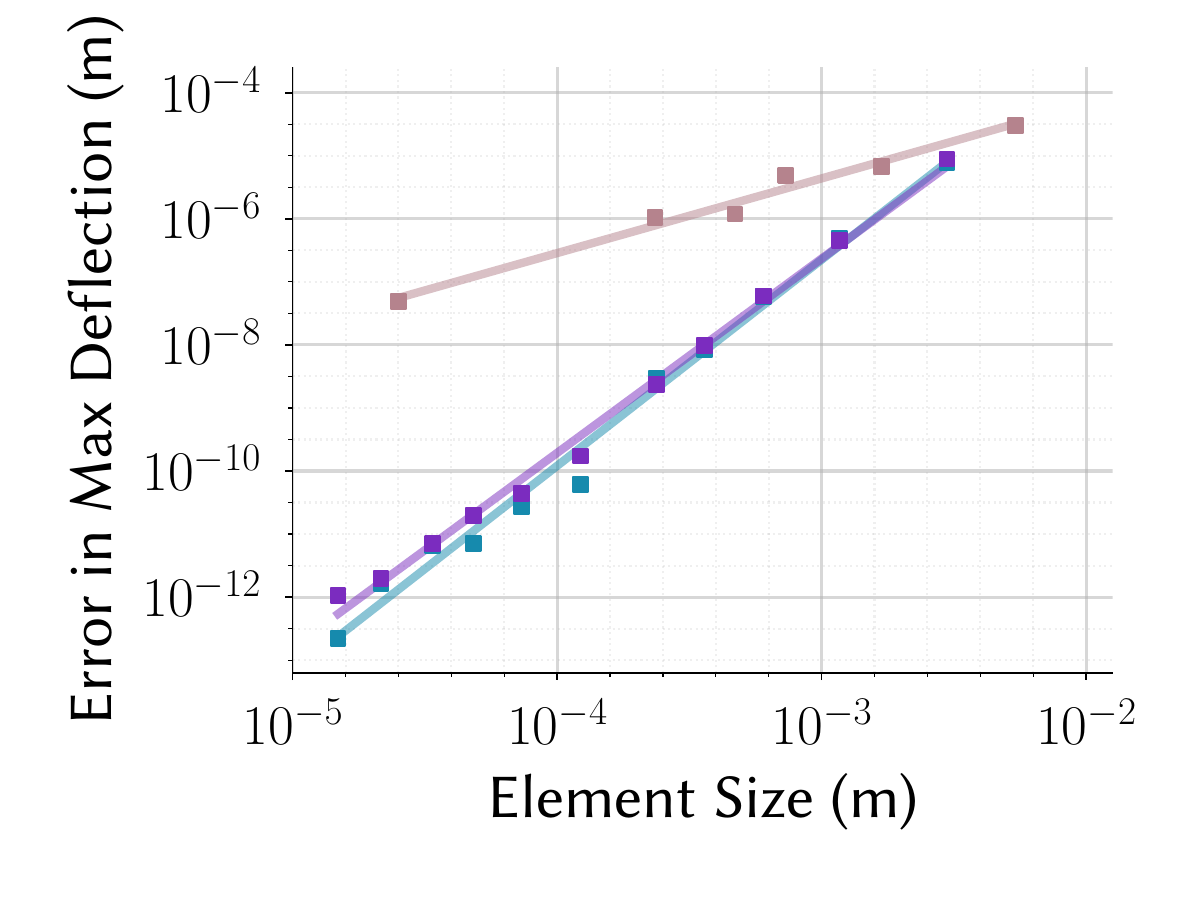}
    \end{minipage}
    \hspace{-0.02\textwidth}
    \begin{minipage}[b]{0.24\textwidth}
        \includegraphics[width=\textwidth]{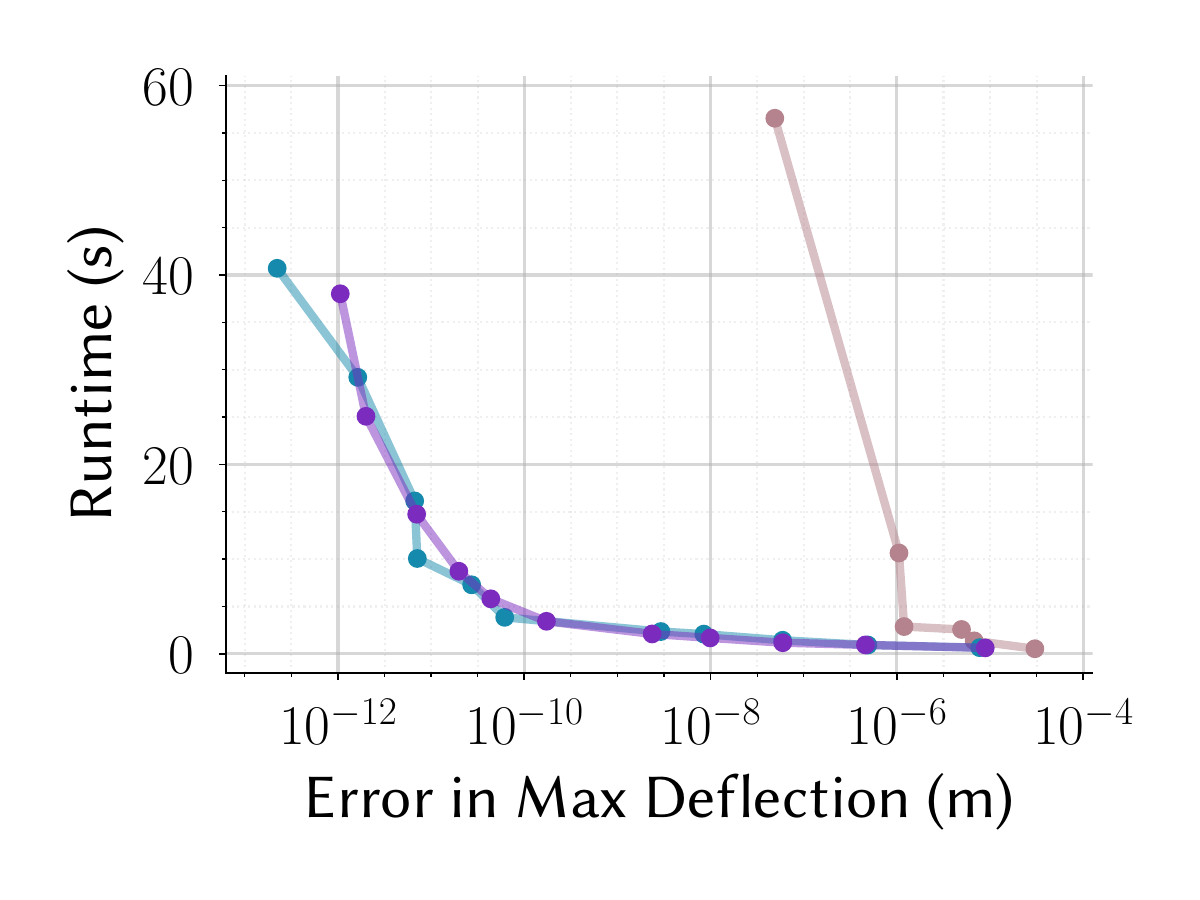}
    \end{minipage}
    \begin{minipage}[b]{0.48\textwidth}
        \includegraphics[width=\textwidth]{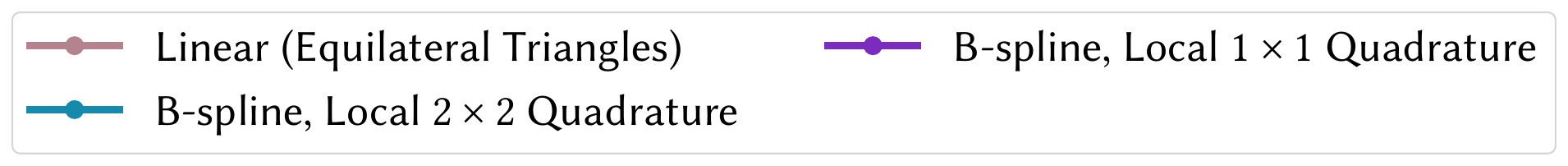}
    \end{minipage}
    \caption{\textbf{Convergence under refinement (bending).} Left: Convergence of maximum deflection error w.r.t. element size for linear FEM and our B-spline FEM with $1 \times 1$ and $2 \times 2$ quadrature. Both B-spline schemes converge to the analytic solution with significantly higher accuracy.  
    Right: Runtime versus error. Our B-spline FEM with reduced $1 \times 1$ quadrature achieves comparable or better accuracy with orders of magnitude lower runtime than linear FEM.}
    \label{fig: bending calibration}
\end{figure}

\paragraph{Convergence under Refinement (Bending).}
\label{sec: bending calibration}
We evaluate the accuracy of our bending model using the linear plate bending benchmark, and compare its convergence behavior to that of linear FEM under mesh refinement. In this experiment, we simulate a flat square plate with all sides fixed, only bending energy enabled, and a uniform load $B=9.81$ applied across the surface.
The maximum deflection of the plate should be $0.048744 B a^4 (1 - \nu^2)/(Eh^3)$, where $a$ is the side length, $\nu$ is the Poisson's ratio, $E$ is the Young's modulus, and $h$ is the thickness of the plate. We set $E=2$ MPa, $\nu=0.03, h=0.01$ m, and simulate with a large timestep size $\Delta t = 0.05$ s until equilibrium is reached. {To ensure a fair comparison, we implemented the same quadratic bending model within the linear FEM codebase and conducted all tests on equilateral triangular meshes, where linear FEM often exhibits best accuracy~\cite{liang2025corotational}.} As shown in~\autoref{fig: bending calibration}, our reduced integration scheme with $1 \times 1$ quadrature achieves accuracy comparable to linear FEM on a mesh of 1.5K regular triangles while providing a $50\times$ speedup. When constrained to the same runtime, our method achieves errors several orders of magnitude lower. Increasing our quadrature density to $2 \times 2$ yields only marginal improvements in accuracy for this example, despite higher computational cost.
In contrast, linear FEM with right-triangle meshes containing alternating diagonal orientations fails to converge to the analytic solution, highlighting its sensitivity to mesh quality, particularly the configuration of edges.

\begin{figure}[htbp]
  \centering
  \includegraphics[width=\linewidth]{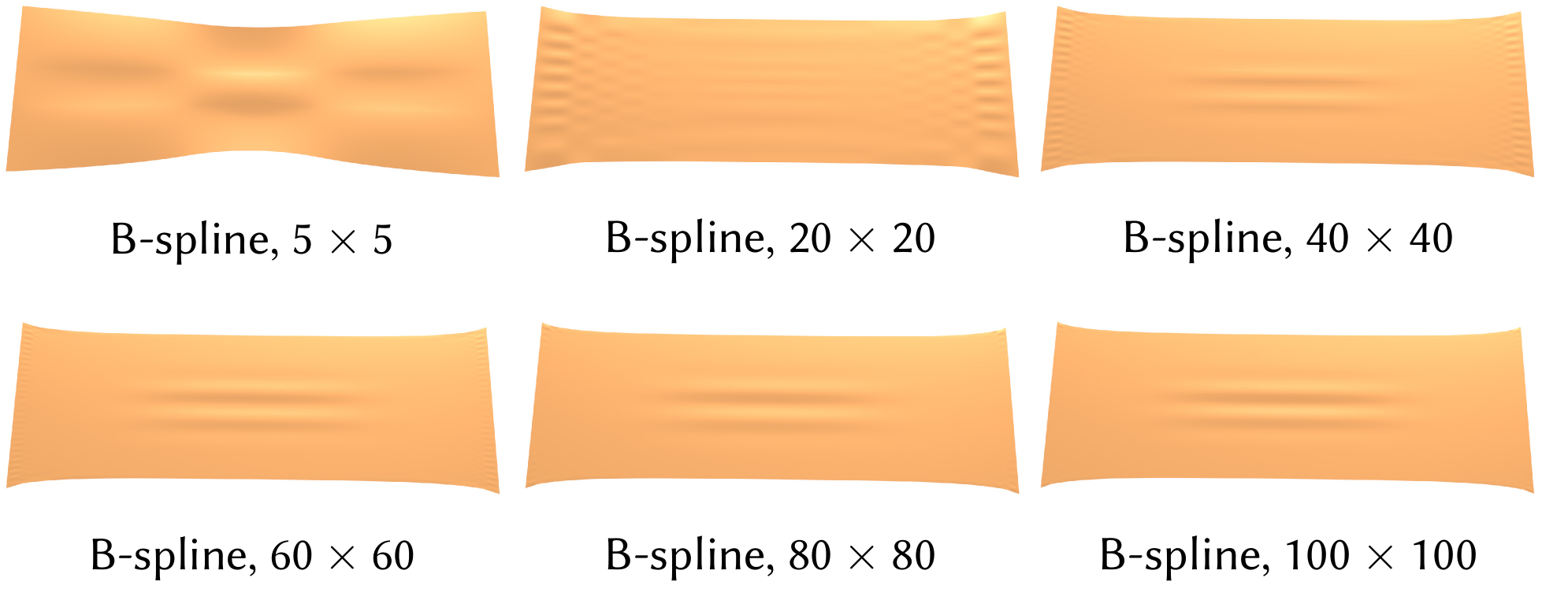}
  \vspace*{-2em}
  \caption{\textbf{Wrinkled sheets.} Significant discretization artifacts appear for mesh resolutions below $20 \times 20$. Wrinkling behavior first emerges at $40 \times 40$, with subsequent refinements producing only minor changes in wrinkle morphology after resolution reaches $60 \times 60$. Convergence of the wrinkle pattern is observed for resolutions beyond $80 \times 80$.}
  \label{fig: wrinkling figure}
\end{figure}

\addtocounter{figure}{1}

\begin{figure*}[h]
    \captionsetup{skip=2pt}
    \centering
    \includegraphics[width=\linewidth]{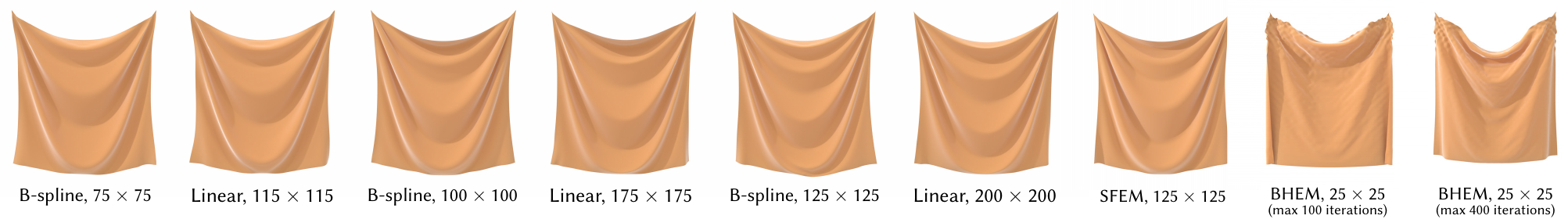}
    \caption{\textbf{Upright hanging cloth.} Simulation of a hanging cloth with top corners pinned. As resolution increases, the wrinkles gradually evolved from three layers to four. Our method captures similar wrinkling behavior with fewer DOFs than linear FEM, achieving a $2\times$ speedup compared to linear FEM, $5\times$ speedup compared to SFEM, and much better convergence compared to BHEM which fails to resolve the wrinkling on the corner after 400 Newton iterations. 
}\label{fig: upright hanging cloth}
\end{figure*}

\begin{figure*}[h]
    \centering
    \captionsetup{skip=2pt}
    \includegraphics[width=\linewidth]{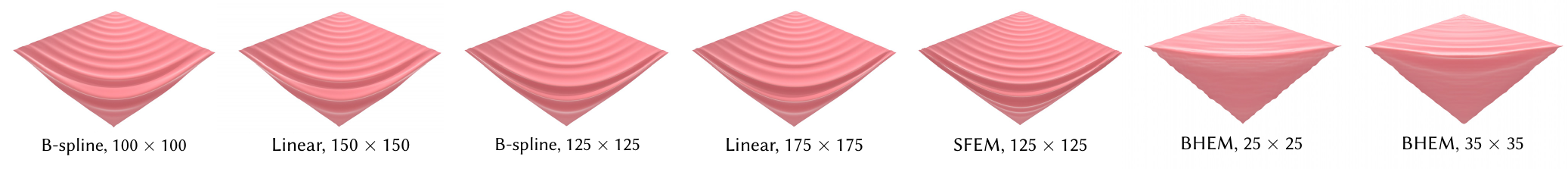}
    \caption{\textbf{Drape test.} Simulation of a piece of cloth laid flat and hanging on two adjacent edges. Our method captures similar wrinkle details using only half the DOFs compared to linear FEM, resulting in a $2 \times$ speedup. On the high-resolution cases, we achieve at least a $10\times$ speedup compared to other high-order FEM methods, where their convergence becomes slow.}
    \label{fig: drape test}
\end{figure*}

\begin{figure*}[h]
    \centering
    \captionsetup{skip=2pt}
    \includegraphics[width=\linewidth]{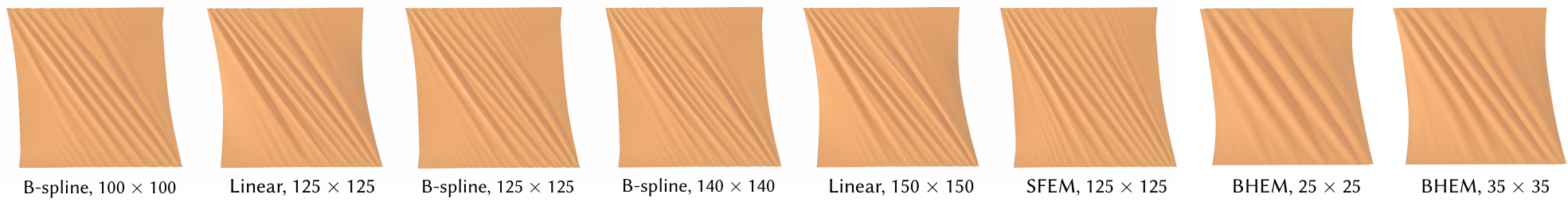}
    \caption{\textbf{Shear test.} Simulation of cloth sheet laid flat, with opposite edges pulled, exerting force in shearing directions. Our method accurately captures sharp shearing wrinkles while maintaining $2\times$ speedup compared to all other methods.}
\label{fig: shear test}
\end{figure*}

\addtocounter{figure}{-4}

\begin{figure}[htbp]
  \centering
  \includegraphics[width=\linewidth]{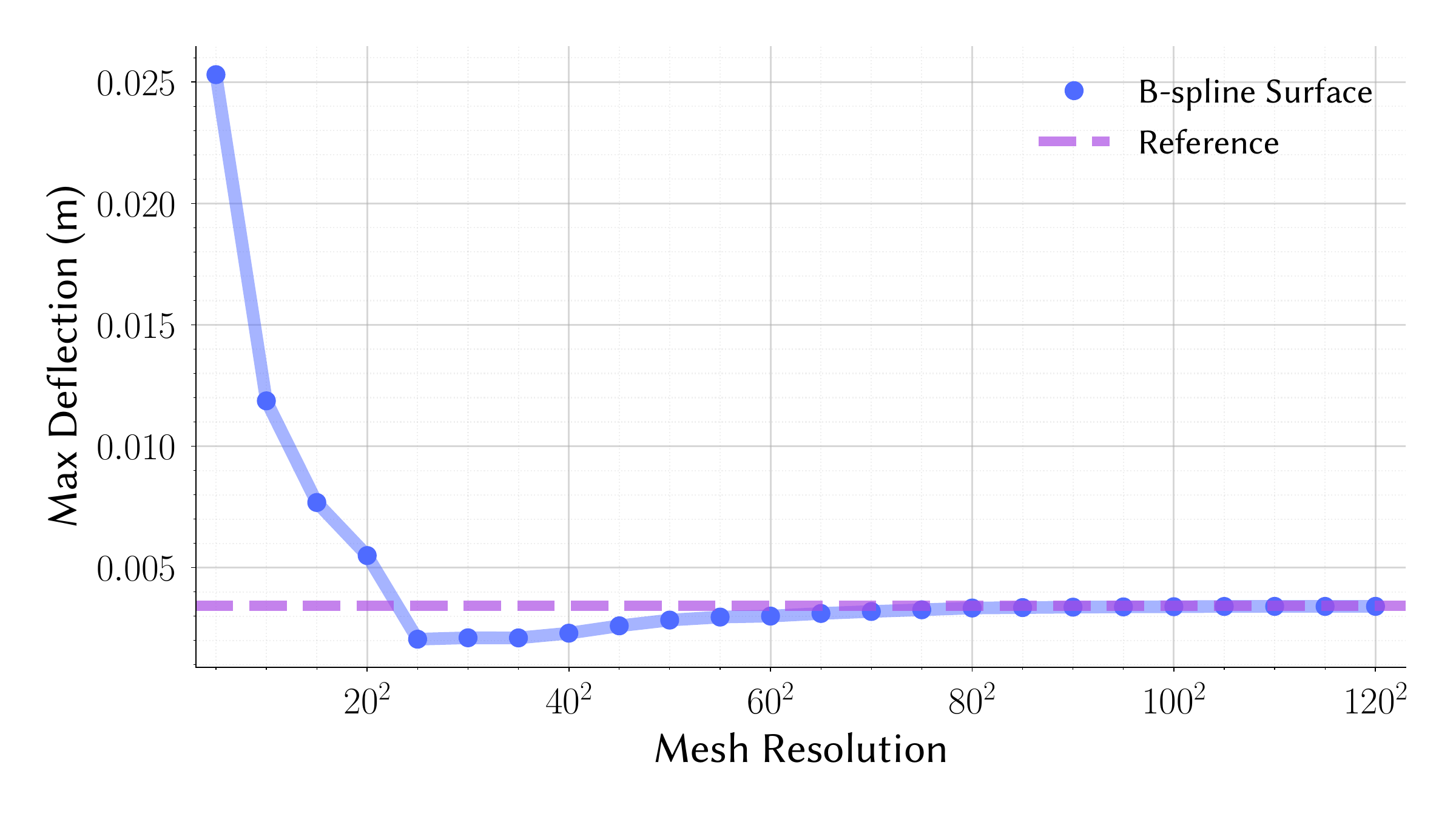}
  \vspace*{-2em}
  \caption{\textbf{Convergence under refinements (membrane).} \rev{Both reduced integration and the local \(2 \times 2\) quadrature rule converge to the analytic maximum deflection under mesh refinement. The \(2 \times 2\) rule is more accurate on coarse meshes, while the convergence of reduced integration confirms its consistency for membrane elasticity.}}
  \label{fig: wrinkling convergence}
\end{figure}

\paragraph{Convergence under Refinement (Membrane).}
\label{sec: membrane calibration}
{We further validate the accuracy of our model's membrane energy discretization using the wrinkling sheet benchmark~\cite{Wang2019Wrinkling}. This problem involves a quasi-static simulation of a flat rectangular plate with an aspect ratio of 2.5, where the shorter edges are fixed and the longer edges are stretched by a factor of 1.1. Gravity is disabled, and material parameters are set to $E = 10^6$~Pa and $\nu = 0.5$. When sufficiently stretched, the planar configuration of the sheet becomes unstable, undergoing a bifurcation to an out-of-plane wrinkled state. For a $1.0~\mathrm{m} \times 2.5~\mathrm{m}$ plate, the maximum out-of-plane deflection is expected to reach 3.45~mm with St. Venant-Kirchhoff (StVK) energy model. To capture this symmetry-breaking behavior numerically and initiate buckling, a small sinusoidal perturbation of amplitude 0.02~m is applied along the thickness direction in the initial configuration~\cite{ni2024simulating,Chen2021wrinkling}. With increasing mesh resolution, our method exhibits the same wrinkling behavior as the reference solution (\autoref{fig: wrinkling figure}) and converges toward a physically realistic maximum deflection (\autoref{fig: wrinkling convergence}).} \rev{Our reduced integration scheme converges to the same analytical solution as the local $2\times 2$ Gaussian quadrature, with comparable convergence rate in the fine-mesh regime.}

\addtocounter{figure}{3}

\addtocounter{figure}{1}

\begin{figure*}[h]
    \centering
    \includegraphics[width=\linewidth]{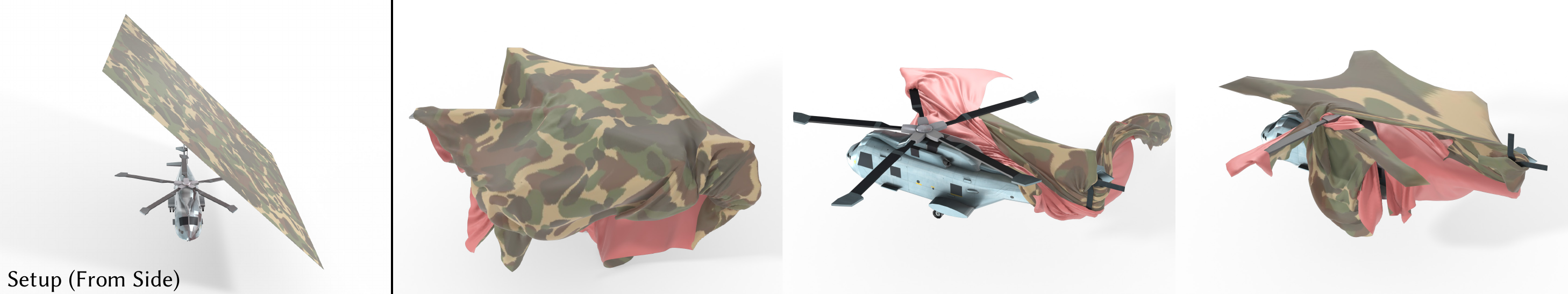}
    \caption{\textbf{Helicopter.} A square cloth sheet falls onto the helicopter from an oblique angle and folds along the helicopter geometry as the propellors rotate. The front and back of the cloth sheet are distinguished by color (camouflage/pink). Our method reproduces rich twists and wrinkles with a $200 \times 200$ B-spline surface and linear triangular contact mesh, while remaining stable under stiff contact.}
    \label{fig: helicopter}
\end{figure*}

\paragraph{\rev{Support of Non-Rectangular Domains}}
\rev{Our method supports simulation on non-rectangular domains by constructing structured B-spline surfaces from irregularly shaped triangular meshes. The mesh-conforming pipeline extracts boundary curves, fits segmented boundaries using B-spline curves, and generates a control net in the interior, with optional refinement to improve resolution. This enables simulation on irregular patches with imposed boundary constraints, which are commonly used to represent garment domains in linear FEM frameworks. Details of the conforming process are provided in the supplemental document.}

\begin{figure}
  \centering
  \includegraphics[width=\linewidth]{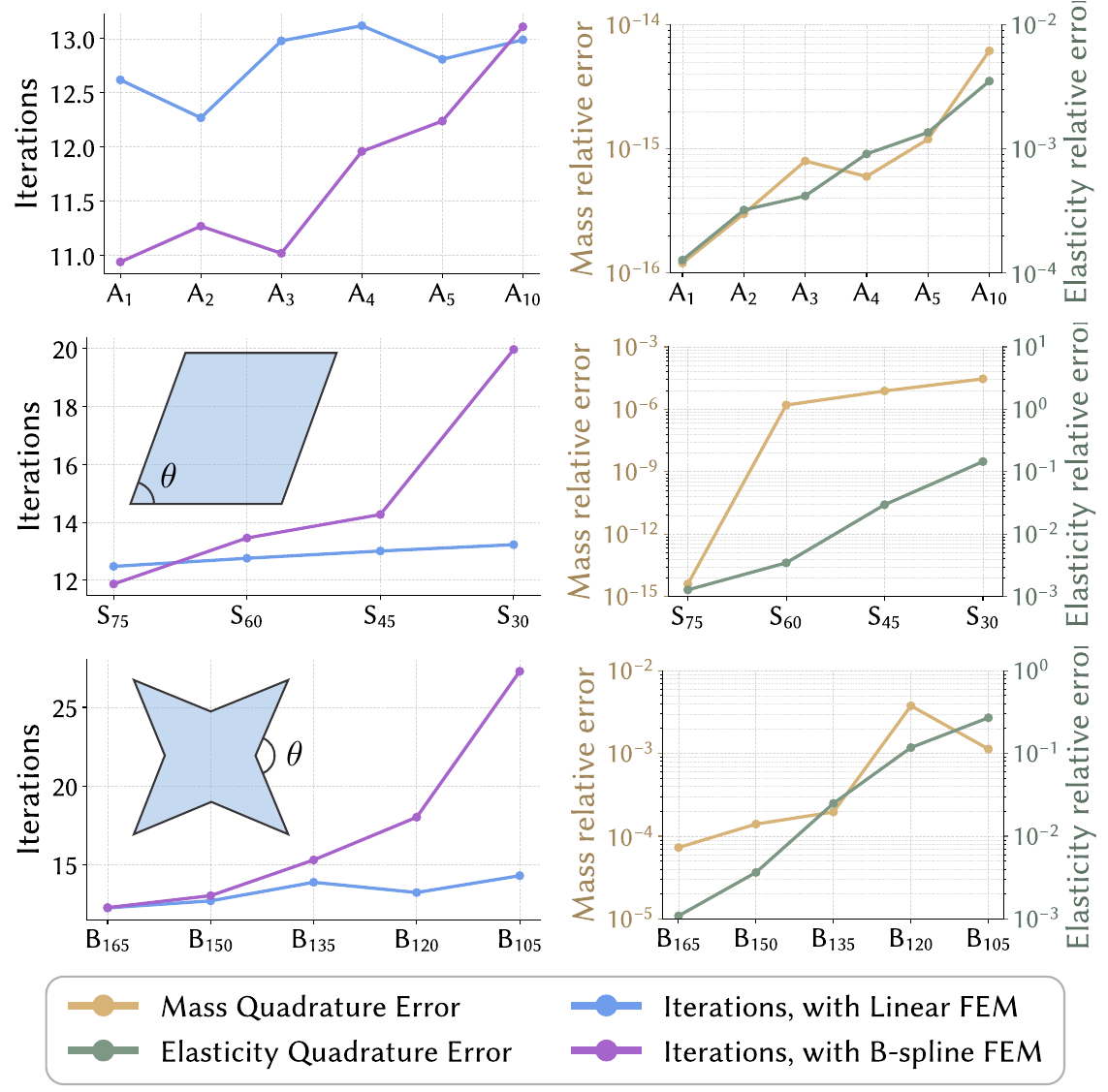}
  \caption{\textbf{Integration error on non-rectangular domains.} Three categories of geometric distortion tested on a contact scenario of falling onto a bunny model: aspect ratios from 1:1 to 1:10 ($A_1$--$A_{10}$), parallelograms with minimum interior angle from $75^\circ$ down to $30^\circ$ ($S_{75}$--$S_{30}$), and boundaries with a sharp mid-edge bend from $165^\circ$ down to $105^\circ$ ($B_{165}$--$B_{105}$). Left: averaged solver iteration counts, compared against linear FEM on the same mesh. Right: relative quadrature error of the mass matrix and elasticity energy, measured against a high resolution linear FEM reference. Quadrature errors grow with distortion but remain small across the range representative of real garment patches. Iteration counts rise sharply for extreme distortions but remain close to the rectangular baseline for mildly irregular shapes, which are representative of typical garment patches.}
  \label{fig: skew measure}
\end{figure}

\rev{As the target domain deviates from a rectangular parameterization, the mapping $\bm{X}(u, v)$ becomes increasingly anisotropic, which can introduce additional quadrature error. To evaluate the robustness of our reduced integration scheme under such conditions, we consider three categories of geometric distortions: (i) varying aspect ratios, (ii) skewed corner angles, and (iii) sharp turns within edge (\autoref{fig: skew measure}). Across all three categories, mass and elasticity quadrature errors remain below $10^{-3}$, and solver iteration counts stay within a small constant factor of the rectangular baseline. Aspect ratio has the mildest effect, with the Jacobian remaining well-conditioned even at 1:10 elongation; elongated patches can additionally be accommodated by assigning different numbers of control points to each side. Skewed corners and sharp boundary bends show more pronounced error growth below $45^\circ$ and $120^\circ$ respectively, which are more extreme than typical garment panels.}

\subsection{Comparison on Square Cloth Sheet}\label{sec: comparison on square cloth sheet}

We compare the visual quality and efficiency of our method over linear FEM, SFEM~\citep{SecondOrderFEM2023} and BHEM~\citep{ni2024simulating} using the benchmark problems in \citet{Kim2020}. In these tests, we vary the resolution of all methods, and compare the performance for pairs that result in similar stable configurations. Because the BHEM implementation adopts an StVK energy formulation, its material parameters cannot be matched exactly; instead, they are tuned to produce simulation results comparable to those of the other methods. The runtime is detailed in the supplemental document.

In the first experiment, we simulate a piece of cloth sheet hanging upright, with its top corners fixed, and moving towards each other at 0.5~m/s in the first 0.1 seconds. Upon reaching static equilibrium, the cloth sheet forms several layers of wrinkle. The pairing is done via identifying similar wrinkle layer and quality (\autoref{fig: upright hanging cloth}). Although much fewer DOFs are used, BHEM fails to resolve the strain near the pinned points even after 400 Newton iterations, producing artifacts near the corners. Our method gives a $2\times$ speedup averaged over all pairs. 

In the second experiment, we test how different FEM formulations perform in configurations involving both stretching and shearing. We simulate a piece of cloth laid flat, with two adjacent edges fixed. The cloth swings and gradually reaches a static configuration. We pair the results based on the wrinkles formed between the connected edges (\autoref{fig: drape test}). Our method is on average $2\times$ faster than linear FEM, $4\times$ faster compared to SFEM, and $6\times$ faster compared to BHEM.

In the third experiment, we evaluate deformation under shearing, by laying flat a piece of 1~m $\times$ 1~m cloth, fixing a pair of opposite edges, and pulling them in opposite shearing directions at 0.5~m/s for 0.1~s. We set the shearing stiffness to be the same as stretching stiffness to produce sharp shearing wrinkles (\autoref{fig: shear test}). Our method effectively captures the sharp wrinkles, with a $2\times$ speedup compared to linear FEM and BHEM, and a $4\times$ speedup compared to SFEM.

\subsection{Stress Tests}

We present additional simulation results that stress the method through scaling and contact-rich configurations, and demonstrate its ability to handle non-rectangular garment geometries.

\addtocounter{figure}{-2}

\begin{figure}[htbp]
    \centering
    \includegraphics[width=\linewidth]{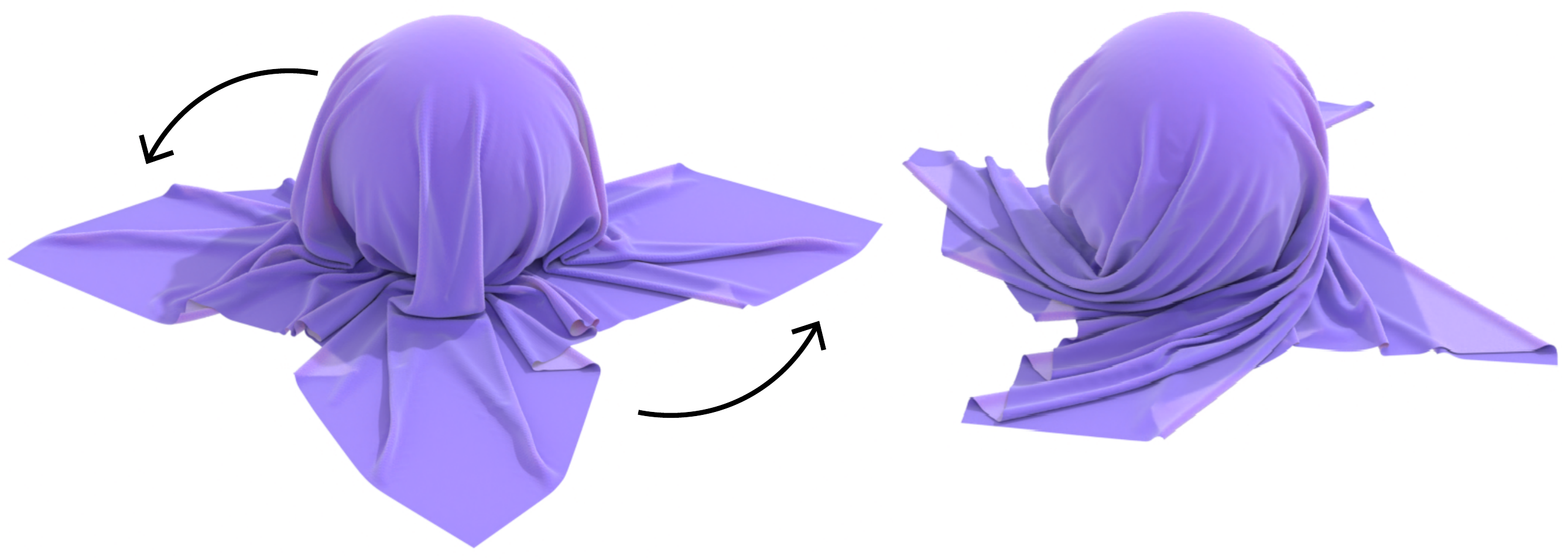}
    \caption{\textbf{Cloth on rotating sphere.} A square cloth sheet falls onto a static sphere and settles into a steady configuration (left). The sphere then rotates counterclockwise, causing the cloth to fold inwards (right). Our method faithfully resolves the fine wrinkles with a $200 \times 200$ B-spline surface and $200 \times 200$ linear triangular mesh.}
\label{fig: cloth on rotating sphere}
\end{figure}

\addtocounter{figure}{1}

\paragraph{Cloth on Rotating Sphere.}
To demonstrate compatibility with frictional contact handling in IPC~\cite{Li2020IPC} at large mesh resolutions, we simulate a square cloth sheet falling onto a static sphere with friction of $\mu = 0.4$ and $\epsilon_v = 10^{-2}$. After the cloth sheet settles into a steady configuration, the sphere rotates counterclockwise, inducing intricate wrinkles on the cloth due to frictional contact with both the sphere and the ground (\autoref{fig: cloth on rotating sphere}). Our method captures these rich wrinkle patterns using a $200 \times 200$ B-spline surface and a $300 \times 300$ linear triangular mesh for contact.

\paragraph{Helicopter.}
To further stress-test our integration with IPC, we consider more challenging contact scenarios involving dense contact and high-resolution discretizations. We simulate a square cloth sheet of size 3~m $\times$ 3~m falling from an oblique angle of $45^{\circ}$ onto a helicopter model 1~m above the ground~(\autoref{fig: helicopter}). The helicopter propeller and back rotor geometry act as a moving boundary, rolling and folding the square cloth sheet into the target model shape, and slowly coming to rest as the cloth sheet is fully stretched. Our method faithfully resolves the fine wrinkles with a $200 \times 200$ B-spline surface and a $200 \times 200$ linear triangular mesh under dense contact.

\begin{figure*}[t]
  \centering
  \includegraphics[width=\linewidth]{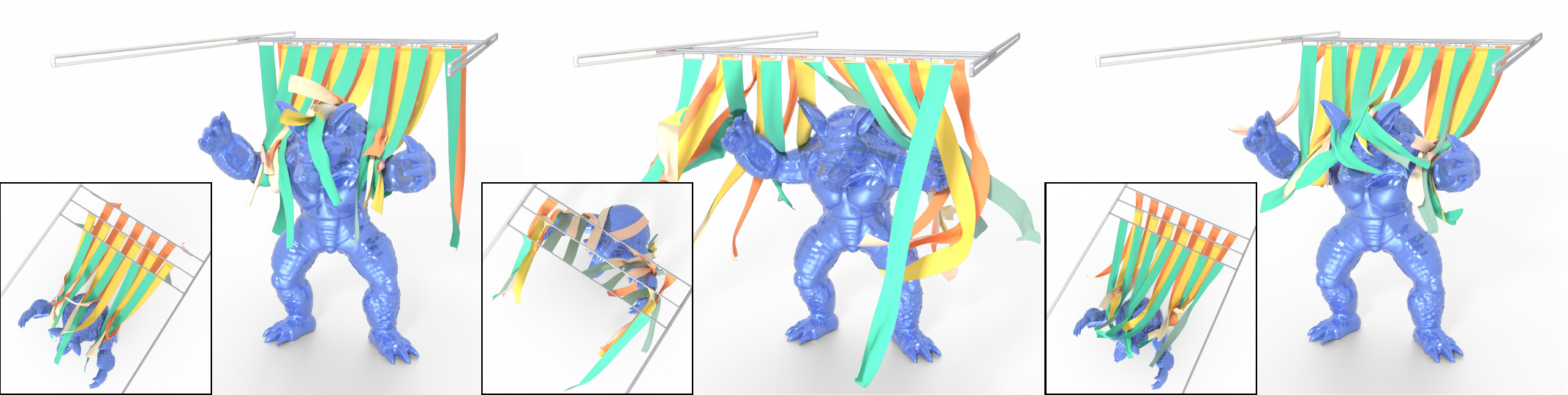}
  \caption{\textbf{Stripes sweeping over armadillo.} Three rows of stripes are attached to the rack and sweep over the armadillo model as the rack moves back and forth, bending under contact with the armadillo and twisting as the stripes collide with each other. The simulation remains free of artifacts under the high aspect ratio. The smoothness afforded by quadratic B-spline continuity captures complex curvature and twist along the short edge using only 10 control points.}
  \label{fig: brush over armadillo}
\end{figure*}

\paragraph{Stripes.}
We perform assessments using rectangular meshes with high aspect ratios to verify that simulation robustness is independent of square mesh shapes. In this scenario, 20 stripes, modeled as B-spline surfaces of resolution $10 \times 150$, are attached to a rack along their top edge. The rack moves back and forth around an armadillo model, causing the stripes to sweep across its surface and twist around one another (\autoref{fig: brush over armadillo}). The $C^1$-continuity of quadratic B-spline basis functions captures smooth bending and twisting behavior along the short edge of each stripe with only 10 control points.

\addtocounter{figure}{1}

\paragraph{Garment Simulation on Character Animation.} \rev{B-spline surface patches generated by our mesh-conforming pipeline can be readily used for garment simulation.} We evaluate our method on three sets of character animations with different garment types (\autoref{fig: character animation}). Our method accurately captures garment dynamics and fine-scale wrinkling induced by large-amplitude body motion, while remaining robust on non-rectangular mesh patches and near patch seam boundaries, without introducing visual or geometric artifacts.

\paragraph{Garment Colliding and Piling.}
We further test the method's capability to handle contact among multiple B-spline surfaces of non-rectangular, high-resolution meshes. In this scenario, the garments from the previous testcase collide with each other and accumulate inside a cubic basket with fixed corners, whose faces are also modeled as B-spline surfaces (\autoref{fig: teaser}). With IPC integration, our method robustly handles dynamic contact among the garments and between the garments and the basket, capturing the resulting fabric deformations induced by collision.

\subsection{Ablation Study on Solver Optimization}\label{sec: ablation}

\begin{figure}[htbp]
    \centering
    \hspace*{-1em}
    \includegraphics[width=0.49\textwidth]{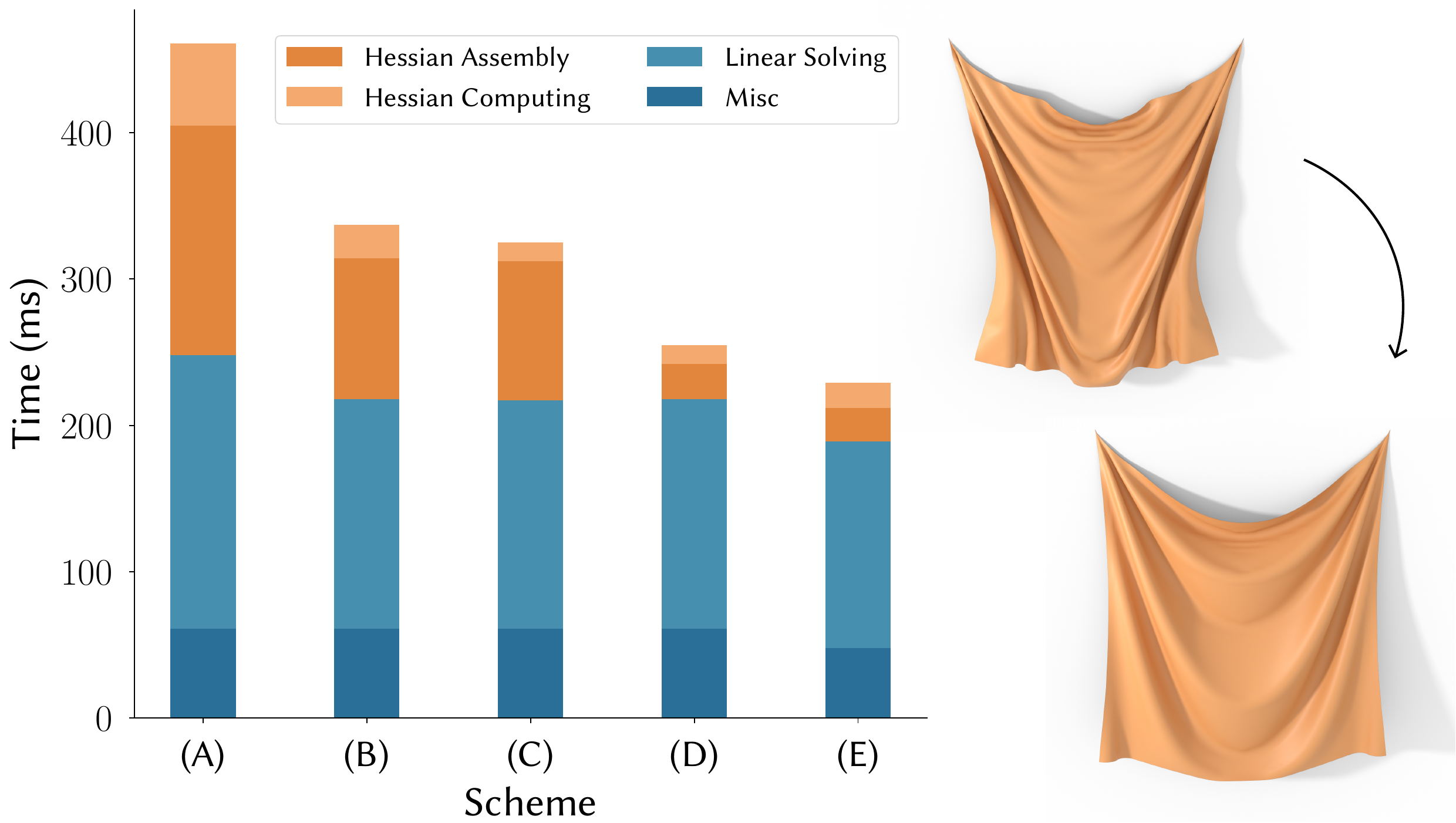}
    \caption{\textbf{Timing breakdown (reduced integration).} Runtime decomposition per Newton iteration. (A) Standard $2 \times 2$ quadrature on interior knot spans. (B) Our reduced integration scheme. (C) Adds bending Hessian precomputation. (D) Further adds parallel Hessian assembly. (E) Linear FEM baseline. Our optimizations (B-D) significantly reduce runtime, approaching that of linear FEM at the same resolution.}\label{fig:time_breakdown}
\end{figure}

\paragraph{Reduced Integration Runtime Breakdown.}\label{sec: reduced integration runtime breakdown}
We evaluate the impact of our performance optimizations by progressively incorporating reduced integration, Hessian precomputation, and parallel Hessian assembly, and measure the resulting runtime improvements (\autoref{fig:time_breakdown}). The simulation uses a $100 \times 100$ B-spline surface for ablation and a $100 \times 100$ triangular mesh as a reference case, shown in \autoref{fig: upright hanging cloth}, with statistics averaged over 100 Newton iterations.

\begin{figure}[htbp]
    \centering
    \setlength{\abovecaptionskip}{2pt} 
    \hspace{-0.02\textwidth}
    \begin{minipage}[b]{0.24\textwidth}
        \includegraphics[width=\textwidth]{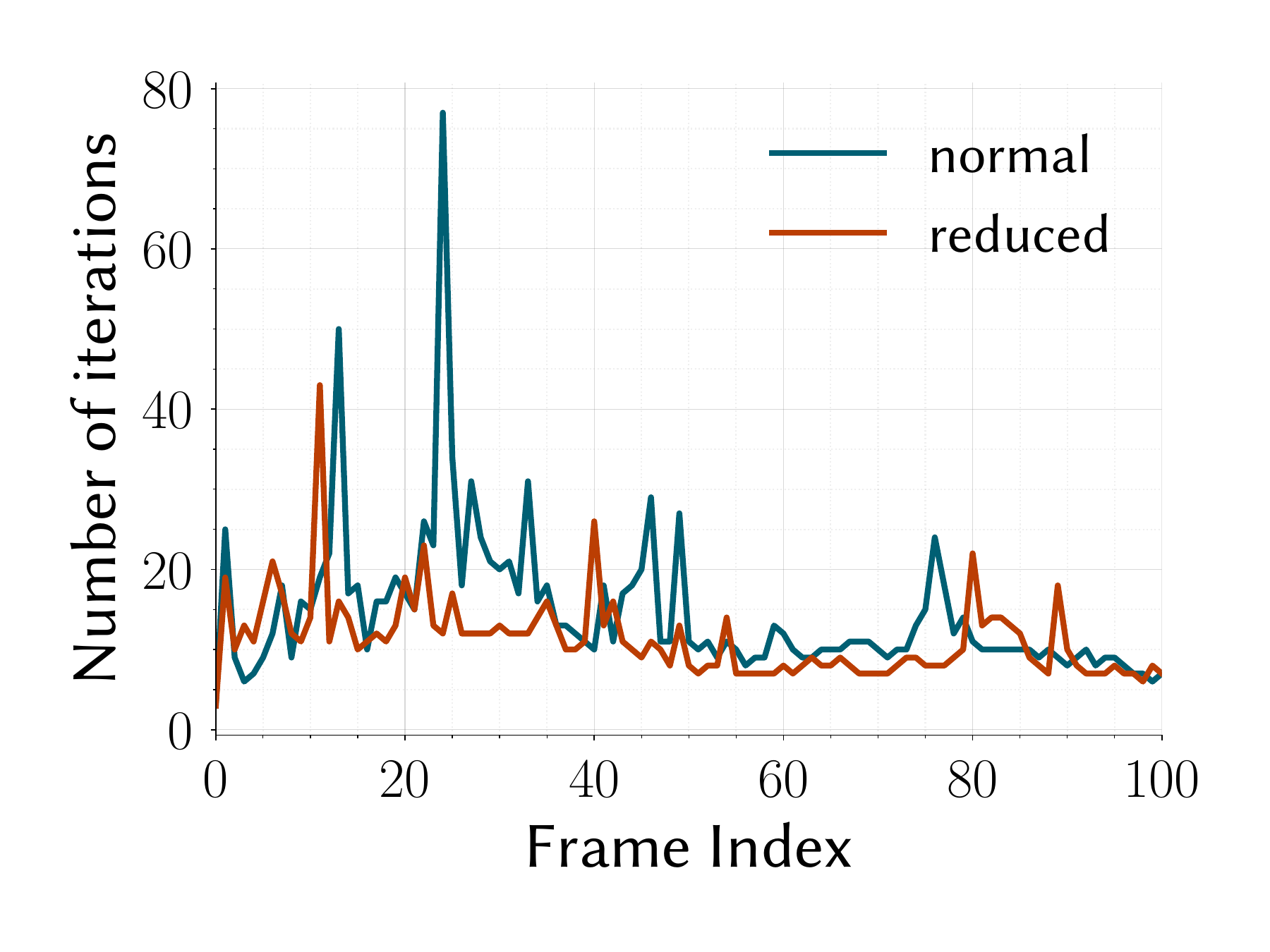}
    \end{minipage}
    \hspace{-0.02\textwidth}
    \begin{minipage}[b]{0.24\textwidth}
        \includegraphics[width=\textwidth]{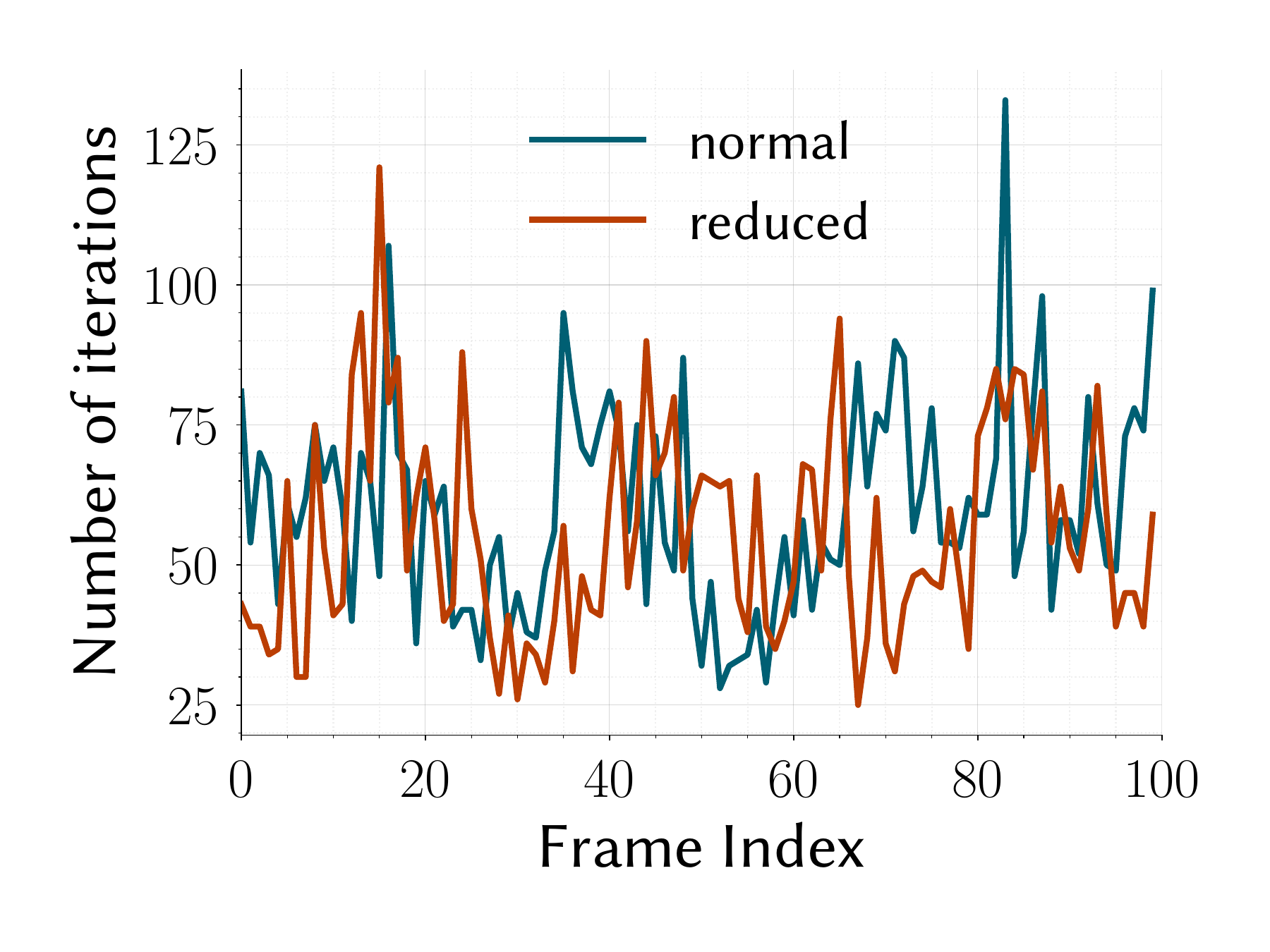}
    \end{minipage}
    \caption{\textbf{Newton convergence with and without reduced integration.} Newton iteration counts over the first 100 timesteps of simulation in upright hanging cloth as in~\autoref{fig: upright hanging cloth} (left) and in character animation as in case (a) in~\autoref{fig: character animation} (right). Reduced integration does not negatively impact the convergence rate of the Newton solver, with or without contact.}
    \label{fig: reduced integration convergence}
\end{figure}

In (A), we apply a uniform $2 \times 2$ quadrature rule for both membrane and bending energies for interior knot spans, and boundary placements as specified in~\autoref{fig: quadrature}. In (B), we switch to our reduced integration scheme, which significantly reduces Hessian related costs. In (C) and (D), we further incorporate bending Hessian precomputation and parallel Hessian assembly, respectively. The resulting performance approaches that of linear FEM (E) at the same resolution.
As shown in ~\autoref{fig: reduced integration convergence} reduced integration does not impair Newton convergence, and that using denser quadrature placement for membrane energy on boundary knot spans is necessary to avoid artifacts.

\paragraph{Effectiveness and Minimality of Reduced Integration.}
Our reduced integration scheme lowers the computational cost for Hessian assembly, and improves the performance of linear solving. We verify that this per-Newton-iteration optimization does not impair convergence; and that further reducing the number of quadrature points leads to significant artifacts. We compare the Newton convergence between reduced integration scheme and the standard per-knot-span $2 \times 2$ quadrature rule, in testcases with and without contact. Under the same material parameters, residue threshold $\Delta C_{\max} / \Delta t = 10^{-2}$, and simulation timestep $\Delta t = 0.01$~s, our reduced integration scheme converges in a similar number of iterations as the standard scheme (\autoref{fig: reduced integration convergence}).

\setlength{\columnsep}{10pt}
\begin{wrapfigure}{r}{0.5\linewidth}
  \includegraphics[width=\linewidth]{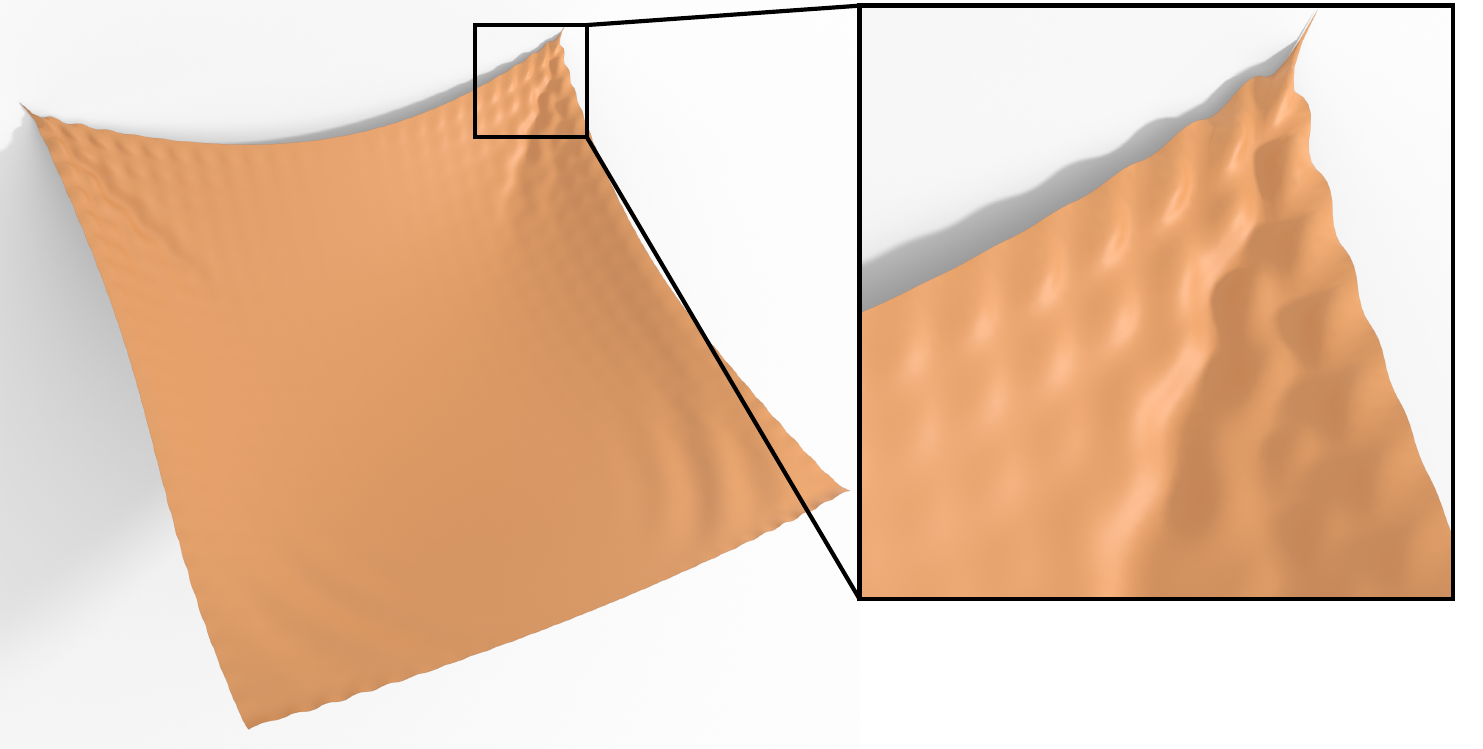}
  \caption{ \textbf{$1 \times 1$ quadrature placement in the interior.} Further reducing the quadrature rule from alternating $1 \times 2$ and $2 \times 1$ patterns to a local $1 \times 1$ scheme introduces severe numerical artifacts due to the presence of hourglass modes.}\label{fig: interior quadrature placement}
\end{wrapfigure}

\begin{figure}[htbp]
    \centering
    \includegraphics[width=\linewidth]{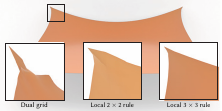}
    \caption{\textbf{Quadrature placement at corners.} 
Left: Cloth hanging from two corner points undergoing deformation.
Right: Zoom-in on the corner marked by the blue box, comparing different quadrature schemes.  
Dual grid one-point quadrature placement leads to severe tangling.  
Local $2 \times 2$ quadrature reduces tangling but leaves noticeable artifacts.  
Local $3 \times 3$ quadrature produces a smooth and physically plausible result.}
\label{fig: boundary quadrature placement}
\end{figure}

We further validate that, in general, the number of quadrature points cannot be reduced beyond our proposed scheme, either on the boundary or in the interior of the surface. For boundary quadrature, we simulate a piece of cloth sheet laid flat, hanging on two of its corners on the same edge, and swinging due to gravity (\autoref{fig: boundary quadrature placement}). Applying reduced integration uniformly to all knot spans or using a local $2 \times 2$ rule at the corners both produce noticeable artifacts. Increasing the local quadrature order to $3 \times 3$ largely mitigates the artifacts. For interior quadrature, further reducing the per-knot-span quadrature order to $1 \times 1$ results in visual artifacts caused by hourglass modes due to insufficient quadrature (\autoref{fig: interior quadrature placement}).

\begin{figure}[htbp]
  \centering
  \includegraphics[width=\linewidth]{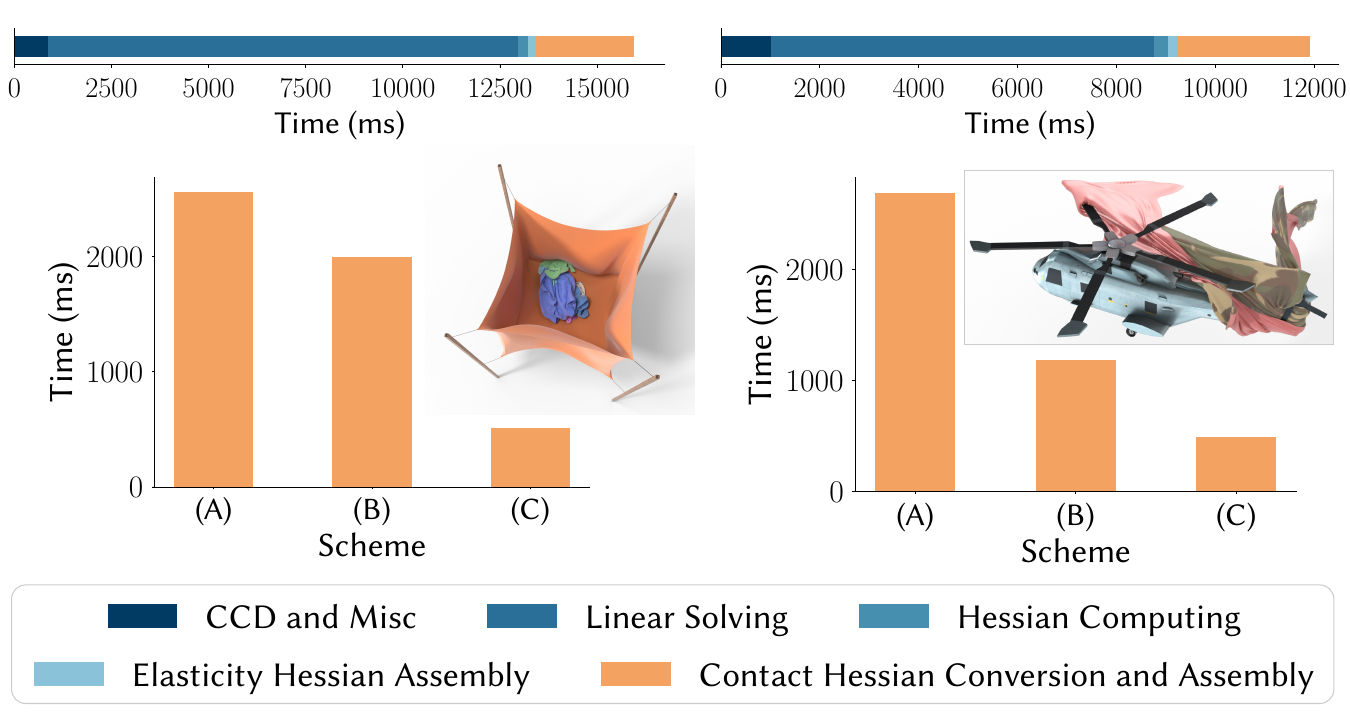}
  \caption{\textbf{Timing breakdown (contact Hessian conversion).} Runtime decomposition averaged over a contact-heavy timestep (left: frame 140 of~\autoref{fig: teaser}; right: frame 300 of~\autoref{fig: helicopter}). For each case, the top bar shows the timing breakdown under scheme~(A). (A) Triplet-based conversion and assembly. (B) Manually parallelized Hessian construction. (C) Further incorporates spatial partitioning and cache-aware optimization for better cache locality. Our method greatly reduces Hessian assembly time, preventing Hessian conversion from becoming a runtime bottleneck in contact-dense regimes where naive parallelization offers limited gains.}
  \label{fig: contact hessian conversion}
\end{figure}

\paragraph{Contact Barrier Hessian Assembly.}
We evaluate the effectiveness of our contact Hessian conversion and assembly algorithm by comparing it against triplet-based construction and unoptimized manual parallel assembly under dense contact scenarios (\autoref{fig: contact hessian conversion}). In (A), the Hessian conversion is done by constructing first the contact Hessian on the linear triangle mesh using triplets, and then propagating each entry of the linear contact Hessian to $9 \times 9 = 81$ triplets associated with the B-spline surface contact Hessian, which often leads to more than 100M triplets in contact-dense scenarios. In (B), we switch to parallel construction without spatial partitioning, where performance gains become limited due to frequent write contention. In (C) we adopt spatial partition and implementation optimization which extracts entries from hash map to improve cache coherency and thus avoids frequent iterations over hash maps. By reducing write contentions, our method preserves good speedups in contact-dense regimes.

\begin{figure}[htbp]
  \centering
  \includegraphics[width=\linewidth]{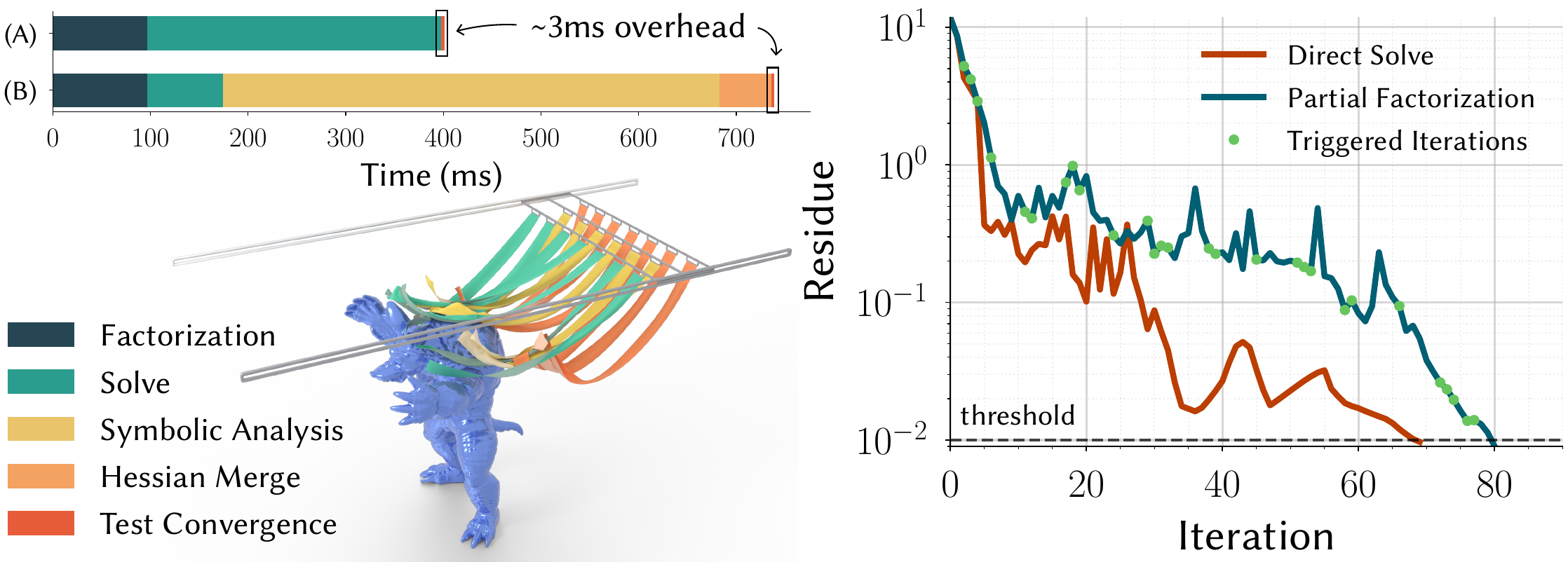}
  \caption{\textbf{Runtime and convergence of partial factorization.} Per-iteration runtime breakdown (left) and per-timestep Newton convergence (right) for frame 555 of~\autoref{fig: brush over armadillo} (low-contact scenario). Residue here is computed as $\norm{\Delta C}_{\infty} / h$. (A) uses partial factorization, reusing symbolic analysis at the cost of multiple base-matrix solves. (B) performs full merges and re-analysis each iteration. This achieves $\sim\!\!2\times$ per-iteration speedup when triggered. Minor convergence degradation due to slightly more Newton iterations is offset by frequent triggering, yielding net performance gains.}
  \label{fig: partial factorization}
\end{figure}

\setcounter{figure}{20}

\begin{figure*}[t]
  \centering
  \includegraphics[width=\linewidth]{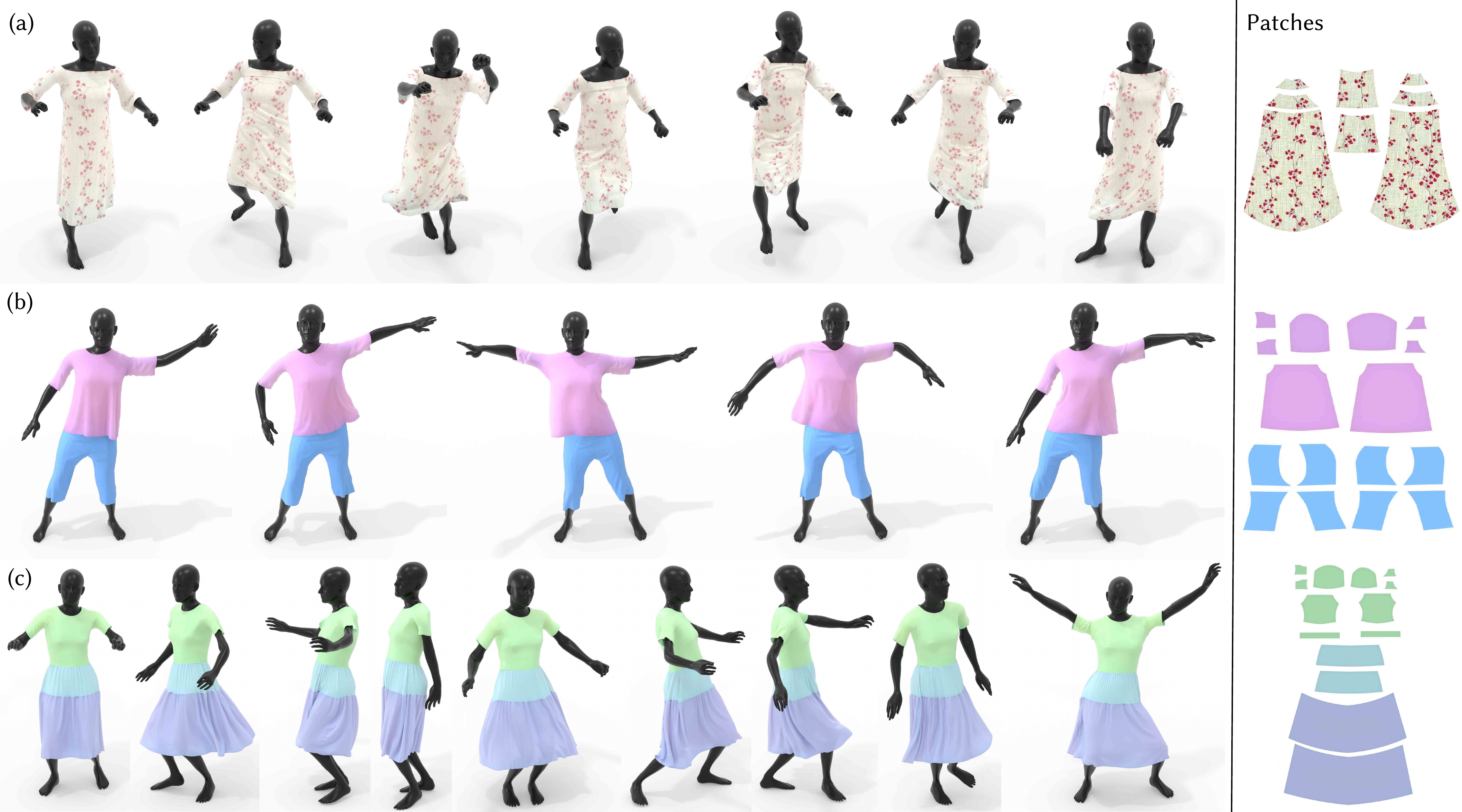}
  \caption{\textbf{Garments on character animation.} Snapshots from three simulation sequences of garments on character animations, \rev{along with the rest shapes of cloth patches (right)}. In (a), the character performs a running motion wearing a long floral dress, generating smooth fabric swing. In (b), the character raises and rotates the arms while wearing a fitted T-shirt and shorts, producing deformations around the shoulders, elbows, and torso. In (c), the character executes twisting and side-stepping motions in a fitted top and a vertically segmented skirt, exhibiting lateral sway and rotational folds in the fabric. With our mesh-conforming pipeline, the method robustly handles non-rectangular garment meshes under dense contact while producing intricate and realistic wrinkle patterns.}
  \label{fig: character animation}
\end{figure*}

\paragraph{Linear Solves with Partial Factorization.}
We evaluate the per-iteration speedup and its impact on per-timestep Newton convergence of our partial factorization scheme in scenarios with few active contact pairs (frame 555 of~\autoref{fig: brush over armadillo}), where the scheme is triggered more frequently (see~\autoref{fig: partial factorization}). (A) shows the runtime decomposition when partial factorization is activated: by reusing the fixed symbolic factorization of the elasticity energy Hessian and avoiding costly Hessian merges due to varying contact sparsity, thereby eliminating the expensive symbolic reanalysis phase. In contrast, the baseline direct solve in (B) incurs this overhead. This results in an approximate $2 \times$ reduction in per-iteration solve time. The scheme is triggered in roughly 40\% of iterations under the tested contact conditions. Although partial factorization introduces a minor increase in the number of nonlinear iterations per timestep (due to approximations in linear solves), the substantial per-iteration savings lead to improved or comparable overall per-timestep runtime.

\section{Conclusion}
We present a B-spline finite element method for cloth simulation that combines the geometric smoothness of quadratic B-spline bases with the efficiency of reduced integration and fast parallel Hessian assembly. Our method achieves high accuracy and visual fidelity while reducing computational cost compared to linear and second-order FEM approaches. Through extensive evaluations, we demonstrate its robustness across a wide range of scenarios, including frictional contact interactions and varying material settings. We believe our work opens new opportunities for smooth, efficient, and artifact-free cloth simulation, and offers a compelling spatial discretization framework for future advancements.

\paragraph{Limitations and Future Work} Our method inherits several limitations from the use of B-spline surfaces. 
First, the surface boundary generally does not interpolate control points, which may complicate enforcing Dirichlet boundary conditions under highly curved configurations.
Second, the tensor-product structure limits the ability to represent complex topologies. {Although with our mesh conversion pipeline we can fit B-spline surfaces to target linear triangle meshes, accuracy of quadrature may be compromised if the target mesh deviates a lot from rectilinear shapes. This could potentially result in ill-conditioned systems or harm the convergence of Newton's method. } 
Third, although membrane locking is effectively mitigated and no visual artifacts like sharp creases exist compared to linear FEM, our method is still not locking-free. It would be meaningful to investigate how strain limiting further improves our results. {Our choice of B-spline basis may also influence the resulting geometry. Because each B-spline basis supports three knot spans, the formulation inherently favors low-frequency deformation modes, which can lead to the suppression of fine geometric details.}
Currently, contact handling and rendering are performed on the same triangle mesh embedded in the B-spline surface. {While our optimized Hessian assembly algorithm significantly reduces the associated overhead, this approach still incurs additional cost during linear system construction. Exploring contact handling directly on the B-spline surface remains an interesting and meaningful direction. }
\rev{Finally, our current method assumes a flat rest shape with near-isometric in-plane deformation. This restricts its applicability to garments whose natural configuration is approximately developable, excluding for example cone-shaped skirts or garments with wrinkled, creased, or otherwise non-flat rest shapes. Lifting this assumption would require incorporating non-zero reference curvature into the shell energy formulation and allowing for non-isometric in-plane deformation, which we leave as important directions for future work.}

\begin{acks}
We sincerely thank Qiqin Le, Qixin Liang, and Xingyu Ni for their insightful discussions and assistance. Yihao Shi was partially supported by a visiting scholarship from Zhejiang University. Minchen Li acknowledges partial support from a Junior Faculty Startup Fund from Carnegie Mellon University and gift funding from Genesis AI. Yin Yang acknowledges partial support from NSF Award No. 2301040. Taku Komura acknowledges partial support from the Innovation and Technology Commission of the HKSAR Government under the ITSP-Platform grants (Refs. ITS/335/23FP and ITS/469/24FP).
\end{acks}

\bibliographystyle{ACM-Reference-Format}
\bibliography{main}

\includepdf[pages=-]{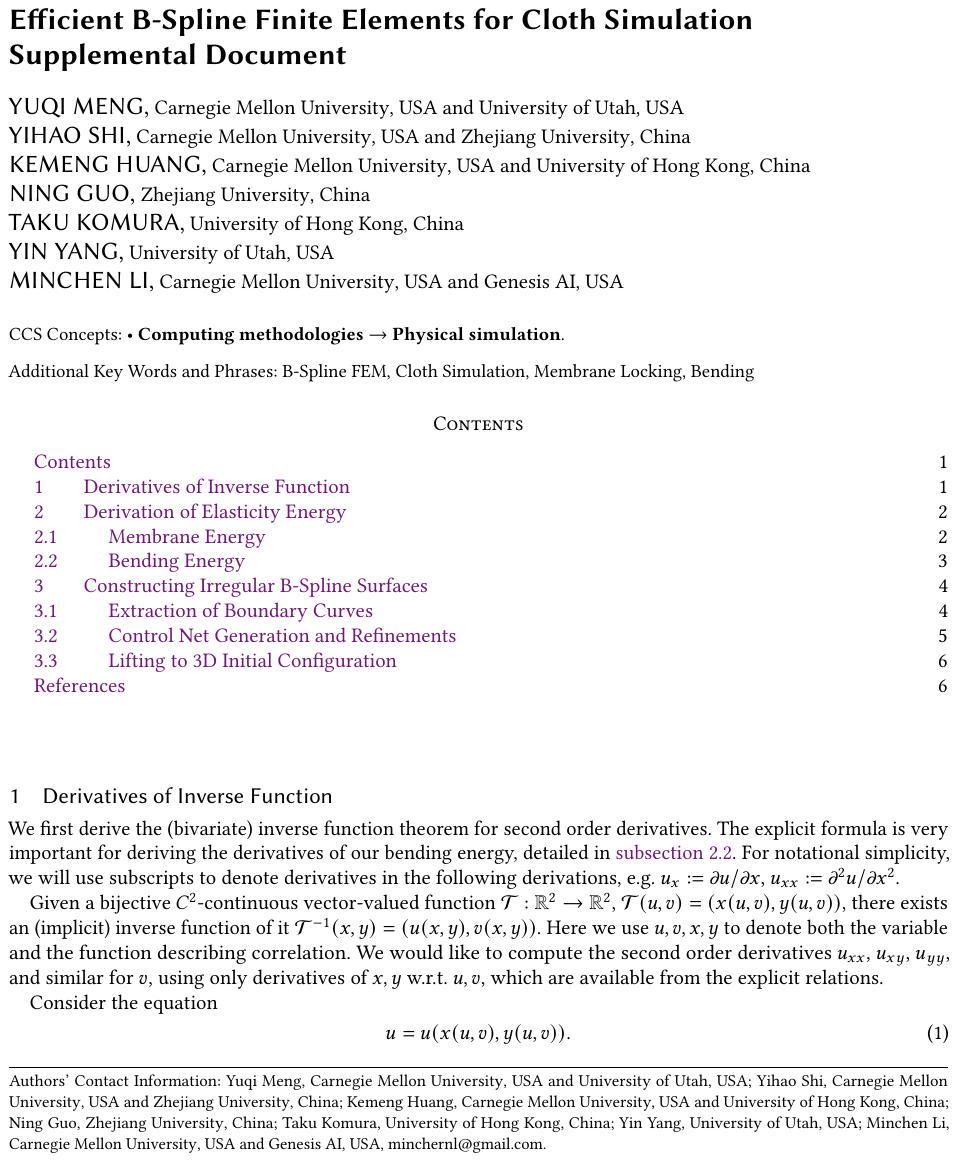}



\end{document}

%% file: tikz-scripts/Quadrature/MembraneQuadrature.tex
\definecolor{node}{HTML}{c77dff}
\definecolor{patchw}{HTML}{00b4d8}
\definecolor{patchb}{HTML}{0077b6}
\definecolor{redgrid}{HTML}{e76f51}
\definecolor{bluegrid}{HTML}{4361ee}

\draw[->, color=black!90, thick] (-0.2,0)--(5.3,0) node[right]{\LARGE $u$};
\draw[->, color=black!90, thick] (0,-0.2)--(0,5.3) node[above]{\LARGE $v$};

\foreach \x in {0, 1, ..., 5} {
    \draw[color=black!50] (\x, 0) -- (\x, 5);
    \draw[color=black!50] (0, \x) -- (5, \x);
}
\foreach \x in {0, 1, ..., 4} {
    \foreach \y in {0, 1, ..., 4} {
        \fill[color=node] (\x + 0.5, \y + 0.5) circle (2.5pt);
    }
    \fill[color=node] (\x + 0.5, 0) circle (2.5pt);
    \fill[color=node] (\x + 0.5, 5) circle (2.5pt);
    \fill[color=node] (0, \x + 0.5) circle (2.5pt);
    \fill[color=node] (5, \x + 0.5) circle (2.5pt);
}
\fill[color=node] (0, 0) circle (2.5pt);
\fill[color=node] (0, 5) circle (2.5pt);
\fill[color=node] (5, 0) circle (2.5pt);
\fill[color=node] (5, 5) circle (2.5pt);
\foreach \x in {0,1,...,4} {
    \foreach \y in {0,1,...,4} {
        \ifodd\x
            \ifodd\y
                \fill[patchb, opacity=0.5] (\x,\y) rectangle ++(1,1);
            \else
                \fill[patchw, opacity=0.5] (\x,\y) rectangle ++(1,1);
            \fi
        \else
            \ifodd\y
                \fill[patchw, opacity=0.5] (\x,\y) rectangle ++(1,1);
            \else
                \fill[patchb, opacity=0.5] (\x,\y) rectangle ++(1,1);
            \fi
        \fi
    }
}
\draw[thick, color=black] (0, 0) rectangle ++ (1, 1);
\draw[thick, color=black] (0, 4) rectangle ++ (1, 1);
\draw[thick, color=black] (4, 0) rectangle ++ (1, 1);
\draw[thick, color=black] (4, 4) rectangle ++ (1, 1);
\fill[color=black!50, opacity=0.5] (0, 0) rectangle ++ (4, 1);
\fill[color=black!50, opacity=0.5] (1, 4) rectangle ++ (4, 1);
\fill[color=black!50, opacity=0.5] (0, 1) rectangle ++ (1, 4);
\fill[color=black!50, opacity=0.5] (4, 0) rectangle ++ (1, 4);
\draw[thick, color=redgrid] (0, 1) -- (0, 4);
\draw[thick, color=redgrid] (1, 1) -- (1, 4);
\draw[thick, color=redgrid] (0, 2) -- (1, 2);
\draw[thick, color=redgrid] (0, 3) -- (1, 3);
\draw[thick, color=redgrid, opacity=0.5] (0, 1) -- (1, 1);
\draw[thick, color=redgrid, opacity=0.5] (0, 4) -- (1, 4);
\draw[thick, color=redgrid] (4, 1) -- (4, 4);
\draw[thick, color=redgrid] (5, 1) -- (5, 4);
\draw[thick, color=redgrid] (4, 2) -- (5, 2);
\draw[thick, color=redgrid] (4, 3) -- (5, 3);
\draw[thick, color=redgrid, opacity=0.5] (4, 1) -- (5, 1);
\draw[thick, color=redgrid, opacity=0.5] (4, 4) -- (5, 4);
\draw[thick, color=bluegrid] (1, 0) -- (4, 0);
\draw[thick, color=bluegrid] (1, 1) -- (4, 1);
\draw[thick, color=bluegrid] (2, 0) -- (2, 1);
\draw[thick, color=bluegrid] (3, 0) -- (3, 1);
\draw[thick, color=bluegrid, opacity=0.5] (1, 0) -- (1, 1);
\draw[thick, color=bluegrid, opacity=0.5] (4, 0) -- (4, 1);
\draw[thick, color=bluegrid] (1, 4) -- (4, 4);
\draw[thick, color=bluegrid] (1, 5) -- (4, 5);
\draw[thick, color=bluegrid] (2, 4) -- (2, 5);
\draw[thick, color=bluegrid] (3, 4) -- (3, 5);
\draw[thick, color=bluegrid, opacity=0.5] (1, 4) -- (1, 5);
\draw[thick, color=bluegrid, opacity=0.5] (4, 4) -- (4, 5);
\node at (1, 1) {$\large \bm{\times}$};
\node at (1, 4) {$\large \bm{\times}$};
\node at (4, 1) {$\large \bm{\times}$};
\node at (4, 4) {$\large \bm{\times}$};
\node at (1, 1.711) {$\large \bm{\times}$};
\node at (1, 2.289) {$\large \bm{\times}$};
\node at (1, 2.711) {$\large \bm{\times}$};
\node at (1, 3.289) {$\large \bm{\times}$};
\node at (4, 1.711) {$\large \bm{\times}$};
\node at (4, 2.289) {$\large \bm{\times}$};
\node at (4, 2.711) {$\large \bm{\times}$};
\node at (4, 3.289) {$\large \bm{\times}$};
\node at (1.711, 1) {$\large \bm{\times}$};
\node at (2.289, 1) {$\large \bm{\times}$};
\node at (2.711, 1) {$\large \bm{\times}$};
\node at (3.289, 1) {$\large \bm{\times}$};
\node at (1.711, 4) {$\large \bm{\times}$};
\node at (2.289, 4) {$\large \bm{\times}$};
\node at (2.711, 4) {$\large \bm{\times}$};
\node at (3.289, 4) {$\large \bm{\times}$};
\node at (1.711, 3) {$\large \bm{\times}$};
\node at (2.289, 3) {$\large \bm{\times}$};
\node at (2.711, 2) {$\large \bm{\times}$};
\node at (3.289, 2) {$\large \bm{\times}$};
\node at (2, 1.711) {$\large \bm{\times}$};
\node at (2, 2.289) {$\large \bm{\times}$};
\node at (3, 2.711) {$\large \bm{\times}$};
\node at (3, 3.289) {$\large \bm{\times}$};
\fill[color=black!50, opacity=0.5] (6, 4) rectangle ++ (1.25, 1.25);
\fill[color=black!50, opacity=0.5] (6, 2) rectangle ++ (1.25, 1.25);
\fill[color=black!50, opacity=0.5] (6, 0) rectangle ++ (1.25, 1.25);
\draw[thick, color=black] (6, 4) rectangle ++ (1.25, 1.25);
\draw[thick, color=bluegrid] (6, 2) rectangle ++ (1.25, 1.25);
\draw[thick, color=redgrid] (6, 0) rectangle ++ (1.25, 1.25);
\node at (6.14, 4.14) {$\large \bm{\times}$};
\node at (6.14, 4.625) {$\large \bm{\times}$};
\node at (6.14, 5.11) {$\large \bm{\times}$};
\node at (6.625, 4.14) {$\large \bm{\times}$};
\node at (6.625, 4.625) {$\large \bm{\times}$};
\node at (6.625, 5.11) {$\large \bm{\times}$};
\node at (7.11, 4.14) {$\large \bm{\times}$};
\node at (7.11, 4.625) {$\large \bm{\times}$};
\node at (7.11, 5.11) {$\large \bm{\times}$};
\node at (6.264, 2.14) {$\large \bm{\times}$};
\node at (6.986, 2.14) {$\large \bm{\times}$};
\node at (6.264, 2.625) {$\large \bm{\times}$};
\node at (6.986, 2.625) {$\large \bm{\times}$};
\node at (6.264, 3.11) {$\large \bm{\times}$};
\node at (6.986, 3.11) {$\large \bm{\times}$};
\node at (6.14, 0.264) {$\large \bm{\times}$};
\node at (6.625, 0.264) {$\large \bm{\times}$};
\node at (7.11, 0.264) {$\large \bm{\times}$};
\node at (6.14, 0.986) {$\large \bm{\times}$};
\node at (6.625, 0.986) {$\large \bm{\times}$};
\node at (7.11, 0.986) {$\large \bm{\times}$};
\foreach \x in {1, ..., 3} {
    \draw[dashed, color=black!50] (\x + 0.5, 1) -- (\x + 0.5, 4);
    \draw[dashed, color=black!50] (1, \x + 0.5) -- (4, \x + 0.5);
}

\node[below] at (1, -0.05) {\LARGE 1};
\node[below] at (2, -0.05) {\LARGE 2};
\node[below] at (3, -0.05) {\LARGE 3};
\node[below] at (4, -0.05) {\LARGE 4};
\node[below] at (5, -0.05) {\LARGE 5};
\node[left] at (-0.05, 1) {\LARGE 1};
\node[left] at (-0.05, 2) {\LARGE 2};
\node[left] at (-0.05, 3) {\LARGE 3};
\node[left] at (-0.05, 4) {\LARGE 4};
\node[left] at (-0.05, 5) {\LARGE 5};

\node at (3.625, -1.3) {\LARGE Membrane Quadrature Points};

%% file: tikz-scripts/Quadrature/BendingQuadrature.tex
\definecolor{node}{HTML}{c77dff}
\definecolor{patchw}{HTML}{00b4d8}
\definecolor{patchb}{HTML}{0077b6}
\definecolor{redgrid}{HTML}{e76f51}
\definecolor{bluegrid}{HTML}{4361ee}

\draw[->, color=black!90, thick] (-0.2,0)--(5.3,0) node[right]{\LARGE $u$};
\draw[->, color=black!90, thick] (0,-0.2)--(0,5.3) node[above]{\LARGE $v$};

\foreach \x in {0, 1, ..., 5} {
    \draw[color=black!50] (\x, 0) -- (\x, 5);
    \draw[color=black!50] (0, \x) -- (5, \x);
}
\foreach \x in {0, 1, ..., 4} {
    \foreach \y in {0, 1, ..., 4} {
        \fill[color=node] (\x + 0.5, \y + 0.5) circle (2.5pt);
    }
    \fill[color=node] (\x + 0.5, 0) circle (2.5pt);
    \fill[color=node] (\x + 0.5, 5) circle (2.5pt);
    \fill[color=node] (0, \x + 0.5) circle (2.5pt);
    \fill[color=node] (5, \x + 0.5) circle (2.5pt);
}
\fill[color=node] (0, 0) circle (2.5pt);
\fill[color=node] (0, 5) circle (2.5pt);
\fill[color=node] (5, 0) circle (2.5pt);
\fill[color=node] (5, 5) circle (2.5pt);
\foreach \x in {0,1,...,4} {
    \foreach \y in {0,1,...,4} {
        \ifodd\x
            \ifodd\y
                \fill[patchb, opacity=0.5] (\x,\y) rectangle ++(1,1);
            \else
                \fill[patchw, opacity=0.5] (\x,\y) rectangle ++(1,1);
            \fi
        \else
            \ifodd\y
                \fill[patchw, opacity=0.5] (\x,\y) rectangle ++(1,1);
            \else
                \fill[patchb, opacity=0.5] (\x,\y) rectangle ++(1,1);
            \fi
        \fi
    }
}
\foreach \x in {1, ..., 3} {
    \draw[dashed, color=black!50] (\x + 0.5, 1) -- (\x + 0.5, 4);
    \draw[dashed, color=black!50] (1, \x + 0.5) -- (4, \x + 0.5);
    
}
\foreach \x in {1, ..., 4} {
    \foreach \y in {1, ..., 4} {
        \node at (\x, \y) {$\large \bm{\times}$};
    }
}
\node at (0.5, 0.5) {$\large \bm{\times}$};
\node at (0.5, 1.5) {$\large \bm{\times}$};
\node at (0.5, 2.5) {$\large \bm{\times}$};
\node at (0.5, 3.5) {$\large \bm{\times}$};
\node at (0.5, 4.5) {$\large \bm{\times}$};
\node at (4.5, 0.5) {$\large \bm{\times}$};
\node at (4.5, 1.5) {$\large \bm{\times}$};
\node at (4.5, 2.5) {$\large \bm{\times}$};
\node at (4.5, 3.5) {$\large \bm{\times}$};
\node at (4.5, 4.5) {$\large \bm{\times}$};
\node at (1.5, 0.5) {$\large \bm{\times}$};
\node at (2.5, 0.5) {$\large \bm{\times}$};
\node at (3.5, 0.5) {$\large \bm{\times}$};
\node at (1.5, 4.5) {$\large \bm{\times}$};
\node at (2.5, 4.5) {$\large \bm{\times}$};
\node at (3.5, 4.5) {$\large \bm{\times}$};

\node[below] at (1, -0.05) {\LARGE 1};
\node[below] at (2, -0.05) {\LARGE 2};
\node[below] at (3, -0.05) {\LARGE 3};
\node[below] at (4, -0.05) {\LARGE 4};
\node[below] at (5, -0.05) {\LARGE 5};
\node[left] at (-0.05, 1) {\LARGE 1};
\node[left] at (-0.05, 2) {\LARGE 2};
\node[left] at (-0.05, 3) {\LARGE 3};
\node[left] at (-0.05, 4) {\LARGE 4};
\node[left] at (-0.05, 5) {\LARGE 5};
\node at (2.625, -1.3) {\LARGE Bending Quadrature Points};

%% file: tikz-scripts/QuadratureSupport/QuadratureSupport.tex
\definecolor{node}{HTML}{5E33AD}
\definecolor{patchw}{HTML}{00b4d8}
\definecolor{patchb}{HTML}{0077b6}

\filldraw[color=patchb, fill opacity=0.5] (0, 0) -- (1, 0) -- (1, 1) -- (0, 1) -- (0, 0);
\filldraw[color=patchb, fill opacity=0.5] (2, 0) -- (3, 0) -- (3, 1) -- (2, 1) -- (2, 1);
\filldraw[color=patchb, fill opacity=0.5] (1, 1) -- (1, 2) -- (2, 2) -- (2, 1) -- (1, 1);
\filldraw[color=patchb, fill opacity=0.5] (2, 2) -- (3, 2) -- (3, 3) -- (2, 3) -- (2, 2);
\filldraw[color=patchb, fill opacity=0.5] (0, 2) -- (1, 2) -- (1, 3) -- (0, 3) -- (0, 2);
\filldraw[color=patchw, fill opacity=0.5] (1, 0) -- (2, 0) -- (2, 1) -- (1, 1) -- (1, 0);
\filldraw[color=patchw, fill opacity=0.5] (1, 2) -- (1, 3) -- (2, 3) -- (2, 2) -- (1, 2);
\filldraw[color=patchw, fill opacity=0.5] (0, 1) -- (1, 1) -- (1, 2) -- (0, 2) -- (0, 1);
\filldraw[color=patchw, fill opacity=0.5] (2, 1) -- (2, 2) -- (3, 2) -- (3, 1) -- (2, 1);
\draw[color=gray!50] (0, 0) -- (3, 0) -- (3, 3) -- (0, 3) -- (0, 0);
\draw[color=gray!50] (1, 0) -- (1, 3);
\draw[color=gray!50] (2, 0) -- (2, 3);
\draw[color=gray!50] (0, 1) -- (3, 1);
\draw[color=gray!50] (0, 2) -- (3, 2);
\fill[color=node, opacity=0.3] (0.5, 0.5) circle (2.5pt);
\fill[color=node, opacity=0.3] (0.5, 1.5) circle (2.5pt);
\fill[color=node, opacity=0.3] (0.5, 2.5) circle (2.5pt);
\fill[color=node, opacity=0.3] (1.5, 0.5) circle (2.5pt);
\fill[color=node, opacity=0.3] (1.5, 2.5) circle (2.5pt);
\fill[color=node, opacity=0.3] (2.5, 0.5) circle (2.5pt);
\fill[color=node, opacity=0.3] (2.5, 1.5) circle (2.5pt);
\fill[color=node, opacity=0.3] (2.5, 2.5) circle (2.5pt);


\fill[color=node] (1.5, 1.5) circle (2.5pt);

\foreach \x in {0, ..., 2} {
    \draw[dashed, color=black!50, opacity=0.5] (\x + 0.5, 0) -- (\x + 0.5, 3);
    \draw[dashed, color=black!50, opacity=0.5] (0, \x + 0.5) -- (3, \x + 0.5);
}
\node[opacity=0.4] at (0.711, 0) {\footnotesize $\bm{\times}$};
\node[opacity=0.4] at (1.289, 0) {\footnotesize $\bm{\times}$};
\node[opacity=0.4] at (2.711, 0) {\footnotesize $\bm{\times}$};
\node[opacity=0.4] at (0.289, 3) {\footnotesize $\bm{\times}$};
\node[opacity=0.4] at (1.711, 3) {\footnotesize $\bm{\times}$};
\node[opacity=0.4] at (2.289, 3) {\footnotesize $\bm{\times}$};
\node[opacity=0.4] at (0, 0.289) {\footnotesize $\bm{\times}$};
\node[opacity=0.4] at (0, 1.711) {\footnotesize $\bm{\times}$};
\node[opacity=0.4] at (0, 2.289) {\footnotesize $\bm{\times}$};
\node[opacity=0.4] at (3, 0.711) {\footnotesize $\bm{\times}$};
\node[opacity=0.4] at (3, 1.289) {\footnotesize $\bm{\times}$};
\node[opacity=0.4] at (3, 2.711) {\footnotesize $\bm{\times}$};
\node at (1, 0.711) {\footnotesize $\bm{\times}$};
\node at (1, 1.289) {\footnotesize $\bm{\times}$};
\node at (1, 2.711) {\footnotesize $\bm{\times}$};
\node at (2, 0.289) {\footnotesize $\bm{\times}$};
\node at (2, 1.711) {\footnotesize $\bm{\times}$};
\node at (2, 2.289) {\footnotesize $\bm{\times}$};
\node at (0.289, 1) {\footnotesize $\bm{\times}$};
\node at (1.711, 1) {\footnotesize $\bm{\times}$};
\node at (2.289, 1) {\footnotesize $\bm{\times}$};
\node at (0.711, 2) {\footnotesize $\bm{\times}$};
\node at (1.289, 2) {\footnotesize $\bm{\times}$};
\node at (2.711, 2) {\footnotesize $\bm{\times}$};

%% file: tikz-scripts/QuadratureSupport/VertexSupport.tex
\definecolor{node}{HTML}{5E33AD}
\definecolor{patchw}{HTML}{00b4d8}
\definecolor{patchb}{HTML}{0077b6}

\filldraw[color=patchb, fill opacity=0.5] (0, 0) -- (1, 0) -- (1, 1) -- (0, 1) -- (0, 0);
\filldraw[color=patchb, fill opacity=0.5] (2, 0) -- (3, 0) -- (3, 1) -- (2, 1) -- (2, 1);
\filldraw[color=patchb, fill opacity=0.5] (1, 1) -- (1, 2) -- (2, 2) -- (2, 1) -- (1, 1);
\filldraw[color=patchb, fill opacity=0.5] (2, 2) -- (3, 2) -- (3, 3) -- (2, 3) -- (2, 2);
\filldraw[color=patchb, fill opacity=0.5] (0, 2) -- (1, 2) -- (1, 3) -- (0, 3) -- (0, 2);
\filldraw[color=patchw, fill opacity=0.5] (1, 0) -- (2, 0) -- (2, 1) -- (1, 1) -- (1, 0);
\filldraw[color=patchw, fill opacity=0.5] (1, 2) -- (1, 3) -- (2, 3) -- (2, 2) -- (1, 2);
\filldraw[color=patchw, fill opacity=0.5] (0, 1) -- (1, 1) -- (1, 2) -- (0, 2) -- (0, 1);
\filldraw[color=patchw, fill opacity=0.5] (2, 1) -- (2, 2) -- (3, 2) -- (3, 1) -- (2, 1);
\draw[color=gray!50] (0, 0) -- (3, 0) -- (3, 3) -- (0, 3) -- (0, 0);
\draw[color=gray!50] (1, 0) -- (1, 3);
\draw[color=gray!50] (2, 0) -- (2, 3);
\draw[color=gray!50] (0, 1) -- (3, 1);
\draw[color=gray!50] (0, 2) -- (3, 2);
\fill[color=node, opacity=0.3] (0, 0) circle (2pt);
\fill[color=node, opacity=0.3] (0, 0.5) circle (2pt);
\fill[color=node, opacity=0.3] (0, 1.5) circle (2pt);
\fill[color=node, opacity=0.3] (0, 2.5) circle (2pt);
\fill[color=node, opacity=0.3] (0, 3) circle (2pt);
\fill[color=node, opacity=0.3] (0.5, 0) circle (2pt);
\fill[color=node, opacity=0.3] (0.5, 0.5) circle (2pt);
\fill[color=node, opacity=0.3] (0.5, 1.5) circle (2pt);
\fill[color=node, opacity=0.3] (0.5, 2.5) circle (2pt);
\fill[color=node, opacity=0.3] (0.5, 3) circle (2pt);
\fill[color=node, opacity=0.3] (1.5, 0) circle (2pt);
\fill[color=node] (1.5, 0.5) circle (2pt);
\fill[color=node] (1.5, 1.5) circle (2pt);
\fill[color=node] (1.5, 2.5) circle (2pt);
\fill[color=node, opacity=0.3] (1.5, 3) circle (2pt);
\fill[color=node, opacity=0.3] (2.5, 0) circle (2pt);
\fill[color=node] (2.5, 0.5) circle (2pt);
\fill[color=node] (2.5, 1.5) circle (2pt);
\fill[color=node] (2.5, 2.5) circle (2pt);
\fill[color=node, opacity=0.3] (2.5, 3) circle (2pt);
\fill[color=node, opacity=0.3] (3, 0) circle (2pt);
\fill[color=node] (3, 0.5) circle (2pt);
\fill[color=node] (3, 1.5) circle (2pt);
\fill[color=node] (3, 2.5) circle (2pt);
\fill[color=node, opacity=0.3] (3, 3) circle (2pt);

\foreach \x in {0, ..., 4} {
    \draw[opacity=0.2, color=black!80] (\x * 0.75, 0) -- (\x * 0.75, 3);
    \draw[opacity=0.2, color=black!80] (0, \x * 0.75) -- (3, \x * 0.75);
    \foreach \y in {0, ..., 4} {
        \fill[opacity=0.25, color=black!80] (\x * 0.75 - 0.075, \y * 0.75 - 0.075) rectangle ++(0.15, 0.15);
    }
}
\draw[opacity=0.2, color=black!80] (0.0, 0.0) -- (0.75, 0.75);
\draw[opacity=0.2, color=black!80] (1.5, 0.0) -- (0.75, 0.75);
\draw[opacity=0.2, color=black!80] (1.5, 0.0) -- (2.25, 0.75);
\draw[opacity=0.2, color=black!80] (3.0, 0.0) -- (2.25, 0.75);
\draw[opacity=0.2, color=black!80] (0.0, 1.5) -- (0.75, 0.75);
\draw[opacity=0.2, color=black!80] (1.5, 1.5) -- (0.75, 0.75);
\draw[opacity=0.2, color=black!80] (1.5, 1.5) -- (2.25, 0.75);
\draw[opacity=0.2, color=black!80] (3.0, 1.5) -- (2.25, 0.75);
\draw[opacity=0.2, color=black!80] (0.0, 1.5) -- (0.75, 2.25);
\draw[opacity=0.2, color=black!80] (1.5, 1.5) -- (0.75, 2.25);
\draw[opacity=0.2, color=black!80] (1.5, 1.5) -- (2.25, 2.25);
\draw[opacity=0.2, color=black!80] (3.0, 1.5) -- (2.25, 2.25);
\draw[opacity=0.2, color=black!80] (0.0, 3.0) -- (0.75, 2.25);
\draw[opacity=0.2, color=black!80] (1.5, 3.0) -- (0.75, 2.25);
\draw[opacity=0.2, color=black!80] (1.5, 3.0) -- (2.25, 2.25);
\draw[opacity=0.2, color=black!80] (3.0, 3.0) -- (2.25, 2.25);

\fill[opacity=0.9, color=black!80] (2.175, 1.425) rectangle ++(0.15, 0.15);
